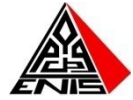

# MEMOIRE

*Présenté à*

## L'École Nationale d'Ingénieurs de Sfax

*en vue de l'obtention du*

## MASTÈRE

### INFORMATIQUE
### *NTSID*

**Par**

## Farah FOURATI

---

### Une approche IDM
### de transformation exogène
### de Wright vers Ada

---

**Soutenu le 11 Juin 2010, devant le jury composé de :**

| | | | |
|---|---|---|---|
| **M.** | **Ahmed HADJ  KACEM** | (MC, FSEGS) | *Président* |
| **M.** | **Kais HADDAR** | (MA, FSS) | *Membre* |
| **M.** | **Mohamed Tahar BHIRI** | (MA, FSS) | *Encadreur* |

# Table des matières

# Liste des figures

# Glossaire

**ADL**

Architecture Description Language

**AL**

architecture logicielle

**ATL**

ATLAS Transformation Language

**CIM**

Computational Independent Model

**CSP**

Communicating Sequential
Processes

**DSL**

Domain Specific Languages

**EBNF**

Extended Backus-Naur Form

**EMF**

Eclipse Modeling Framework

**EMFT**

EMF Technology

**EMOF**

Essential MOF

**IDE**

Integrated Development
Environment

**IDM**

Ingénierie Dirigée par les Modèles

**JVM**

Java Virtual Machine

**MDA**

Model Driven Architecture

**MDE**

Model-Driven Engineering

**MOF**

Meta-Object Facility

**MWE**

Modeling Workflow Engine

**OCL**

Object Constraint Language

**PDM**

Platform Description Model

**PIM**

Platform Independent Model

**PSM**

Platform Specific Model

**QVT**

Querry View Transformation

**TMF**

Textual Modeling Framework

**XMI**

XML Metadata Interchange

**XML**

Extensible Markup Language

*À tous ceux qui m'aiment,*
*À tous ceux que j'aime,*
*À mes chers parents et à mon cher frère,*
*À la mémoire de mes grands parents.*

# Remerciements

**Merci à** monsieur **Mohamed Tahar Bhiri**, maître assistant à la Faculté des Sciences  de Sfax, pour avoir accepté de m'encadrer tout au long de ce travail. Il m'a donnée l'opportinuté de travailler sur un sujet passionnant. Il a dirigé mes recherches dans les moindres détails. Ses conseils ont été d'un apport considérable dans un souci de mieux s'inverstir dans le fond du sujet.

**Merci à** monsieur **Ahmed Hadj Kacem**, maître de conférences à la Faculté des Sciences Economiques et de Gestion de Sfax, qui m'a fait l'honneur de présider le jury de ce mémoire.

**Merci à** monsieur **Kais Haddar**, maître assistant à la Faculté des Sciences  de Sfax,  qui m'a fait l'honneur d'évaluer mon mémoire.

**Merci à** tout le **corps enseignant** qui a contribué à ma formation.

Enfin, **merci à** tous les membres de ma **famille** et à mes **amis** pour leur soutien tout au long de mon travail.

# Introduction

Le domaine de l'architecture logicielle est devenu un champ à part entière au niveau du génie logiciel. L'architecture logicielle (AL) fournit une description de haut niveau de la structure d'un système. Elle est définie par des composants, des connecteurs et des configurations. Elle offre de nombreux avantages tout au long du cycle de vie du logiciel. Elle facilite le passage de l'étape de conception à l'étape d'implémentation. De même, lors de l'étape de maintenance (corrective et évolutive), elle facilite la localisation des erreurs et l'extensibilité du logiciel.

La communauté scientifique a développé plusieurs formalismes permettant la spécification plus ou moins précise des descriptions architecturales: langages de description d'architectures (ADL) : (Oussalah, 2005) et (Haddad, 2006).

L'ADL Wright est un ADL formel. Il propose des concepts tels que composant, connecteur, port, rôle et configuration permettant de décrire les aspects structuraux d'une AL. De même, il supporte CSP de Hoare (Hoare, 1985) afin de décrire les aspects comportementaux d'une AL. La sémantique formelle de Wright est basée sur celle de CSP.

Des travaux (Graiet, 2007) (Bhiri, 2008) dirigés par l'encadreur de ce mastère ont permis d'ouvrir l'ADL Wright sur le langage Ada. Le travail décrit dans (Bhiri, 2008) constitue la référence de base de ce mémoire. Il propose des règles systématiques permettant de transformer une architecture logicielle décrite en Wright vers un programme concurrent Ada comportant plusieurs tâches exécutées en parallèle. L'ouverture de Wright sur Ada a un double objectif : permettre de raffiner une architecture logicielle décrite en Wright et d'utiliser des outils de vérification formelle associés à Ada tels que FLAVERS (Cobleigh, 2002).

L'objectif de ce mémoire est de proposer une approche IDM basée sur le principe « tout est modèle » (Bézivin, 2004) afin de valider et d'automatiser les règles de transformation de Wright vers Ada décrites dans (Bhiri, 2008). Pour y parvenir, nous avons élaboré deux méta-modèles : le méta-modèle de Wright jouant le rôle de méta-modèle source et le méta-modèle partiel d'Ada jouant le rôle de méta-modèle cible pour l'opération de transformation exogène de Wright vers Ada. De plus, nous avons conçu et réalisé un programme Wright2Ada écrit en ATL (Jouault, 2006) (ATL) permettant de transformer un modèle source conforme au méta-modèle Wright vers un modèle cible conforme au méta-modèle partiel d'Ada. En outre, nous avons doté notre programme Wright2Ada par des interfaces conviviales afin de l'utiliser dans un contexte réel. Pour ce faire, nous avons utilisé avec profit les outils de modélisation Xtext (Xtext), Xpand et Check (Xpand). Enfin, nous avons testé notre programme Wright2Ada en utilisant une approche orientée tests syntaxiques (Xanthakis, 1999).

Ce mémoire comporte sept chapitres. Le premier chapitre présente l'ADL Wright, la concurrence en Ada et les règles de transformation de Wright vers Ada. Le second chapitre présente les principes généraux de l'IDM et les outils de modélisation utilisés dans ce travail à savoir : Ecore, XMI, Check, Xtext, ATL et Xpand. Les deux chapitres trois et quatre proposent respectivement les deux méta-modèles source et cible : le méta-modèle de Wright et le méta-modèle partiel d'Ada. Le cinquième chapitre propose un programme Wright2Ada permettant de transformer une architecture logicielle décrite en Wright vers un programme concurrent Ada. Le sixième chapitre propose des transformations IDM permettant d'avoir des interfaces conviviales afin d'utiliser notre programme Wright2Ada dans un contexte réel. Enfin le dernier chapitre préconise une approche basée sur le test fonctionnel permettant d'augmenter la confiance dans notre programme Wright2Ada.

Les six annexes de ce mémoire indiquent; en A, le programme ATL de transformation Wright2Ada, en B, un exemple d'utilisation de Wright2Ada, en C, la grammaire Xtext de Wright, en D, le modèle XMI conforme à la grammaire de Wright, en E, le programme ATL de transformation GrammaireWright2Wright, en F, le template Xpand de génération de code Ada.

# Chapitre 1 : Traduction d'une architecture logicielle Wright vers Ada

L'ADL (Architectural Description language) formel Wright permet de décrire des architectures abstraites ayant des propriétés bien définires. Mais, il ne supporte pas d'outils permettant d'implémenter ces architectures abstraites. (Bhiri, 2008) et (Graiet, 2007) proposent une approche favorisant une traduction systématique de Wright vers Ada en considérant une architecture abstraite décrite en Wright comme un programme concurrent Ada. Ainsi, l'architecture abstraite obtenue en Ada peut être raffinée step-by-step en utilisant des outils de vérification formelle associés à Ada tel que FLAVERS (Cobleigh, 2002).

Ce chapitre est composé de trois sections. La première section présente les aspects structuraux et comportementaux de l'ADL formel Wright. La deuxième section présente la concurrence en Ada. Dans ces deux sections, nous nous limitons à la présentation des concepts utilisés dans la traduction d'une architecture Wright vers un programme concurrent Ada. Enfin, la troisième section, présente la traduction de Wright vers Ada présentée dans (Bhiri, 2008).

## 1.1 L'ADL Wright

Le langage Wright a été développé à l'université Carnegie Mellon en 1997 (Allen, 1997). Il décrit les composants et les connecteurs comme étant des éléments de première classe d'une architecture. « Il fournit une base formelle pour la description des configurations et des styles architecturaux » (Allen, 1997). En effet, il utilise une notation basée sur CSP (Hoare, 1985) pour décrire le comportement et les interactions, ce qui permet de définir une sémantique formelle et de rendre possible un certain nombre d'analyse et de vérification (El Boussaidi, 2006) (Graiet, 2007).

Dans la suite, nous présentons les aspects structuraux et comportementaux de l'ADL Wright

### 1.1.1 Les aspects structuraux de Wright

L'ADL Wright fournit une spécification explicite pour le composant, le connecteur et la configuration. Ces derniers seront détaillés dans les différentes parties de cette sous-section.

#### 1.1.1.1 Un composant Wright

Un composant est une unité abstraite et indépendante. Il possède un type. Il comporte deux parties: la partie interface composée d'un ensemble de ports qui fournissent les points d'interactions entre le composant et son environnement, et la partie calcul qui décrit le comportement réel du composant.

Les ports *port* sont basés sur un ensemble d'événements émis et reçus.

La partie calcul *computation* décrit les traitements réalisés et les événements produits en réaction aux sollicitations reçues des ports.

Exemple :

Dans une architecture Filtre et Canal (*pipe and filter*), considérons un exemple de deux filtres (Filtre1, Filtre2), le premier reçoit un flux de caractères en entrée et le transmet en sortie au second filtre.

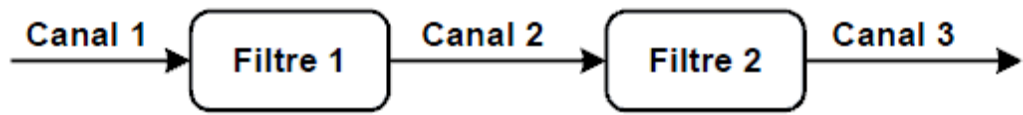

**Figure 1**: *Exemple de filtre et canal*

Le filtre correspond à un composant qui peut être simplement décrit comme suit :

**Component** Filtre
    **Port** Entrée *[Lire les données jusqu'à ce que la fin des données soit atteinte]*
    **Port** Sortie *[sortir les données de manière permanente]*
**Computation** *[Lire continuellement les données à partir du port* Entrée*, puis les envoyer sur le port* Sortie*]*

Les spécifications du comportement des ports et de la partie calcul sont écrites ici de manière informelle pour plus de compréhension. Elles sont normalement écrites avec le langage CSP.

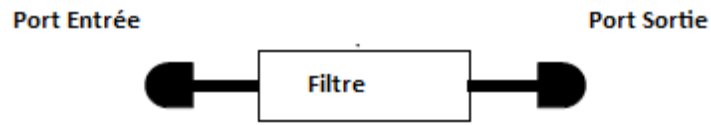

**Figure 2** *: Schéma illustratif du composant Filtre.*

### 1.1.1.2 Un connecteur Wright

Un connecteur représente une interaction explicite et abstraite entre une collection de composants. Il possède un type. Il comporte deux parties : une interface constituée de points d'interactions appelés rôles (*Role*) et une partie qui représente la spécification d'assemblages (*Glue*).

Le rôle indique comment se comporte un composant qui participe à l'interaction.

La glu spécifie les règles d'assemblage entre un ensemble de composants pour former une interaction.

Dans l'exemple précédent, le canal a deux rôles (source et récepteur), la partie «Glue» décrit comment les données sont délivrées du rôle Source au rôle Récepteur. La description avec Wright est comme suit :

**Connector** Canal
    **Role** Source *[produire continuellement les données, signaler la fin par une fermeture]*
    **Role** Récepteur *[Lire continuellement les données, fermer à la fin ou avant la fin des données]*
**Glue** *[*Récepteur *reçoit les données dans le même ordre qu'elles sont produites par* Source*]*

Les spécifications des rôles et de la partie assemblage sont écrites ici de manière informelle pour plus de compréhension. Elles sont normalement écrites avec le langage CSP.

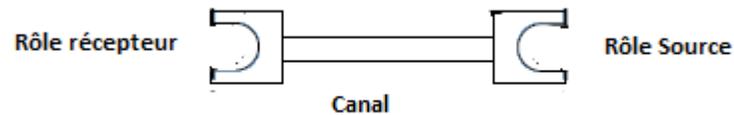

**Figure 3** *: Schéma illustratif du composant Filtre.*

### 1.1.1.3 Une configuration Wright

« Une Configuration exprime la composition des briques logicielles, celles-ci pouvant être horizontales (interconnections) ou verticales (abstraction / encapsulation)» (Déplanche, 2005).

Une configuration représente donc, l'architecture complète du système, celle-ci est décomposée en trois parties. La première définit les types des composants et connecteurs. La deuxième (*Instances*) spécifie l'ensemble des instances de composants et de connecteurs nommées de façon explicite et unique. La troisième (*Attachments*) décrit les liaisons entre les ports des instances de composants et les rôles des instances de connecteurs formant ainsi un graphe biparti.

L'application correspondante à l'exemple présenté précédemment a la configuration suivante :

**Configuration** Filtre-Canal
       **Component** Filtre
       **Connector** Canal
**Instances**
       Filtre1, Filtre2 : Filtre
       Canal1 : Canal
**Attachments**
       Filtre1.Sortie **as** C1.Source
       Filtre2.Entrée **as** C1.Récepteur
**End Configuration**

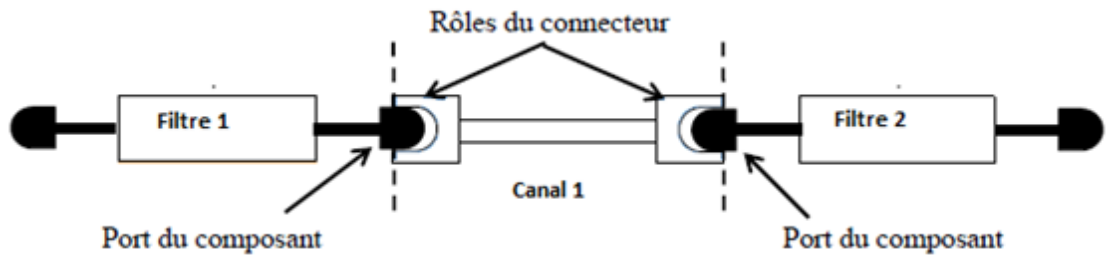

**Figure 4**: *Schéma illustratif de la configuration Filtre-Canal.*

Les ports et les rôles associés dans la troisième partie *Attachments* doivent être compatibles. Cela signifie que le port est conforme aux règles imposées par le rôle pour pouvoir participer à l'interaction spécifiée par le connecteur. Les spécifications des ports et des rôles sont nécessaires pour la vérification de la compatibilité. (El Boussaidi, 2006) (Graiet, 2007).

## 1.1.2 Les aspects comportementaux de Wright

La formulation du comportement des composants et des connecteurs de façon informelle ne permet pas de prouver des propriétés non triviales sur l'architecture d'un système. Ainsi, pour spécifier le comportement et la coordination des composants et des connecteurs, Wright utilise une notation formelle basée sur le processus CSP (Communicating Sequential Processes) (Hoare, 1985).

CSP est un modèle mathématique qui a pour but de formaliser la conception et le comportement de systèmes qui interagissent avec leur environnement de manière permanente. Il est basé sur des solides fondements mathématiques qui permettent une analyse rigoureuse.

Nous ne présentons que les notions essentielles de CSP utilisés dans Wright.

### 1.1.2.1 Les événements

Dans le modèle CSP, tout est représenté par des événements. Un événement correspond à un moment où une action qui présente un intérêt. CSP ne fait pas la distinction entre les événements initialisés et observés. Mais, CSP pour Wright le fait

: Un événement initialisé s'écrit sous la forme ē ou _e. Un événement observé est noté e. e représente le nom de l'événement. De plus, les événements peuvent transmettre des données : e?x et e!x, représentent respectivement les données d'entrée et de sortie.

CSP définit un événement particulier noté √, qui indique la terminaison de l'exécution avec succès.

### 1.1.2.2 Les processus

Pour définir un comportement, il faut pouvoir combiner les événements. Un processus correspond à la modélisation du comportement d'un objet par une combinaison d'événements et d'autre processus simples. Les principaux opérateurs fournis par CSP sont :

- L'opérateur préfixe noté →: Le séquencement ou le préfixage est la façon la plus simple de combiner des événements. Un processus qui s'engage dans un événement e, puis se comporte comme le processus P, est noté « e→P ».

- La récursion : Par la possibilité de nommer un processus, il devient possible de décrire les comportements répétitifs très facilement. Nous décrivons par exemple le processus qui ne s'engage que dans l'événement e et qui ne s'arrête jamais par : P=e→P

- L'opérateur de choix externe ou déterministe noté □: Si nous avons le processus e→P□u→Q et que l'environnement s'engage dans l'événement u, alors le processus s'engagera dans cet événement et se comportera comme le processus Q. Ce choix est typiquement utilisé entre des événements observés. L'environnement se réfère à d'autres processus qui interagissent avec le processus en question.

- L'opérateur de choix interne ou non déterministe noté ⊓: A l'inverse du choix déterministe, c'est le processus qui choisit de façon non déterministe le comportement à choisir parmi plusieurs. Cette fois le processus ē→P⊓ū→Q va choisir entre initialiser l'événement e et continuer comme P ou initialiser u et continuer comme Q. Il décide lui-même de ce choix sans se préoccuper de l'environnement.

- L'alphabet : l'alphabet fait référence à un processus et est noté αP, pour le processus P. L'alphabet d'un processus est l'ensemble des événements sur lequel le processus a une influence.

**Remarque :** Le symbole § désigne le processus de terminaison avec succès, ce qui veut dire que le processus s'est engagé dans un événement succès √ et s'est arrêté. Formellement, §= √→STOP (En CSP il est généralement noté «SKIP»).

## 1.2 La concurrence en Ada

Le langage Ada (Booch, 1991) (Le Verrand, 1982) a été conçu par l'équipe dirigée par Jean Ichbiah pour répondre aux besoins du Département de la Défense américain (DoD). Une révision d'Ada a conduit à la création d'un nouveau langage, appelé Ada 95 (DE Bondeli, 1998) (Rousseau, 1994). Tout au long de ce mémoire, le nom Ada sans qualification supplémentaire fait référence à la version précédente, Ada83, qui est la plus couramment utilisée de nos jours.

Le langage Ada offre des possibilités importantes vis-à-vis de la programmation structurée, la programmation modulaire, la programmation générique et la gestion des exceptions (Graiet, 2007).

### 1.2.1 Les tâches en Ada

Les tâches Ada permettent d'avoir des entités qui s'exécutent parallèlement et qui coopèrent selon les besoins pour résoudre les problèmes concurrents (ou non séquentiels). Une tâche (**task**) est constituée d'une spécification décrivant l'interface présentée aux autres tâches et d'un corps décrivant son comportement dynamique.

```
with Text_Io;                          task T2;
use Text_Io;                           task body T2 is
procedure Ord is                       begin
task T1;                                    loop
task body T1 is                                  Put("T2");
```

| | |
|---|---|
| **begin**<br>      **loop**<br>          Put("T1");<br>      **end loop;**<br>**end** T1; |       **end loop;**<br>**end** T2;<br>**begin**<br>      null ;<br>**end** Ord; |

**Figure 5**:*Exemple illustrant les tâches indépendantes en Ada*

L'activation d'une tâche est automatique. Dans l'exemple précédent, les tâches locales T1 et T2 sont deux tâches actrices c. à d. qu'elles ne proposent pas des possibilités de rendez-vous. Elles deviennent actives quand l'unité parente atteint le **begin** qui suit la déclaration des tâches. Ainsi, ces tâches, dépendantes, sont activées en parallèle avec la tâche principale, qui est le programme principal. La tâche principale attend que les tâches dépendantes s'achèvent pour terminer. La terminaison comporte donc deux étapes : elle s'achève quand elle atteint le **end** final, puis elle ne deviendra terminée que lorsque les tâches dépendantes, s'il y en a, seront terminées.

## 1.2.2 Mécanisme de rendez-vous simple

Les tâches Ada peuvent interagir entre elles au cours de leurs existences, ce mécanisme est appelé rendez-vous. Un rendez-vous entre deux tâches arrive comme conséquence d'un appel de l'entrée (**entry**) d'une tâche par une autre. L'acceptation du rendez-vous se fait par l'instruction **accept**.

Sur le plan conceptuel, les tâches peuvent être classées en trois catégories :

- Les tâches serveuses qui proposent des rendez-vous et n'en demandent pas : par exemple la tâche Allocateur (voir figure 6).

- Les tâches actrices qui ne proposent pas des rendez-vous. Mais elles en demandent : par exemple la tâche T1 ou T2.

- Les tâches actrices/serveuses qui proposent et demandent des rendez-vous (voir figure 6).

| | |
|---|---|
| **With** Text_Io; | **task body** T1 **is** |

```
use Text_Io;                              begin
procedure Ecran is                               Allocateur.Prendre;
task T1;                                          for I in 1 .. 5
task T2;                                          loop
task Allocateur is                                       Put("T1");
        entry Prendre;                            end loop;
        entry Relacher;                           Allocateur.Relacher;
end Allocateur;                           end T1;
task body Allocateur is                   task body T2 is
begin                                     begin
        loop                                      Allocateur.Prendre;
                select                            for I in 1 .. 5
                        accept Prendre;           loop
                        accept Relacher;                  Put("T2");
                or                                end loop;
                        terminate;                Allocateur.Relacher;
                end select;               end T2;
        end loop;                         begin
end Allocateur;                                   null;
                                          end Ecran;
```

**Figure 6**: *Exemple illustrant les rendez-vous en Ada*

Les entrées exportées par une tâche Ada indiquent des possibilités des rendez-vous (ici Allocateur propose deux types de rendez-vous : Prendre et Relacher). Un rendez-vous nécessite deux partenaires : la tâche appelante (ici T1 ou T2) et la tâche appelé (ici Allocateur). La tâche appelante désigne explicitement la tâche qu'elle va appeler (*nom_tâche_Appelée.entrée_tâche_appelée*), ce qui n'est pas le cas pour la tâche appelé (**entry** *nom_entrée*). Un tel mécanisme de rendez-vous est appelé désignation asymétrique.

Une tâche Ada ne peut traiter qu'un seul rendez-vous à la fois, c'est le principe d'exclusion mutuelle sur les rendez-vous. Dans l'exemple précédent, l'instruction **for** utilise l'écran à accès exclusif ; c'est une section critique. La tâche Allocateur est bloquée à l'infini sur le rendez-vous Prendre. La terminaison de cette tâche doit être programmée explicitement par l'instruction **terminate** dans l'instruction **select**.

Généralement, l'instruction **select** permet à la tâche appelée de sélectionner arbitrairement une demande de rendez-vous parmi les rendez-vous disponibles. C'est le non déterminisme dans le langage Ada. Le format de cette instruction se présente comme suit : **select**

      **accept** ENTREE_A;

      …

**or**

      **accept** ENTREE_B;

      …

**or**

      **accept** ENTREE_C;

      …

-- etc.

**end select**;

      Chaque entrée (ou type de rendez-vous) proposée par une tâche est dotée d'une file d'attente FIFO (Structure de Données First In First Out) gérée par l'exécutif d'Ada. Cette file FIFO permet de stoker les demandes sur cette entrée dans l'ordre d'arrivée.

## 1.3 De l'ADL Wright vers le programme concurrent Ada

      Dans cette section nous présentons la contribution de (Bhiri, 2008) permettant de traduire d'une façon systématique une architecture logicielle formalisée en Wright vers Ada. Une telle contribution comporte un ensemble de règles permettant de traduire les constructions Wright (configuration, composant, connecteur et processus CSP) en Ada. Le code Ada à générer correspond à l'architecture de l'application. Nous allons suivre une démarche descendante pour présenter le processus de traduction de Wright vers Ada.

### 1.3.1 Traduction d'une configuration

      Une configuration Wright est traduite en Ada par un programme concurrent dans lequel :

– chaque instance de type composant est traduite par une tâche Ada.

– chaque instance de type connecteur est traduite également par une tâche Ada.

– les tâches de même type ne communiquent pas entre elles.

La figure 7 illustre le principe de la traduction d'une configuration Wright en Ada. Pour des raisons de traçabilité, nous gardons les mêmes identificateurs utilisés dans la spécification Wright. En plus, pour favoriser des retours en arrière, − d'Ada vers Wright − nous transportons la nature de chaque instance soit Component, soit Connector.

| *Spécification en Wright* | *Code Ada* |
|---|---|
| **Configuration** ClientServeur | **procedure** ClientServeur **is** |
| **Component** Client | **task** Component_c **is** |
| **Component** Serveur | **end** Component_c ; |
| **Connector** CS | **task** Component_s **is** |
| **Instances** | **end** Component_s; |
| c : Client | **task** Connector_cls **is** |
| s : Serveur | **end** Connector_cls; |
| cls : CS | **task body** Component_c **is** |
| **Attachments** | **end** Component_c; |
| … | **task body** Component_s **is** |
| **End Configuration** | **end** Component_s; |
| | **task body** Connector_cls **is** |
| | **end** Connector_cls; |
| | **begin** |
| | **null;** |
| | **end** ClientServeur; |

**Figure 7** *: Traduction d'une configuration Wright*

La traduction proposée possède un avantage majeur : elle permet de conserver la sémantique d'une configuration Wright. En effet, celle-ci est définie formellement en CSP comme la composition parallèle des processus modélisant les composants et les connecteurs formant cette configuration (Graiet, 2007). De plus, un programme concurrent en Ada peut être modélisé en CSP comme la composition parallèle des tâches formant ce programme.

## 1.3.2 Traduction des événements

Nous ignorons les données portées par les événements CSP. De telles données seront introduites progressivement en appliquant un processus de raffinement sur le code Ada généré. Ainsi, nous distinguons :

– un événement observé de la forme e.

– une émission ou encore événement initialisé de la forme _e.

### 1.3.2.1 Traduction d'un événement observé

Un événement observé de la forme e est traduit par une entrée (**entry**) et par une acceptation de rendez-vous (instruction **accept**).

La figure 8 illustre le principe de la traduction d'une réception CSP en Ada.

| Spécification en Wright | Code Ada |
|---|---|
| **Component** Client<br>**Port** appelant =<br>      _request → result → appelant  ∏ §<br>**Instances**<br>c : Client<br>… | **task** Component_c **is**<br>    **entry** result;<br>**end** Component_c ;<br>**task body** Component_c **is**<br>    …<br>    **accept** result;<br>    …<br>**end** Component_c; |

**Figure 8**: *Traduction d'une réception*

### 1.3.2.2 Traduction d'un événement initialisé

Un événement initialisé de la forme _e est traduit par une demande de rendez-vous sur l'entrée e exportée par une tâche de type différent (seules les tâches de types différents communiquent) à identifier. Pour y parvenir, il faut analyser la partie **Attachments** de la configuration. La figure 9 illustre le principe de la traduction d'une émission.

| Spécification en Wright | Code Ada |
|---|---|
| **Component** Client<br>**Port** appelant =<br>    _request → result → appelant  ∏ §<br>**Connector** cs<br>**Role** client = _request → result → client ∏ §<br>**Role** serveur = request →_result → serveur □ §<br>**Instances**<br>    c : Client<br>    cls: cs<br>**Attachments**<br>Client**.** appelant **As** cls**.**client | **task** Component_c **is**<br>    **entry** result;<br>**end** Component_c ;<br>**task** Connector_cls **is**<br>    **entry** request;<br>    **entry** result;<br>**end** Connector_cls;<br>**task body** Component_c **is**<br>**begin**<br>    Connector_cls**.**request;<br>**end** Component_c; |

**Figure 9**: *Traduction d'une émission*

## 1.3.3 Traduction de l'interface d'un composant

L'interface d'un composant Wright est traduite par une interface d'une tâche

Ada. Cette interface est obtenue de la manière suivante :

*Pour chaque port appartenant au composant Wright*
*Faire*
       *Pour chaque événement appartenant au port*
       *Faire*
              *Si événement est un événement observé de la forme e*
                     *Alors créer une entrée ayant le nom suivant : port_e*
              *Finsi*
       *Finfaire*
*Finfaire*

La figure 10 illustre le principe de la traduction de l'interface d'un composant Wright.

| *Spécification en Wright* | *Code Ada* |
|---|---|
| **Component** Client<br>**Port** appelant =<br>       _request → result → appelant  ∏ §<br>**Instances**<br>c : Client<br>… | **task** Component_c **is**<br><br>    **entry** appelant_result;<br><br>**end** Component_c ; |

**Figure 10**: *Traduction de l'interface d'un composant*

## 1.3.4 Traduction des interfaces des connecteurs

L'interface d'un connecteur Wright est traduite par une interface d'une tâche Ada. Cette interface est obtenue de la manière suivante :

*Pour chaque rôle appartenant au connecteur Wright*
*Faire*
       *Pour chaque événement appartenant au rôle*
       *Faire*
              *Si événement est un événement initialisé de la forme_e*
              *Alors*
                     *Créer une entrée ayant le nom suivant : rôle_e*
              *Finsi*
       *Finfaire*
*Finfaire*

La figure 11 illustre le principe de la traduction de l'interface d'un connecteur Wright.

| *Spécification en Wright* | *Code Ada* |
|---|---|
| **Connector** cs <br> **Role** client = _request → result → client ∏ § <br> **Role** serveur = request →_result → serveur □ § <br> **Instances** <br>       cls: cs | **task** Connector_cls **is** <br>       **entry** client_request; <br>       **entry** serveur_result; <br> **end** Connector_cls; |

**Figure 11***: Traduction de l'interface d'un connecteur*

## 1.3.5 De CSP Wright vers Ada

Dans cette section, nous décrivons les règles permettant de traduire en Ada les opérateurs CSP couramment utilisés en Wright.

### 1.3.5.1 Traduction de l'opérateur de préfixage

Nous distinguons les deux cas :

| CSP | Traduction Ada |
|---|---|
| Cas 1 : a → P | accept a ; <br> traiter P |
| Cas 2 : _a → P | nom_ tache.a; <br> traiter P |

**Figure 12***: Traduction de l'opérateur de préfixage*

### 1.3.5.2 Traduction de l'opérateur de récursion

La récursion en CSP permet la description des entités qui continueront d'agir et d'interagir avec leur environnement aussi longtemps qu'il le faudra. Nous distinguons les cas suivants :

| CSP | Traduction Ada |
|---|---|
| Cas 1 : P= a → Q → P | loop <br>       accept a; <br>       traiter Q <br> end loop; |
| Cas 2 : P=_a → Q → P | loop <br>       nom_ tache.a; <br>       traiter Q <br> end loop; |
| Cas 3 : P= a → Q → P ∏ § | loop <br>       exit when condition_interne ; <br>       accept a; <br>       traiter Q <br> end loop; |

**Figure 13***: Traduction de l'opérateur de récursion*

### 1.3.5.3 Traduction de l'opérateur de choix non déterministe

La notation P $\prod$ Q avec P $\neq$ Q, dénote un processus qui se comporte soit comme P soit comme Q, la sélection étant réalisée de façon arbitraire, hors du contrôle ou de la connaissance de l'environnement extérieur. Nous distingons les cas fournis par la figure 14.

| *CSP* | *Traduction Ada* |
|---|---|
| Cas 1 : a → P $\prod$ b → Q<br><br>avec a et b quelconques. | **if** condition_interne **then**<br>    **accept** a;<br>    traiter P<br>**else**<br>    **accept** b;<br>    traiter Q<br>**end if;** |
| Cas 2 : _a → P $\prod$ § | **if** condition_interne **then**<br>    nom_tache.a;<br>    traiter P<br>**else**<br>    **exit;**<br>**end if;** |
| Cas 3 : _a → P $\prod$ _b → Q | **if** condition_interne **then**<br>    nom_tache.a;<br>    traiter P<br>**else**<br>    nom_tache.b;<br>    traiter Q<br>**end if;** |
| Cas 4 : _a → P $\prod$ b → Q | **if** condition_interne **then**<br>    nom_tache.a;<br>    traiter P<br>**else**<br>    **accept** b;<br>    traiter Q<br>**end if;** |
| Cas 5 : _a → P $\prod$ _a → Q | nom_tache.a;<br>**if** condition_interne **then**<br>    traiter P<br>**else**<br>    traiter Q<br>**end if;** |

**Figure 14** : *Traduction de l'opérateur de choix non déterministe*

**1.3.5.4 Traduction de l'opérateur de choix déterministe**

Le processus P □ Q avec P ≠ Q, introduit une opération par laquelle l'environnement peut contrôler celui de P ou de Q qui sera sélectionné, étant entendu que ce contrôle s'exerce sur la toute première action ou événement. Nous donnons la traduction suivante :

| *CSP* | *Traduction Ada* |
|---|---|
| a → P □ b → Q<br><br>avec a et b quelconques. | **select**<br>    **accept** a;<br>    traiter P<br>**or**<br>    **accept** b;<br>    traiter Q<br>**end select;** |

**Figure 15***: Traduction de l'opérateur de choix déterministe*

**Remarque:** Le traitement modélisé par condition_interne utilisé dans les deux traductions précédentes traduit un non déterminisme lié à l'opérateur ∏. Il cache souvent des fonctionnalités à fournir par le futur logiciel. Ces fonctionnalités peuvent être précisées en réduisant progressivement le non déterminisme.

## 1.4 Conclusion

Après avoir présenté les aspects structuraux et comportementaux de Wright, la concurrence en Ada et le processus de traduction de Wright vers Ada, nous étudions dans le chapitre suivant les principes généraux et les outils IDM permettant l'automatisation de Wright vers Ada. Les outils IDM retenus sont EMF, Xtext, Check, ATL et Xpand.

# Chapitre 2 : Les bases de l'Ingénierie Dirigée par les Modèles

L'IDM (Ingénierie Dirigée par les Modèles) basée sur MDA définie par l'OMG est une nouvelle approche du génie logiciel permettant de représenter et manipuler des modèles comme des entités de première classe. Dans ce chapitre nous présentons successivement les principes généraux de l'IDM, et les outils, de la plate-forme Eclipse, utilisés dans ce mémoire EMF, XMI, Xtext, Check, ATL et Xpand.

## 2.1 L'ingénierie dirigée par les modèles

Dans cette section nous commençons par une présentation des principes généraux de l'IDM (Bézivin, 2004) ou MDE (Model Driven Engineering). Ensuite, nous donnerons un aperçu sur les origines de l'IDM, qui est l'architecture dirigée par les modèles.

### 2.1.1 Les principes généraux de l'IDM

L'ingénierie dirigée par les modèles se base sur le principe « tout est modèle ».
Un modèle est une abstraction de la réalité (le système). Il aide à répondre aux questions que l'on peut se poser sur le système modélisé. Pour qu'un modèle soit productif, il doit pouvoir être manipulé par une machine. Le langage de modélisation a pris la forme d'un modèle, appelé méta-modèle. Un méta-modèle est un modèle qui définit le langage d'expression d'un modèle (OMG, 2006). Autrement dit, un méta-modèle est un modèle d'un ensemble de modèles. La figure 16 inspirée de (Jouault, 2006) (Bézivin, 2004) représente la relation entre le système et le modèle, ainsi que, la relation entre le modèle et le méta-modèle.

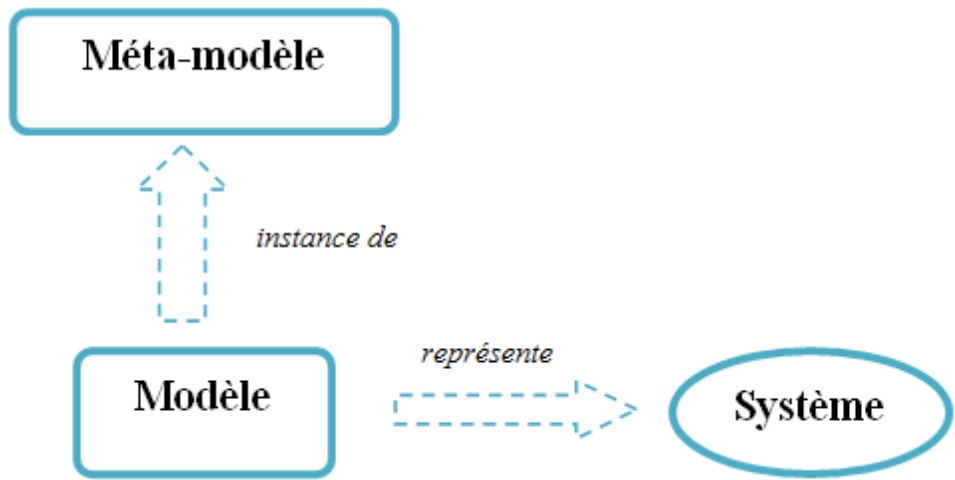

**Figure 16***: Les relations de bases dans l'IDM.*

*Dans la figure 16, la relation* représente *dénote d'un modèle est une représentation d'un système, tant disque la relation* instance de *dénote qu'un modèle est conforme à un méta-modèle si ce modèle appartient à l'ensemble modélisé par ce méta-modèle.*

La figure 17 montre un exemple des relations de l'IDM inspiré de (Favre, 2006).

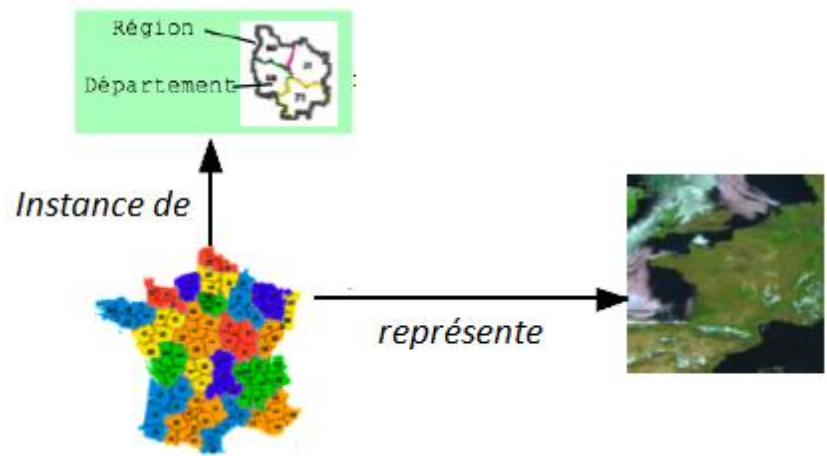

**Figure 17***: Un exemple montrant les relations de base dans l'IDM*

## 2.1.2 Architecture dirigée par les modèles MDA (Model Driven Architecture)

Après l'acceptation du concept clé de méta-modèle comme langage de description de modèle, de nombreux méta-modèles ont émergés afin d'apporter chacun leurs spécificités dans un domaine particulier. Devant le danger de voir émerger indépendamment et de manière incompatible cette grande variété de méta-modèles, il y avait un besoin urgent de donner un cadre général pour leur description. La réponse logique fut donc d'offrir un langage de définition de méta-modèles qui prit lui-même la forme d'un modèle : ce fut le méta-méta-modèle MOF (Meta-Object Facility) (OMG, 2006). En tant que modèle, il doit également être défini à partir d'un langage de modélisation. Pour limiter le nombre de niveaux d'abstraction, il doit alors avoir la propriété de méta-circularité, c'est à dire la capacité de se décrire lui-même (Combemale, 2008).

*C'est sur ces principes que se base l'organisation de la modélisation de l'OMG généralement décrite sous une forme pyramidale représentée par la figure 18 (Bézivin, 2003).*

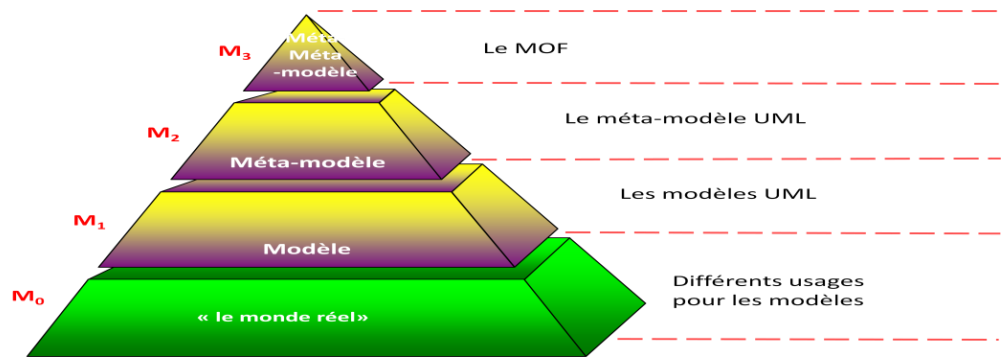

**Figure 18**: *Pyramide de modélisation de l'OMG*

Le monde réel est représenté à la base de la pyramide (niveau M0). Les modèles représentant cette réalité constituent le niveau M1. Les méta-modèles permettant la définition de ces modèles constituent le niveau M2. Enfin, le méta-méta-modèle, unique et méta-circulaire, est représenté au sommet de la pyramide (niveau M3).

L'idée de base de MDA est de séparer les spécifications fonctionnelles d'un système des détails de son implémentation sur une plate-forme donnée. Pour cela, MDA définit une architecture de spécification structurée en plusieurs types de modèles.

- CIM (Computational Independent Model): aussi connu sous le nom modèle ou modèle métier, il s'agit des modèles indépendants de l'informatisation. Un CIM modélise les exigences d'un système, son but étant d'aider à la compréhension du problème ainsi que de fixer un vocabulaire commun pour un domaine particulier (par exemple le diagramme des cas d'utilisation d'UML).

- PIM (Platform Independent Model): aussi connu sous le nom de modèle d'analyse et de conception. C'est un modèle abstrait indépendant de toute plate-forme d'exécution. Il a pour but de décrire le système d'une vue fonctionnelle.

- PDM (Platform Description Model) : pour les modèles de description de la plate-forme sur laquelle le système va s'exécuter. Il définit les différentes fonctionnalités de la plate-forme et précise comment les utiliser.

- PSM (Platform Specific Model) : pour les modèles spécifiques à une plate-forme donnée. En général il est issu de la combinaison du PIM et du PDM. Il représente une vue détaillée du système.

La figure 19 établie par (obeo) donne une vue générale d'un processus MDA appelé couramment cycle de développement en Y en faisant apparaître les différents niveaux d'abstraction associés aux modèles.

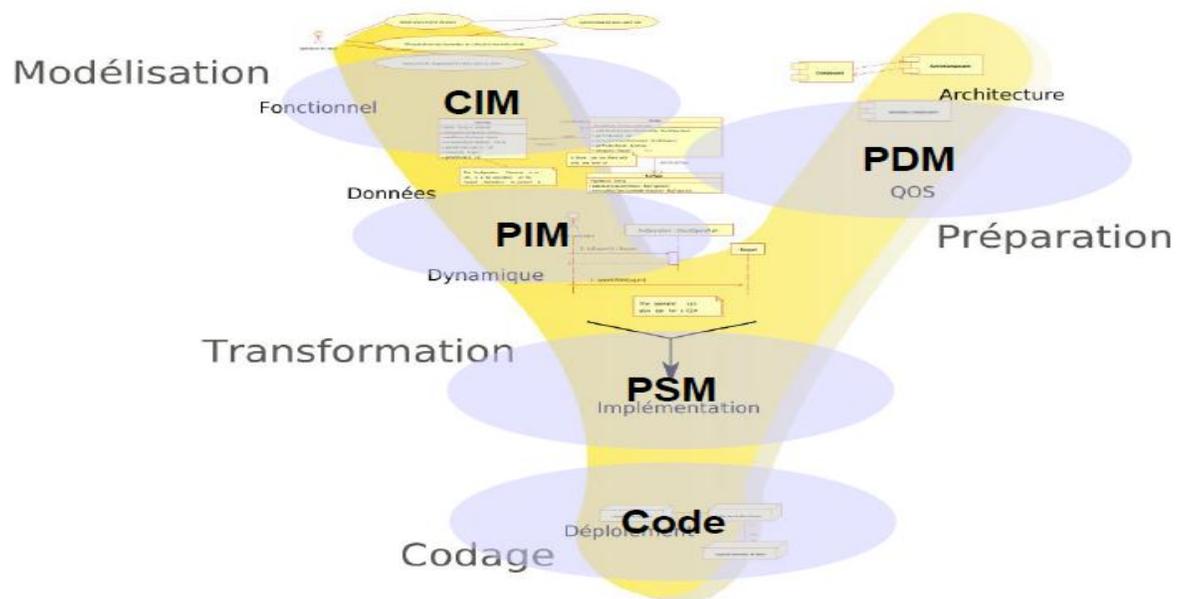

**Figure 19**: *Le processus en Y de l'approche MDA*

### 2.1.3 La transformation des modèles

Les transformations sont au cœur de l'approche MDA. Elles permettent d'obtenir différentes vues d'un modèle, de le raffiner ou de l'abstraire, de plus elles permettent de passer d'un langage vers un autre. Elle assure le passage d'un ou plusieurs modèles d'un niveau d'abstraction donné vers un ou plusieurs autres modèles du même niveau (transformation horizontale) ou d'un niveau différent (transformation verticale). Les transformations horizontales sont de type PIM vers PIM ou bien PSM vers PSM. Les transformations verticales sont de type PIM vers PSM ou bien PSM vers code. Les transformations inverses verticales (rétro-ingénierie) sont type PSM vers PIM ou bien code vers PSM.

La figure 20 (Piel, 2007) donne une vue d'ensemble sur la transformation des modèles.

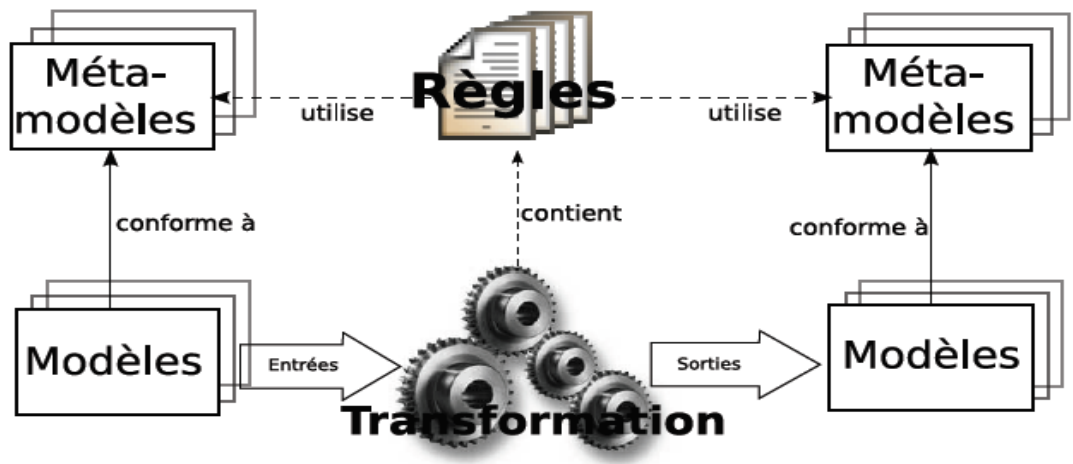

**Figure 20**: *Architecture de la transformation des modèles.*

*Les règles de transformation sont établies entre les méta-modèles source et cible, c'est-à-dire entre l'ensemble des concepts des modèles source et cible. Le processus de transformation prend en entrée un modèle conforme au méta-modèle source et produit en sortie un ou plusieurs autre(s) modèle(s) conforme(s) au méta-modèle cible, en utilisant les règles préalablement établies.*

## 2.2. La plate-forme de modélisation sous Eclipse

Dans cette section nous présentons brièvement les outils de modélisation de la plateforme Eclipse utilisés au cours de notre travail. Cette présentation concerne EMF et XMI.

### 2.2.1 EMF (Eclipse Modeling Framework)

EMF a été conçu pour simplifier le chargement, la manipulation et la sauvegarde de modèles au sein de l'environnement Eclipse. Il repose sur un formalisme de description de méta-modèles nommé Ecore. Ce formalisme est un sous-ensemble de la norme EMOF (Essential MOF), elle même étant un sous-ensemble de MOF2.

Un aperçu d'un fichier Ecore ouvert avec son éditeur est donné par la figure 21.

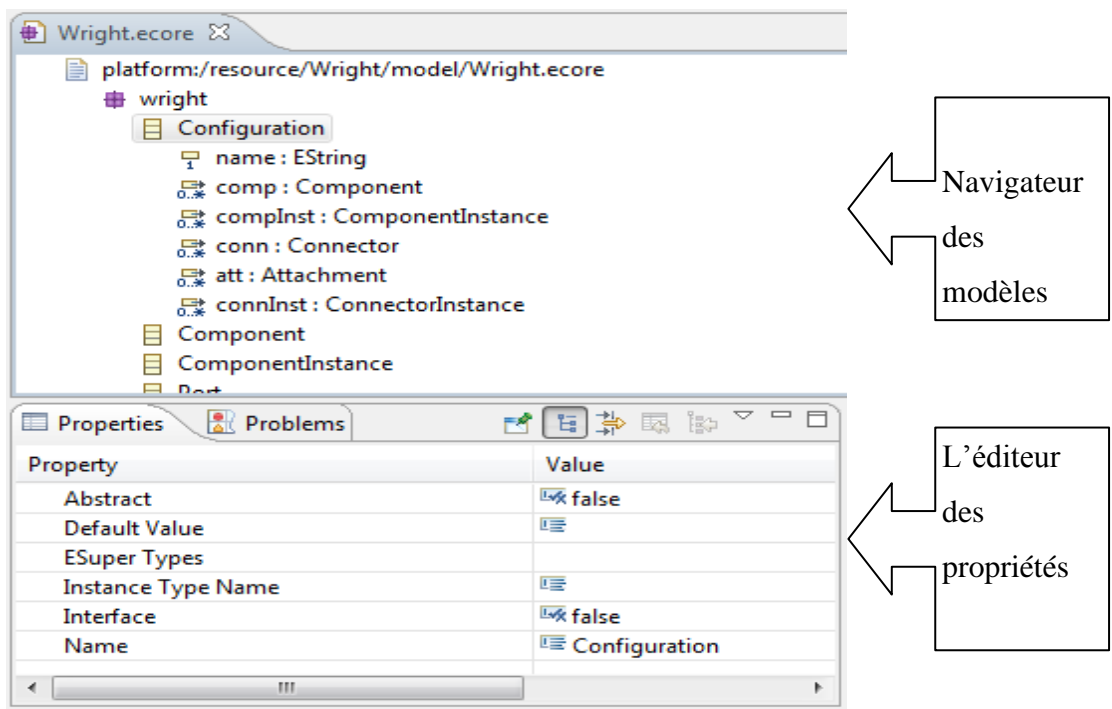

**Figure 21**: *Capture d'écran sur un fichier Ecore*

L'arbre de navigation est utilisé pour naviguer et créer de nouveaux éléments Ecore avec un clic sur l'emplacement souhaité. La deuxième ligne définit le nom du paquetage. L'éditeur des propriétés est utilisé pour modifier des éléments Ecore.

### 2.2.2 XMI (XML Metadata Interchange)

Le format XMI permet la sérialisation des modèles sous un format physique. Comme son nom l'indique, XMI reprend la syntaxe XML (Extensible Markup Language – en français : langage extensible de balisage-) conçue autour du principe de balises.

Un aperçu d'un fichier XMI ouvert avec son éditeur est donné par la figure 22.

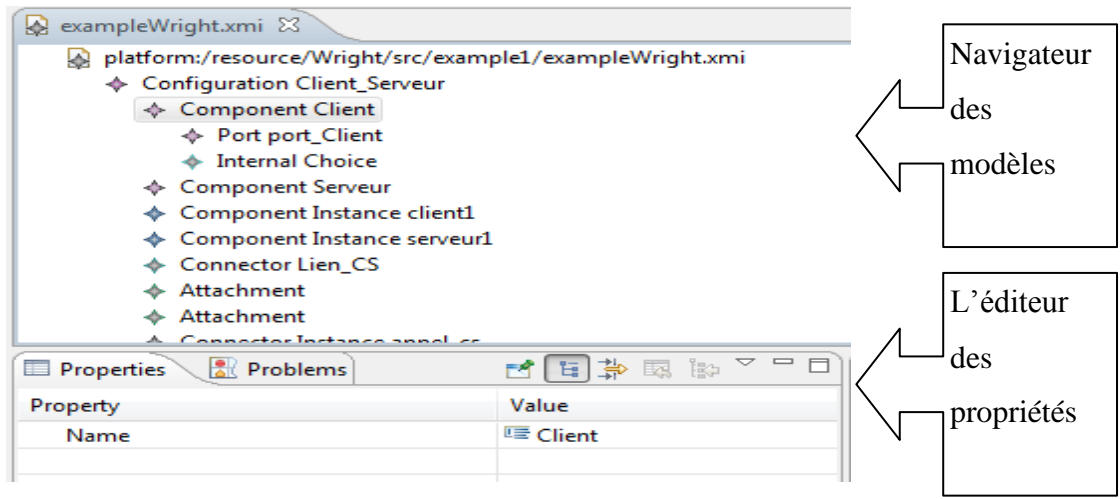

**Figure 22** : *Capture d'écran sur un fichier XMI*

L'arbre de navigation est utilisé pour naviguer et créer de nouvelles instances XMI avec un clic sur l'emplacement souhaité. La deuxième ligne définit l'instance racine. L'éditeur de propriétés est utilisé pour modifier les instances et établir les liens entre les instances.

## 2.3 MWE (Modeling Workflow Engine)

MWE, est une composante de EMFT (EMF Technology), est un moteur de génération déclaratif configurable qui s'exécute sous Eclipse. Il fournit un langage de configuration simple, basé sur les balises XML. Un générateur workflow est composé d'un ensemble de composants qui sont exécutées séquentiellement dans une JVM (Java Virtual Machine) unique.

### 2.3.1 Les propriétés

Ce sont des variables qui contiennent des valeurs utilisables par d'autres éléments. Les propriétés peuvent être déclarées n'importe où dans un fichier workflow. Ils seront disponibles après leurs déclarations.

Il y a deux types de propriétés : les propriétés fichiers et les propriétés simples. L'utilisation d'une propriété se fait selon le format `${nom_prop}`.

La balise `<property>` possède plusieurs attributs :

- name : cet attribut définit le nom de la propriété

- value : cet attribut définit la valeur de la propriété

- file : cet attribut permet de préciser le nom d'un fichier qui contient la définition d'un ensemble de propriétés. Ce fichier sera lu et les propriétés qu'il contient seront définies.

Exemples :

```
<workflow>
        <property file="org/xtext/example/GenerateWright1.properties"/>

        <property name="runtimeProject" value="../${projectName}"/>
…
</workflow>
```

Dans cet exemple `projectName` est le nom d'une propriété simple déclarée dans le fichier de déclaration des propriétés `GenerateWright1.properties`.

### 2.3.2 Les composants

Les composants workflow représentent une partie du processus générateur. Ils représentent généralement les analyseurs modèles, validateurs de modèles, les transformateurs de modèles et les générateurs de codes.

Exemple 1 :

```
<component class="org.eclipse.emf.mwe.utils.DirectoryCleaner">
        <directory="${runtimeProject}/src-gen">
</component>
```

Le composant `DirectoryCleaner` contient, entre autre, la propriété `directory`. Il permet de nettoyer le répertoire, ici `${runtimeProject}/src-gen`, qui contient des artefacts générés avant de (re-) produire ces artefacts.

Exemple 2 :

```
<bean class="org.eclipse.emf.mwe.utils.StandaloneSetup">
     <platformUri="${runtimeProject}/..">
</bean>
```

La classe `StandaloneSetup` n'est pas un composant du workflow au sens étroit. Toutefois, cette classe a besoin d'être référencée dans le workflow afin de mettre en place le méta-modèle EMF défini en mode autonome. Elle contient, entre autre, la propriété `platformUri`.

## 2.4 L'outil de création de DSL Xtext

Dans cette section nous allons présenter l'outil Xtext (Xtext) qui est une composante de TMF (Textual Modeling Framework) intégré dans le framework de modélisation d'Eclipse (Eclipse Modeling Framework : EMF). C'est une partie du projet oAW (open Architectural Ware).

### 2.4.1 Vue d'ensemble de l'outil Xtext

Xtext (Haase, 2007) permet le développement de la grammaire des langages spécifiques aux domaines (DSL : Domain Specific Languages) et d'autres langages textuels en utilisant une grammaire qui ressemble au langage EBNF (Extended Backus-Naur Form). Il est étroitement lié à l'Eclipse Modeling Framework (EMF). Il permet de générer un méta-modèle Ecore, un analyseur syntaxique (parser, en anglais) basé sur le générateur ANTLR ou JavaCC et un éditeur de texte sous la plate-forme Eclipse afin de fournir un environnement de développement intégré IDE spécifique au langage.

La figure 23 fournit une vue d'ensemble de l'outil Xtext. La forme textuelle du DSL constitue le point de départ. La grammaire du DSL est le point d'entrée de l'outil. Xtext produit, en sortie, un analyseur syntaxique, un éditeur et un méta-modèle pour le DSL.

Xtext permet de considèrer la forme textuelle du DSL comme un modèle conforme à la grammaire d'entrée, donc au méta-modèle généré.

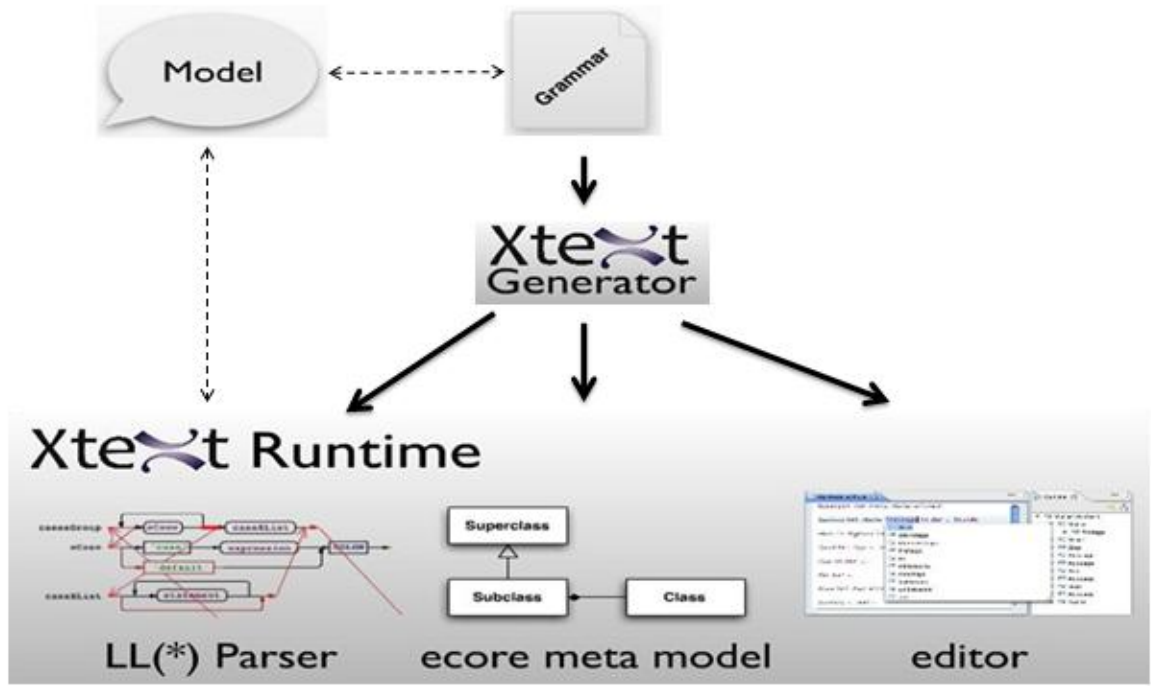

**Figure 23**: *Vue d'ensemble de l'outil Xtext*

## 2.4.2 Présentation de la grammaire de Xtext sur un exemple

Dans cette partie nous allons donner un petit aperçu sur la grammaire de Xtext en utilisant l'exemple qui se trouve par défaut lors de la création d'un projet Xtext. La capture d'écran de la figure 24 donne la syntaxe de cet exemple.

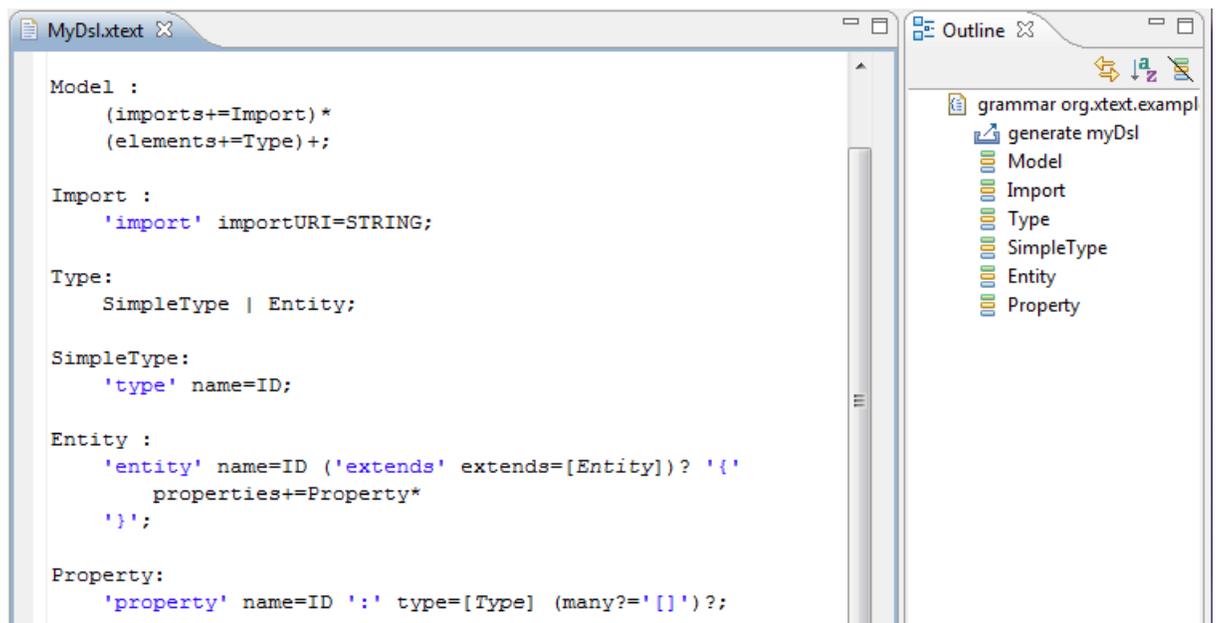

**Figure 24** : *Capture d'écran sur un exemple de grammaire xtext*

La grammaire de xtext est composée d'un ensemble de règles (ici Model, Import, Type…). Une règle décrit l'utilisation d'un ensemble de règles terminales et non terminales. Les règles telles qu'ID et STRING sont des règles terminales.

La première règle Model de notre fichier s'appelle la règle d'entrée. Il s'agit d'un conteneur pour tous les Import et Type. Comme nous pouvons avoir zéro ou plusieurs imports, la cardinalité est « * » et un ou plusieurs types, la cardinalité est « + ». L'opérateur d'affectation « += » dénote qu'il s'agit d'un ensemble d'imports et d'un ensemble d'éléments types.

La deuxième règle est la règle Import, cette règle commence obligatoirement par le mot clef 'import' suivie d'une chaine de caractères. L'opérateur d'affectation « = » dénote qu'il s'agit d'une règle terminale de type chaine de caractères.

La troisième règle Type indique via l'opérateur d'alternance « | » qu'un type peut être soit un type simple (SimpleType) soit une entité (Entity). Un type simple commence par le mot clef 'type' suivi d'un identificateur ; son nom.  Une entité commence par le mot clef  'entity' suivi d'un identificateur (son nom), d'une accolade ouvrante, un certain nombre de propriétés (zéro ou plusieurs) et se termine par une accolade de fermeture. Une entité peut référencer une autre entité de son super type précédé par le mot clef 'extends', Il

s'agit d'une rèfèrence croisée. Ce cas est optionnel, on le note par la cardinalité « ? ».  Pour définir les références croisées il faut utiliser les crochets.

La règle Property commence par le mot clef 'property', suivi du nom de la propriété, d'une référence vers son type (précédé par ' :') et d'un suffixe optionnel '[]'. Ce suffixe est optionnel, il faut donc utiliser l'opérateur d'affectation « ?= » et la cardinalité « ? ».

Après avoir exécuté le générateur (Run As -> MWE Workflow). Nous pouvons avoir l'éditeur de texte de notre DSL, pour cela il suffit d'exécuter notre projet xtext (clic droit sur le projet -> Run As -> Eclipse Application). Cet éditeur permet une coloration syntaxique, un mode plan et une validation du code en temps réel. La figure 25 donne une capture d'écran sur un modèle de l'exemple présenté avec l'éditeur de texte de xtext spécifique au DSL crée.

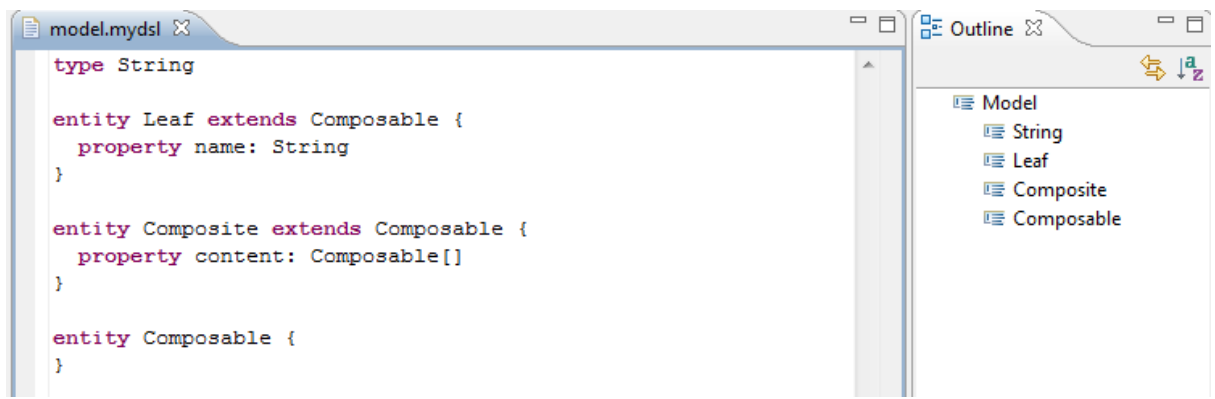

**Figure 25**: *Capture d'écran sur un exemple de modèle dans l'éditeur*

## 2.5 Le langage Check pour la vérification de contraintes

La plate-forme open Architecture Ware fournit aussi un langage formel appelé Check (Xpand) pour la spécification des contraintes que le modèle doit respecter pour être correct. Les contraintes spécifiées avec ce langage doivent être stockées dans des fichiers avec l'extension « *.Chk* ». Ce fichier doit commencer par une importation du méta-modèle selon le format « **import** metamodel; ».

Chaque contrainte est spécifiée dans un contexte, soit une méta-classe du méta-modèle importé, pour lequel la contrainte s'applique. Les contraintes peuvent être de deux types :

- *Warning* : dans ce cas, si la contrainte n'est pas vérifiée un message sera affiché sans que l'exécution s'arrête. A l'instar des erreurs non bloquantes produites par le compilateur C.

- *Error* : dans ce cas, si la contrainte n'est pas vérifiée un message sera affiché et l'exécution sera arrêtée.

Exemple:

```
context Entity WARNING
"Names have to be more than one character long." :
name.length > 1;
```

Cette contrainte avertit que l'attribut name des instances de la méta-classe Entity doit normalement être plus long qu'un caractère. Le langage Check est basé sur OCL (OCL).

## 2.6 Le langage de transformation de modèles ATL

Cette section présente le langage transformation de modèles ATL (ATLAS Transformation Language (ATL) (Jouault, 2006)) qui est une réponse du groupe de recherche INRIA et LINA à l'OMG MOF (Meta Object Facilities (MOF, 2003)) / QVT (Querry View Transformation (QVT, 2010)).

ATL est un langage de transformation de modèles dans le domaine de l'IDM (Ingénierie Dirigée par les Modèles) ou MDE (Model-Driven Engineering). Il fournit aux développeurs un moyen de spécifier la manière de produire un certain nombre de modèles cibles à partir de modèles sources.

### 2.6.1 Vue d'ensemble sur la transformation ATL

Une transformation de modèles définit la façon de générer un modèle Mb, conforme au méta-modèle MMb, à partir du modèle Ma conforme au méta-modèle MMa. Un élément majeur dans l'ingénierie des modèles est de considérer, dans la mesure du possible, tous les objets traités comme des modèles. La transformation de modèles doit être donc, lui-même, définie comme un modèle (MMa2MMb.atl). Ce modèle de transformation doit être conforme au méta-modèle qui définit la sémantique de transformation de modèles (ATL). Tous les méta-modèles doivent être conformes au méta-méta-modèle considérée (Ecore). La figure 26 donne une vue d'ensemble sur la transformation ATL.

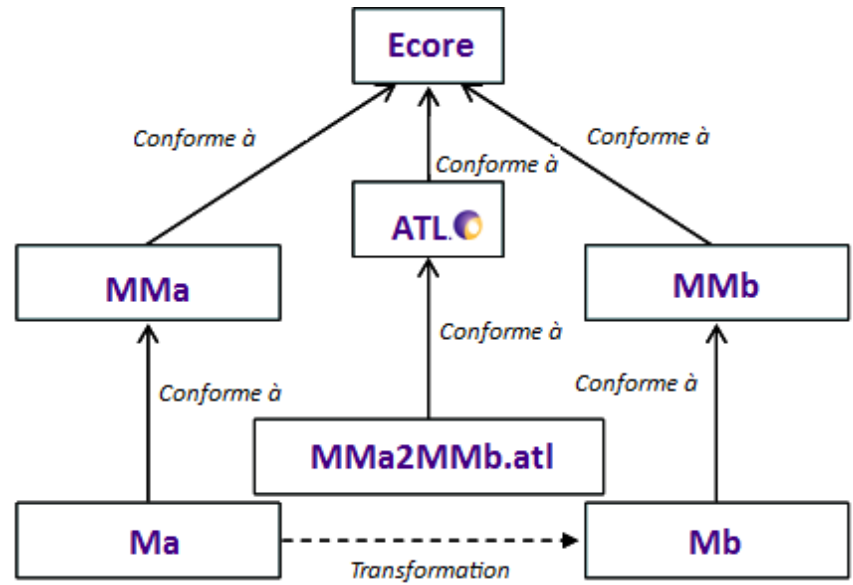

**Figure 26**:*Vue d'ensemble sur la transformation ATL*

Le langage ATL offre différents types d'unités, qui sont définis dans des fichiers d'extension « .atl » distincts. Ces unités sont le module permettant de définir les opérations des transformations des modèles, des requêtes ATL et des bibliothèques qui peuvent être importées par les différents types d'unités ATL, y compris les autres bibliothèques. Un aperçu de ces différents types d'unités est fourni dans la suite.

## 2.6.2 Présentation d'ATL

### 2.6.2.1 Les modules ATL

Un module ATL correspond à la transformation d'un ensemble de modèles sources vers un ensemble de modèles cibles conformes à leurs méta-modèles. Sa structure est composée d'une section d'en-tête, d'une section d'importation facultative, d'un ensemble de helpers et d'un ensemble de règles.

*2.6.2.1.1 La section d'en-tête*

La section d'en-tête définit le nom du module de transformation ainsi que le nom des variables correspondantes aux modèles sources et cibles. Elle encode également le mode d'exécution du module qui peut être soit le mode normal (défini par le mot clef 'from') ou le mode de raffinage (défini par le mot clef 'refining'). La syntaxe de la section d'en-tête est définie comme suit :

```
module MMa2MMb;
create Mb : MMb [from|refining] Ma : MMa ;
```

*2.6.2.1.2 La section d'importation*

Cette section est optionnelle, elle permet de déclarer les bibliothèques ATL qui doivent être importées. La syntaxe de la section d'importation est définie comme suit :

```
uses nom_bibliothèque;
```

*2.6.2.1.3 Les helpers*

Les fonctions ATL sont appelées des *helpers* d'après le standard OCL (Object Constraint Language (OCL)) sur le quel ATL se base. OCL définit deux sortes de *helpers* : opération et attribut.

La syntaxe d'un *helper* opération est définie comme suit :

```
helper [context type_du_contexte]? def : nom_du_helper
( nom_paramètre1 : type_paramètre1 , nom_paramètre2 : type_paramètre2) :
type_de_retour = expression;
```

La syntaxe d'un *helper* attribut est définie comme suit :

```
helper [context type_du_contexte]? def : nom_du_helper : type_de_retour =
expression;
```

Un *helper* peut être spécifié dans le contexte d'un type OCL (par exemple *String* ou *Integer*) ou d'un type venant du méta-modèle source, comme il peut être sans contexte dans ce cas il prendra le contexte par défaut ; celui du module ATL. Les *helpers* peuvent utiliser la récursivité, de plus, le *helper* opération peut être polymorphe.

Il existe une différence dans la sémantique d'exécution du *helper* opération et du *helper* attribut, le premier est calculé à chaque appel tandis que le deuxième est calculé une seule fois selon l'ordre de sa déclaration dans le fichier ATL.

Si nous considérons l'exemple de la transformation du méta-modèle Book vers le méta-modèle Publication (ATL) présentés par la figure 27.

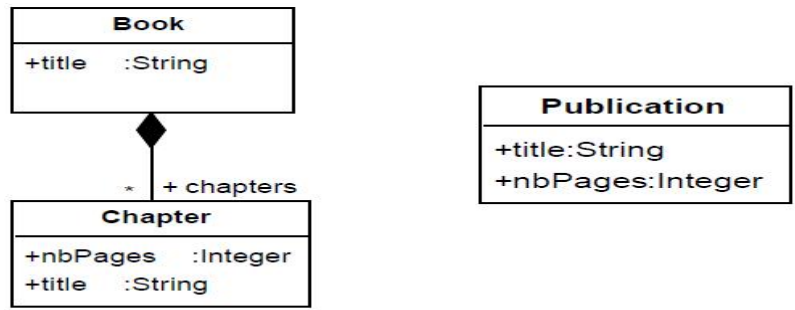

**Figure 27**:*méta-modèles Book et Publication*

Nous pouvons définir un *helper* opération qui permet de calculer la somme des pages du livre en fonction de la somme des pages de ces chapitres, ceci se présente comme suit :

```
helper context Book!Book def : getSumPages() : Integer = self.chapters-
>collect(f|f.nbPages).sum();
```

### *2.6.2.1.4 Les règles*

Le langage ATL est un langage hybride, il contient aussi bien les constructions déclaratives que les constructions impératives. Le style recomdandé est le style déclaratif, cependant, pour faciliter les constructions plus ou moins compliqués il est possible d'avoir recourt au style impératif. Le langage comprend trois types de règles déclaratives et un type de règles impératives.

- Les règles standards (*Matched rules*) : Ils constituent le noyau de la transformation déclarative. Elles sont appelées une seule fois pour chaque tuple correspondant à leurs motifs d'entrée trouvé dans les modèles source. Elles permettent de spécifier, pour quels éléments sources, les éléments cibles sont généré, ainsi que, la façon dont ces éléments cibles sont initialisés. Ces règles ont le format suivant :

```
rule rule_name {
    from
            in_var : in_type [( condition )]?
    [using {      var1 : var_type1 = init_exp1;
            ...
            varn : var_typen = init_expn;}]?
```

```
        to
                out_var1 : out_type1 (bindings1),
                ...
                out_varn : out_typen (bindingsn        )
        [do {  action_block }]?
}
```

Les variables de la section optionnelle *using* sont des variables locales.

Le bloc impératif (*action block*) contient une séquence d'instructions impératives.

La règle de l'exemple précédent se traduit comme suit :

```
rule Book2Publication {
    from b : Book!Book
    to out : Publication!Publication (
        title <- b.title,
        nbPages <- b.getSumPages()
)}
```

Dans cette règle les instances de la méta-classe Book sont traduites en des instances de la méta-classe Publication en gardant le même titre et le même nombre de pages. Ce dernier est calculé par le *helper getSumPages* pour les instances de Book.

- Les règles paresseuses (*lazy rule*): Elles ressemblent aux règles standards, à la différence qu'elles ne sont déclenchées que par d'autres règles. Elles sont appliquées sur chaque tuple autant de fois qu'elles sont référencées.

- Les règles paresseuses uniques (*unique lazy rule*): Identiques aux règles paresseuses non uniques, à la différence qu'elles sont appliquées une unique fois pour chaque tuple.

- Les règles appelées (*called rules*): Ils fournissent le style de programmation impératif. Elles peuvent être vues comme des *helper* particuliers. Elles doivent être déclenchées explicitement. Exemple de règle appelée:

```
entrypoint rule NewPublication (t: String, n: Integer) {
    to    p : Publication!Publication (
              title <- t
          )
    do {  p.nbPages <- n
    }
}
```

*2.6.2.1.5 Les modes d'exécution des modules*

Le moteur d'exécution ATL définit deux modes d'exécution pour les différents modules ATL.

- Le mode normal : C'est le mode d'exécution par défaut, il faut spécifier explicitement la manière dont les éléments de modèle cible doivent être générés à partir des éléments du modèle source. Ce mode est spécifié par le mot clef *from* dans l'en-tête. Il est utilisé dans le cas d'une transformation exogène : le métamodèle source et cible sont différents.
- Le mode de raffinage : Dans ce mode d'exécution la conception de la transformation qui vise à dupliquer le modèle source avec seulement quelques modifications. Concevoir cette transformation dans le mode d'exécution par défaut nécessite la spécification des règles, aussi bien pour les éléments qui vont être modifiés, que ceux ne vont de être modifiés. Ce mode est spécifié par le mot clef *refining* dans l'en-tête. Il est utilisé dans le cas d'une transformation endogène : un même méta-modèle source et cible.

### 2.6.2.2 Les requêtes ATL

Une requête ATL peut être considérée comme une opération qui calcule une valeur primitive d'un ensemble de modèles de source. Ces règles ont le format suivant :

```
query query_name = exp;
```

### 2.6.2.3 Les bibliothèques ATL

Une bibliothèque ATL permet de définir un ensemble de *helpers* qui peuvent être appelés à partir des différentes unités ATL. Ce fichier est défini par l'en-tête :

```
library nom_bib;
```

## 2.7 Le langage de génération de code Xpand

Dans cette section nous allons présenter l'outil Xpand (Xpand) du projet oAW (oAW) intégré dans le framework de modélisation d'eclipse EMF.

### 2.7.1 Structure générale d'un template Xpand

Le template Xpand (Klatt, 2007) permet le contôle de la génération de code correspondant à un modèle. Le modèle doit être conforme à un méta-modèle donnée. Le template est stocké dans un fichier ayant l'extension « .xpt ».

Un fichier template se compose d'une ou plusieurs instructions **IMPORT** afin d'importer les méta-modèles, de zéro ou plusieurs **EXTENSION** avec le langage Xtend et d'un ou plusieurs blocks **DEFINE**.

Exemple:

```
«IMPORT meta::model»
«EXTENSION my::ExtensionFile»
«DEFINE javaClass FOR Entity»
«FILE fileName()»
package «javaPackage()»;
public class «name» {
// implementation
}
«ENDFILE»
«ENDDEFINE»
```

Le template de cet exemple importe la définition du méta-modèle, charge une extension Xtend et définit un template pour des simples classes java appliquées à un élément de la méta-classe Entity. `fileName` est une fonction Xtend appelée pour retourner le nom du fichier de sortie. `Name` est un méta-attribut de la méta-classe Entity.

### 2.7.2 Le bloc DEFINE

Les blocs DEFINE, aussi appelés templates, constituent le concept central du langage Xpand. C'est la plus petite unité du fichier template. La balise DEFINE se compose d'un nom, une liste optionnelle de paramètres et du nom de la méta-classe pour laquelle le template est défini. Les templates peuvent être polymorphes, ils ont le format suivant :

```
«DEFINE templateName(formalParameterList) FOR MetaClass»
a sequence of statements

«ENDDEFINE»
```

### 2.7.3 L'instruction FILE

L'instruction FILE redirige la sortie produite, à partir des instructions de son corps, vers la cible spécifiée. La cible est un fichier dont le nom est spécifié par `expression`. Le format de l'instruction FILE se présente comme suit:

```
«FILE expression »
a sequence of statements
«ENDFILE»
```

### 2.7.4 L'instruction EXPAND

L'instruction EXPAND appelle un bloc DEFINE et insère sa production "*output*" à son emplacement. Il s'agit d'un concept similaire à un appel de sous-routine (méthode). Le format de l'instruction EXPAND se présente comme suit :

```
«EXPAND definitionName [(parameterList)]
      [FOR expression | FOREACH expression [SEPARATOR expression] ]»
```

`definitionName` est le nom du bloc DEFINE appelé. Si FOR ou FOREACH est omise l'autre template est appelé pour l'instance courante *this*. Si FOR est spécifié, la définition est exécutée pour le résultat d'une expression cible. Si FOREACH est spécifiée, l'expression cible doit être évaluée à un type collection. Dans ce cas, la définition spécifiée est exécutée pour chaque élément de cette collection. Il est possible de spécifier un séparateur pour les éléments générés de la collection.

Exemple :

```
«DEFINE javaClass FOR Entity»
..
«EXPAND methodSignature FOR this.methods»
..
«ENDDEFINE»
«DEFINE methodSignature FOR Method»
...
«ENDDEFINE»
```

## 2.8 Conclusion

Dans ce chapitre, nous avons présenté les principes généraux de l'IDM et les outils : Ecore, XMI, MWE, Xtext, Check, ATL et Xpand autour de la plate-forme Eclipse. Dans la suite de ce mémoire, nous allons utiliser avec profit ces outils afin d'établir un programme de transformation d'une architecture logicielle décrite en Wright vers un programme concurrent Ada.

# Chapitre 3: Un méta-modèle de l'ADL Wright

Dans ce Chapitre, nous proposons un méta-modèle Wright représentant la plupart des concepts issus de ce langage : composant, connecteur, configuration et processus CSP. Ce méta-modèle joue, dans notre contexte, le rôle de méta-modèle source dans notre approche IDM de transformation d'une architecture logicielle décrite en Wright vers une programme concurrent Ada. Ce chapitre comporte trois sections. La première section présente le fragment du méta-modèle Wright consaré aux aspects structuraux couvrant les concepts composant, connecteur et configuration. La deuxième section présente le fragment du méta-modèle Wright relatif aux aspects comportements couvrant le concept Processus CSP. Enfin, la troisième section établit les liens entre ces deux fragments.

## 3.1 La partie structurelle

### 3.1.1 Aspects syntaxiques

L'ADL Wright repose essentiellement sur les concepts composant, connecteur et configuration. La figure 28 donne le fragment du méta-modèle Wright permettant de représenter ces concepts. Un tel fragment comporte huit méta-classes et treize méta-associations. La méta-classe Configuration occupe une position centrale. Elle englobe des composants, des instances de composants, des connecteurs, des instances de connecteurs et des attachements. Ceci est traduit par une méta-composition entre Configuration et respectivement Component, ComponentInstance, Connector, ConnectorInstance et Attachment. A un composant Wright -respectivement connecteur- est attaché plusieurs ports –respectivement plusieurs rôle-. Ceci est traduit par une méta-composition entre Component et Port –respectivement entre Connector et Role-. Une instance de composant doit avoir un type de Composant. Ceci est traduit par la méta-association entre Component et ComponentInstance. De même, une instance de connecteur doit avoir un type de connecteur. Ceci est traduit par la méta-association entre Connector et ConnectorInstance. Un attachment concerne un port appartenant à une instance de composant et un rôle appartenant à une instance de connecteur. Ceci est traduit par les méta-associations entre Attachment et respectivement ComponentInstance, Port, ConnectorInstance et Role.

Dans la suite, nous décrivons les contraintes OCL attachées au fragment du méta-modèle relatif aux aspects structuraux de Wright.

**Figure 28**: *Fragment du méta-modèle Wright : Partie structurelle*

## 3.1.2 Les contraintes OCL

Nous avons établi plusieurs propriétés décrivant des contraintes d'utilisation des constructions structurelles de Wright. De telles propriétés  sont décrites d'une façon informelle et formelle en se servant d'OCL.

• Propriété 1 :

Les noms désignant des composants, des instances de composants, des connecteurs, des instances de connecteurs, des ports, des rôles et des configurations doivent être des identificateurs valides au sens de Wright. Une formalisation partielle de cette propriété en OCL est:

```
context Component
```

```
inv identifier_card: name.size()>0
inv letter: --le premier caractère de name doit être une lettre majuscule
--ou miniscule.
inv tail: --les autres caractères doivent être lettres majuscules, ou
--miniscules, ou des chiffres.
```

- Propriété 2 :

Tous les ports attachés à un composant doivent avoir des noms deux à deux différents. Ceci peut être formalisé en OCL comme suit :

```
context Component
inv different_port_names : self.port-> forAll( p1, p2 : Port | p1<>p2
implies p1.name<>p2.name)
```

- Propriété 3 :

Tous les rôles attachés à un connecteur doivent avoir des noms deux à deux différents. Ceci peut être formalisé en OCL comme suit :

```
context Connector
inv different_role_names : self.role-> forAll( r1, r2 : Role | r1<>r2
implies r1.name<>r2.name)
```

- Propriété 4 :

Dans une même configuration un composant, une instance de composant, et une instance de connecteur doivent avoir des noms deux à deux différents. Ceci, est formalisé en OCL comme suit :

```
context Configuration
inv different_names_component : self.comp->forAll(c1, c2 : Component |
c1<>c2 implies c1.name<>c2.name)
inv different_names_connector : self.conn->forAll(c1, c2 : Connector |
c1<>c2 implies c1.name<>c2.name)
inv different_names_componentInstance : self.compInst->forAll(c1, c2 :
ComponentInstance | c1<>c2 implies c1.name<>c2.name)
```

```
inv   different_names_component   :   self.connInst->forAll(c1,   c2   :
ConnectorInstance | c1<>c2 implies c1.name<>c2.name)
inv        different_names_in_configuration        :        self.comp-
>collect(self.comp.name)->excludesAll(self.compInst-
>collect(self.compInst.name))
      and       self.comp->collect(self.comp.name)->excludesAll(self.conn-
>collect(self.conn.name))
      and   self.comp->collect(self.comp.name)->excludesAll(self.connInst-
>collect(self.connInst.name))
      and                 self.compInst->collect(self.compInst.name)-
>excludesAll(self.conn->collect(self.conn.name))
      and                 self.compInst->collect(self.compInst.name)-
>excludesAll(self.connInst->collect(self.connInst.name))
      and   self.conn->collect(self.conn.name)->excludesAll(self.connInst-
>collect(self.connInst.name))
```

- Propriété 5 :

Une configuration privée de composants n'admet ni instance de composant ni attachement. De même, une configuration privée de connecteurs n'admet ni instance de connecteur ni attachement. Ces contraintes peuvent être formalisées en OCL comme suit:

```
context Configuration
inv  component_without : self.comp -> size()= 0 implies ( self.compInst -
> size()= 0 and self.att ->size()= 0)
inv  connector_without : self.conn -> size()= 0 implies ( self.connInst -
> size()= 0 and self.att ->size()= 0)
```

- Propriété 6 :

Chaque instance déclarée au sein d'une configuration doit utiliser un type déclaré au sein de la même configuration. Ceci peut être formalisé en OCL comme suit :

```
context Configuration
inv declared_component : self.compInst -> forAll( i : ComponentInstance|
self.comp ->includes( i.type))
```

```
inv declared_connector : self.connInst -> forAll( i : ConnectorInstance|
self.conn ->includes( i.type))
```

- Propriété 7 :

Tous les attachements utilisent des instances déclarées au sein de la même configuration. Ceci peut être formalisé en OCL comme suit:

```
context Configuration
inv declared_instance : self.att -> forAll( a : Attachment |self.compInst
->        includes(a.originInstance)        and        self.connInst-
>includes(a.targetInstance))
```

- Propriété 8 :

Un attachement est valide si et seulement si le port et le rôle concernés sont bel et bien attachés respectivement à l'instance concernée de type composant et l'instance concernée de type connecteur. Ceci peut être formalisé en OCL comme suit :

```
context Attachment
inv attachment_port_concerns_component : self.originInstance.type.port ->
includes( self.originPort)
inv attachment_role_concerns_connector : self.targetInstance.type.role ->
includes( self.targetRole)
```

- Propriété 9 :

Les instances de composants reliées à un composant donné doivent être de même type.

```
context Component
inv        instance_type_component:        self.compInst        -
>forAll(i:ComponentInstance|i.type=self)
```

- Propriété 10 :

Les instances de connecteurs reliées à un connecteur donné doivent être de même type.

```
context Connector
```

```
inv          instance_type_connector:          self.connInst        -
>forAll(i:ConnectorInstance|i.type=self)
```

## 3.2 La partie comportementale

### 3.2.1 Les aspects syntaxiques

CSP de Hoare repose sur deux concepts essentiels: événement et processus. Il offre plusieurs opérateurs permettant d'enchaîner des événements et par conséquent de construire des processus CSP tels que: préfixage (ou séquencement), récursion, choix déterministe et choix non déterministe. En outre, Wright augmente CSP de Hoare en distinguant entre événement initialisé et observé.

La figure 29 donne le fragment du méta-modèle Wright lié à ses aspects comportementaux. Un tel fragment comporte deux hiérarchies. La hiérarchie ayant comme méta-classe fondatrice ProcessExpression modélise le concept processus en CSP. Les méta-classes descendants Prefix, ExternalChoice, InternalChoice et ProcessName représentent respectivement les opérateurs préfixage, choix externe (ou déterministe), choix interne (ou non déterministe) et le nommage d'un processus (favorisant la récursion) fournis par CSP. L'autre hiérarchie ayant comme méta-classe fondatrice EventExpression représente le concept événement en CSP de Wright. Les méta-classes descendantes EventSignalled, EventObserved, InternalTraitment et SuccesEvent représentent respectivement événement initialisé, événement observé, traitement interne et événement succès fournis par CSP de Wright. Les liens entre ces deux hiérarchies sont traduits par les deux méta-agrégations entre Prefix et EventExpression et ProcessExpression et EventExpression qui exprime l'alphabet d'un processus. Les deux méta-agrégations entre Prefix et respectivement EventExpression et ProcessExpression traduisent fidèlement la structure d'un opérateur de préfixage (e → P): il s'engage dans l'événement e puis se comporte comme P. La structure de l'opérateur de choix déterministe est traduite par la méta-agrégation entre ExternalChoice et Prefix. De même, la struture de l'opérateur de choix non déterministe est traduite par la méta-agrégation entre InternalChoice et Prefix.

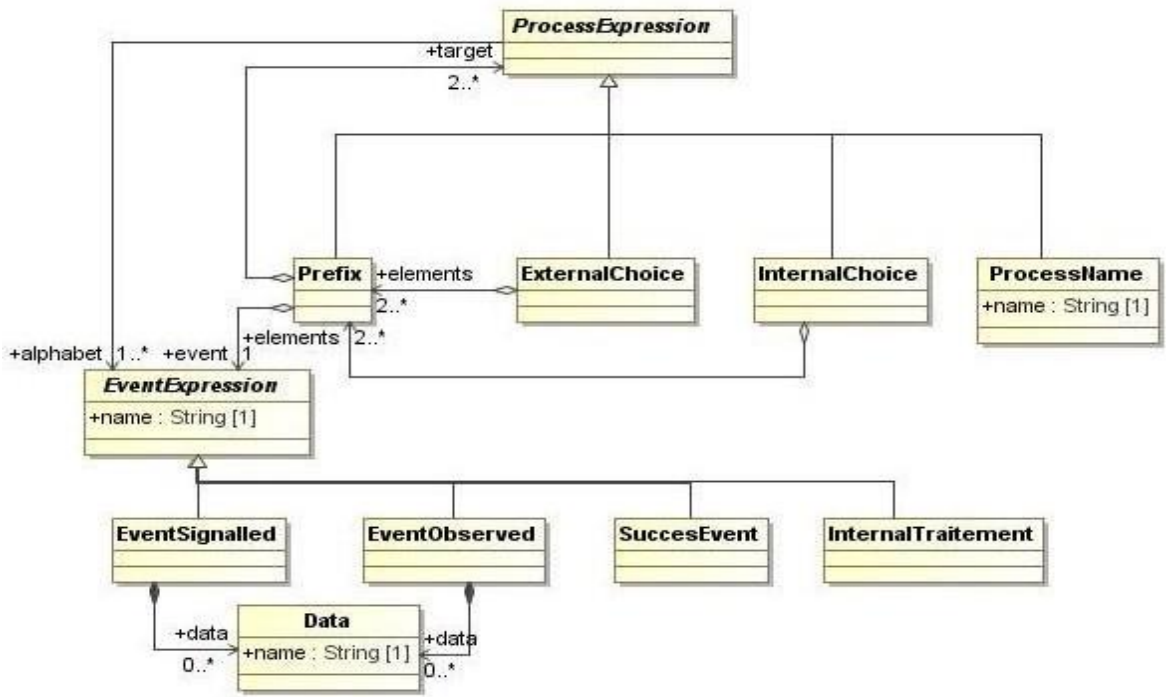

**Figure 29**: *Fragment du méta-modèle Wright: Partie comportementale*

## 3.2.2 Les contraintes OCL

Les propriétés attachées au fragment du méta-modèle décrivant les aspects comportementaux de Wright sont :

- Propriété 11 :

Le méta-attribut name de la méta-classe ProcessName doit stocker un identificateur valide au sens Wright. Ceci peut être formalisé en OCL comme suit :

```
context ProcessName
inv identifier_card: name.size()>0
inv letter: --le premier caractère de name doit être une lettre majuscule
--ou miniscule.
inv tail: --les autres caractères doivent être lettres majuscules, ou
--miniscules, ou des chiffres.
```

- Propriété 12 :

Le méta-attribut name de la méta-classe EventExpression doit stocker un identificateur valide au sens Wright –possibilité d'utiliser la notation qualifiée . -. Ceci peut être formalisé en OCL comme suit :

```
context ProcessName
inv identifier_card: name.size()>0
inv letter: --le premier caractère de name doit être une lettre majuscule
--ou miniscule.
inv tail: --les autres caractères doivent être lettres majuscules, ou
--miniscules, ou des chiffres ou le caractère '.'.
```

- Propriété 13 :

Un choix externe doit être basé uniquement sur des événements observés et succès. Ceci peut être formalisé en OCL comme suit :

```
context ExternalChoice
inv event_observed_in_EC: self.elements -> forAll( e : Prefix |
e.event.oclIsTypeOf(EventObserved) || e.event.oclIsTypeOf(SuccesEvent) )
```

## 3.3 Connexions entre les deux parties structurelle et comportementale

La figure 30 donne les liens entre les deux fragments du méta-modèle Wright présentés dans les deux sections 3.1 et 3.2.

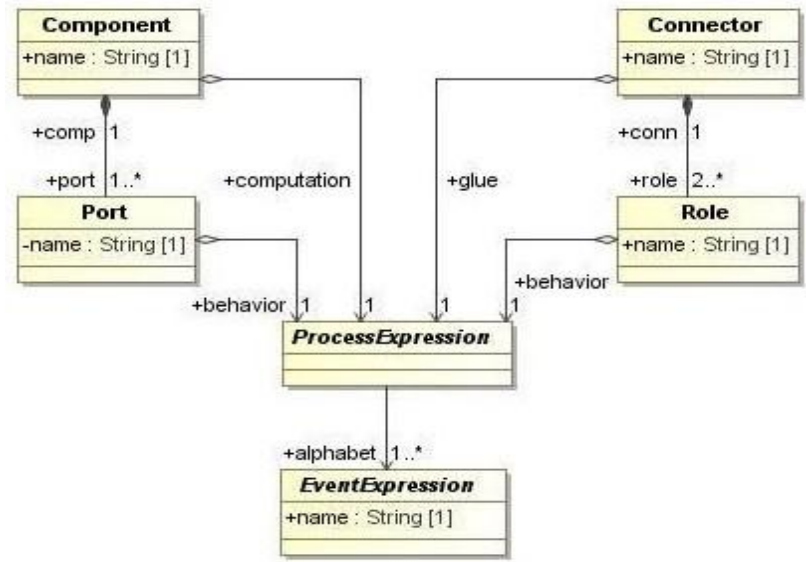

**Figure 30**: *Connexion entre les deux fragments du méta-modèle Wright*

Le comportement d'un port est décrit par un processus CSP. Ceci est traduit par la méta-agrégation entre Port et ProcessExpression. De même, le comportement d'un composant Wright est décrit par un processus CSP. Ceci est traduit par une méta-agrégation entre Component et ProcessExpression. D'une façon symétrique, les aspects comportementaux d'un rôle et d'un connecteur sont décrits respectivement par deux méta-agrégations entre Role et ProcessExpression et Connector et ProcessExpression.

Afin d'apporter plus de précisions à notre méta-modèle Wright, nous avons défini des nouvelles propriétés :

- Propriété 14 :

L'alphabet d'un processus associé à un port ne doit pas inclure des événements décrivant des traitements internes. Ceci peut être formalisé en OCL comme suit :

```
context Port
inv  not_IT_behavior_port  :  self.behavior.alphabet  ->  forAll(
a:EventExpression | not a.oclIsTypeOf(InternalTraitement))
```

- Propriété 15 :

L'alphabet d'un processus associé à un rôle ne doit pas inclure des événements décrivant des traitements internes. Ceci peut être formalisé en OCL comme suit :

```
context Role
inv    not_IT_behavior_role: self.behavior.alphabet    ->    forAll(
a:EventExpression | not a.oclIsTypeOf(InternalTraitement))
```

- Propriété 16 :

Tous les alphabets des processus associés aux ports d'un composant doivent être inclus dans l'alphabet du processus associé au composant. Ceci peut être formalisé en OCL comme suit :

```
context  Component
inv: self.computation.alphabet ->
select(s:EventExpression|s.oclIsTypeOf(EventObserved) or
s.oclIsTypeOf(EventSignalled)) -> collect(o:EventExpression|   o.name)-
>includesAll(self.port -> collect(p:Port|p.behavior.alphabet ->
collect(a:EventExpression|p.name.concat('.').concat(a.name))))
```

- Propriété 17 :

Tous les alphabets des processus associés aux rôles d'un connecteur doivent être inclus dans l'alphabet du processus associé au connecteur. Ceci peut être formalisé en OCL comme suit :

```
context Connector
inv: self.glue.alphabet ->
select(s:EventExpression|s.oclIsTypeOf(EventObserved) or
s.oclIsTypeOf(EventSignalled)) -> collect(o:EventExpression|   o.name) ->
includesAll(self.role -> collect(r:Role|r.behavior.alphabet ->
collect(a:EventExpression|r.name.concat('.').concat(a.name))))
```

## 3.4 Vue d'ensemble sur le méta-modèle Wright

Le méta-modèle de Wright utilisé comme méta-modèle source pour notre approche de transformation de Wright vers Ada est donné par la figure 31.

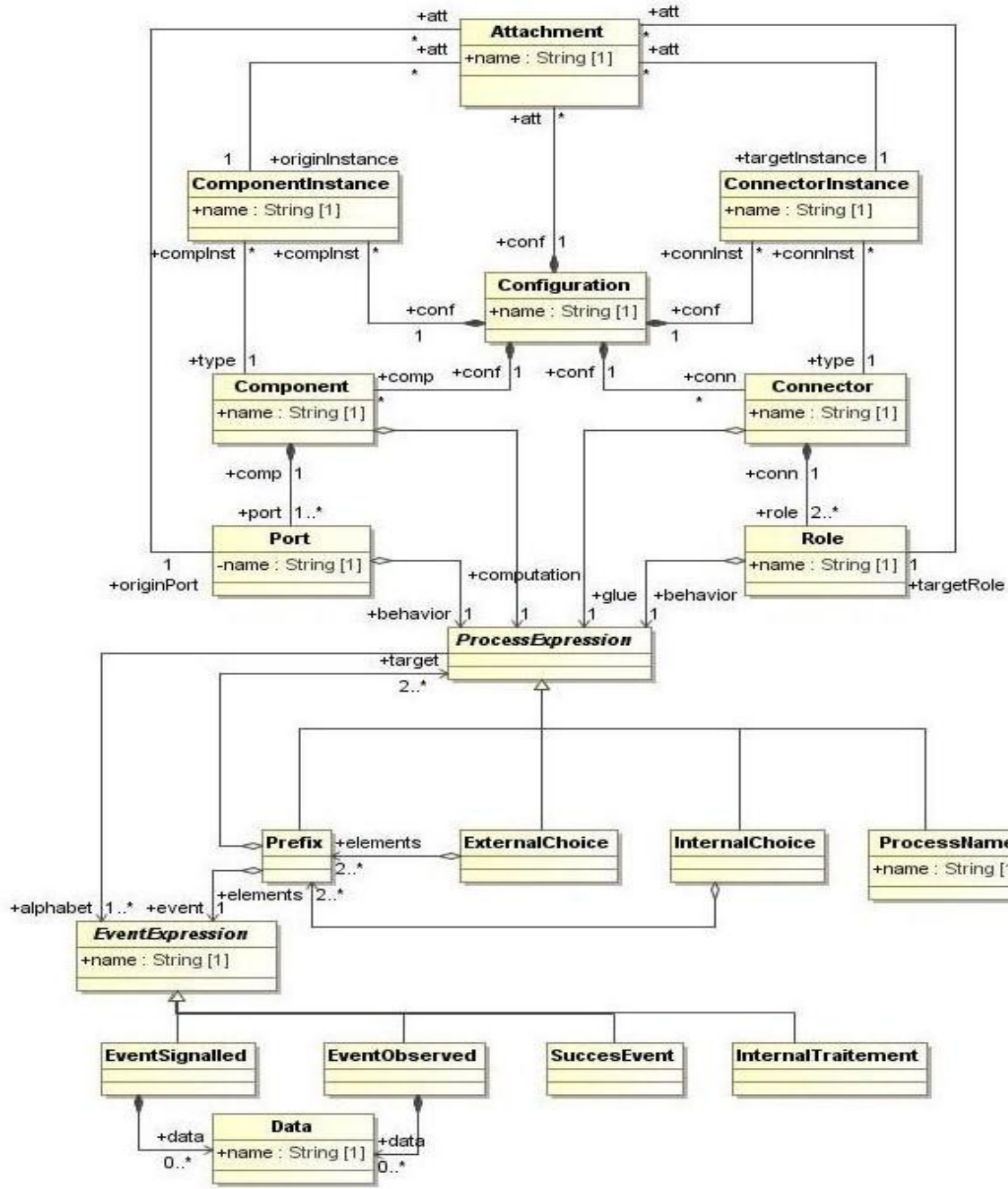

**Figure 31**: *Méta-modèle de Wright*

## 3.5 Conclusion

Dans ce chapitre, nous avons propsé un méta-modèle Wright décrivant les aspects structuraux et comportementaux de ce langage nécessaires et suffisants pour notre approche

IDM de transformation automatique de Wright vers Ada. Notre méta-modèle Wright comporte 18 méta-classes et 17 propriétés décrivant la sémantique statique de Wright. Ces propriétés sont formalisées en OCL. Dans le chapitre suivant, nous allons proposer un méta-modèle partiel Ada jouant le rôle du méta-modèle cible dans notre approche IDM de transformation de Wright vers Ada.

# Chapitre 4: Le méta-modèle partiel d'Ada

Dans ce chapitre, nous proposons un méta-modèle partiel Ada issu de description BNF de ce langage (BNF-Ada) en se limitant aux constructions d'Ada utilisées dans la transformation de Wright vers Ada. Ces Constructions sont : sous-programmes non paramétrés, structures de contrôle, tâches ayant des entrées non paramétrées, instruction non déterministe (select), instruction de demande de rendez-vous sans paramètres, instruction d'acceptation de rendez-vous sans paramètres. Ce chapitre comporte quatre sections. Les trois premières sections proposent des fragments du méta-modèle partiel Ada décrivant la structure des concepts Ada: sous-programmes, tâche et instruction. Ensuite, la quatriéme section donne une vue d'ensemble de notre méta-modèle partiel Ada. Enfin, la cinquiéme section présente des contraintes OCL attachés au méta-modèle.

## 4.1 Les sous-programmes Ada

En Ada, un sous-programme est une unité de programmation comportant deux parties : interface et implémentation. La partie implémentation possède deux parties : partie déclarative et partie exécutive. La partie interface correspond à la signature du sous-programme. En outre Ada distingue nettement les fonctions des procédures aussi bien sur le plan syntaxique que sémantique. En effet, l'appel d'une procédure est considéré comme instruction. Par contre, l'appel d'une fonction doit être inséré au sein d'une expression Ada.

La description BNF d'un sous-programme est donnée comme suit :

```
proper_body ::= subprogram_body | …
subprogram_body ::=
      subprogram_specification "is"
      declarative_part
      "begin"
      handled_sequence_of_statements
      "end" [ designator ] ";"
subprogram_specification ::=
      ( "procedure" defining_program_unit_name [ formal_part ] )
      | ( "function" defining_designator [ formal_part ] "return"
subtype_mark )
```

```
declarative_part ::= { ( basic_declarative_item | body ) }
body ::= proper_body | …
basic_declarative_item ::= basic_declaration |…
basic_declaration ::= object_declaration | subprogram_declaration |…
handled_sequence_of_statements ::= sequence_of_statements [ …]
sequence_of_statements ::= statement { statement }
subprogram_declaration ::= subprogram_specification ";"
```

De cette déscription nous pouvons dériver le méta-modèle de la figure 32.

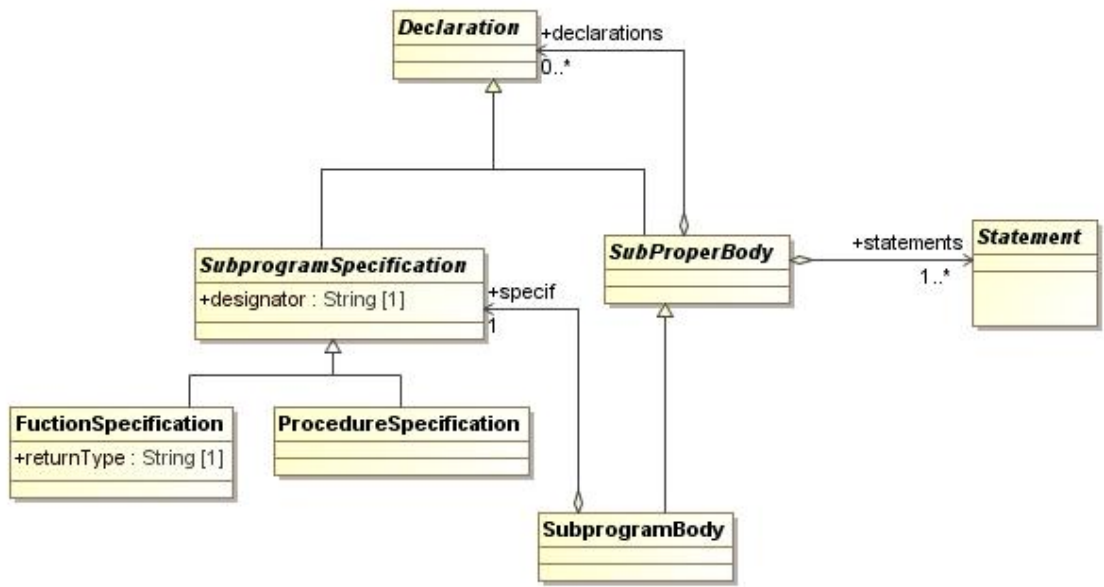

**Figure 32**: *Le méta-modèle d'un sous-programme Ada*

La méta-classe SubprogramBody représente le concept sous-programme ayant trois parties : entête, partie déclarative et partie exécutive. Ces trois parties sont traduites respectivement par trois méta-agrégation entre : SubprogramBody et Declaration, et, SubprogramBody et Statement.

## 4.2 Les tâches Ada

Une tâche en Ada est une unité de programmation comportant deux parties : interface et implémentation. La partie interface offre des services appelés entrées (entry).

Ces services indiquent des possibilités des rendez-vous fournis par la tâche. La partie implémentation comporte deux parties : partie déclarative et partie exécutive. La partie exécutive réalise la politique d'acceptation des rendez-vous adoptée par la tâche.

La déscription BNF d'une tâche Ada est donnée par comme suit :

```
object_declaration ::= single_task_declaration |…
proper_body ::= subprogram_body | task_body |…
single_task_declaration ::=
      "task" defining_identifier [ "is" task_definition ] ";"
task_definition ::= { task_item } [ … ] "end" [ task_identifier ]
task_item ::= entry_declaration | …
entry_declaration ::= "entry" defining_identifier [ … ] ";"
task_body ::= "task" "body" defining_identifier "is"
      declarative_part
      "begin"
      handled_sequence_of_statements
      "end" [ task_identifier ] ";"
```

A ce point nous pouvons enrichir le méta-modèle précédent comme présenté dans la figure 33.

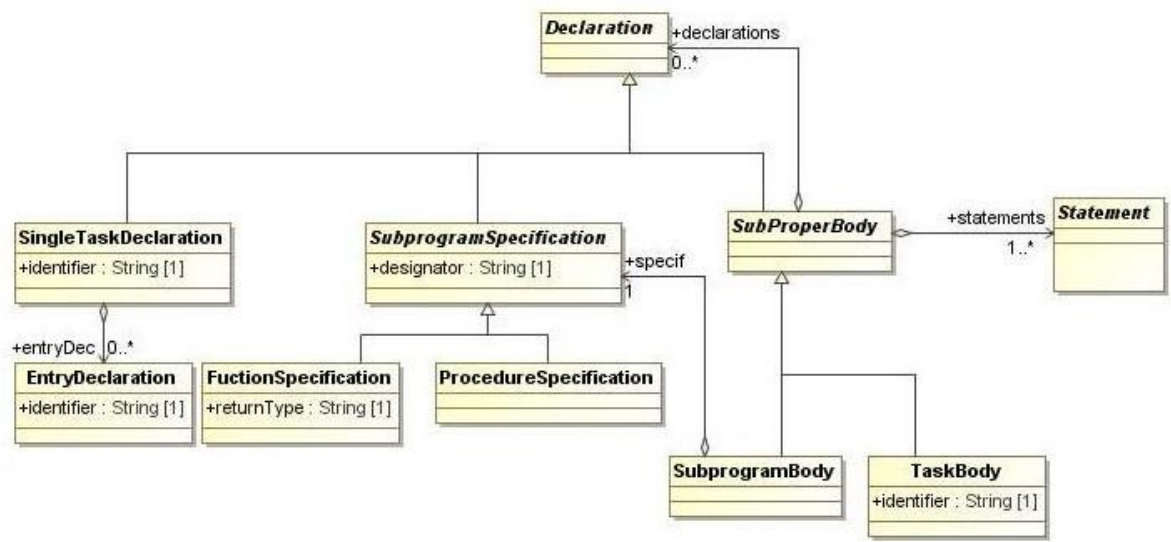

**Figure 33**: *Le méta-modèle représentant un sous-programme et une tâche Ada*

La méta-classe TaskBody représente le concept tâche ayant trois parties : spécification (ou interface), partie déclarative et partie exècutive. Ces trois parties sont traduites respectivement par les deux méta-agrégations entre : TaskBody et Declaration, TaskBody et Statement.

# 4.3 Les instructions Ada

Les instructions concernées sont : les instructions simples et composées.

## 4.3.1 Les instructions simples

- L'instruction nulle :

L'instruction nulle est l'instruction qui ne fait rien. Son écriture BNF :

```
null_statement ::= "null" ";"
```

- L'instruction exit :

L'instruction exit est utilisée pour achever l'exécution de l'instruction loop englobante; l'achèvement est conditionné si elle comprend une garde (une condition). Son écriture BNF :

```
exit_statement ::= "exit" [ loop_name ] [ "when" condition ] ";"
condition ::= expression
```

- L'instruction return :

L'écriture BNF de l'instruction return se présente comme suit:

```
return_statement ::= "return" [ expression ] ";"
```

- L'appel d'une procédure :

L'écriture BNF de l'invocation d'une procédure se présente comme suit :

```
procedure_call_statement ::=  ( procedure_name | prefix )
[ actual_parameter_part ] ";"
```

Dans notre transformation nous nous intéressons pas aux paramètres.

- Les appels d'entrées :

Les appels d'entrées ou encore demandes de rendez-vous peuvent apparaître dans divers contextes. Son écriture BNF se présente comme suit :

```
entry_call_statement ::= entry_name [ actual_parameter_part ] ";"
```

Dans notre transformation nous nous intéressons pas aux paramètres.

Le méta-modèle qui représente les instructions simples est présenté par la figure 34.

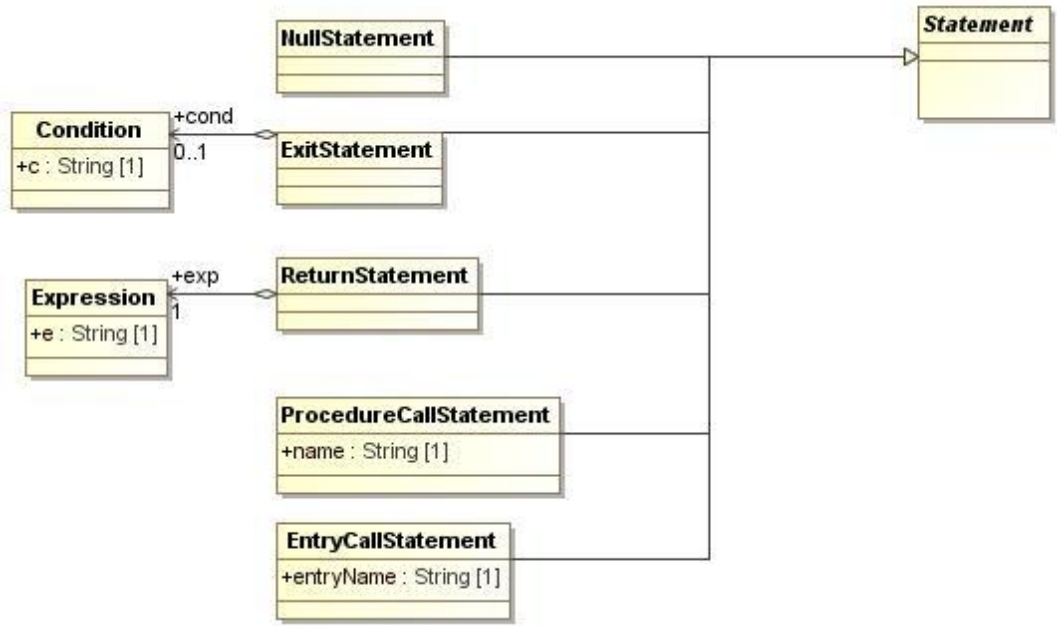

**Figure 34**: *Le méta-modèle des instructions simples*

Le méta-attribut name appartenant à la méta-classe ProcedureCallStatement mémorise l'identificateur de la procédure appelée. Egalement, le méta-attribut entryName stocke le nom de l'entrée appelé. Les deux méta-agrégations ExitStatement et Condition, et ReturnStatement et Expression modèlisent respectivement la condition attachée à l'instruction exit et l'expression associée à return.

## 4.3.2 Les instructions composées

- L'instruction if :

Dans notre cas, nous nous intéressons à un simple if_else. L'écriture BNF de l'instruction if se présente comme suit :

```
if_statement ::=
```

```
"if" condition "then"
sequence_of_statements
{     "elsif"     condition     "then"
sequence_of_statements }
[ "else" sequence_of_statements ]
"end" "if" ";"
condition ::= expression
```

- L'instruction case :

Son écriture BNF se présente comme suit :

```
case_statement ::=   "case" expression "is"
                              case_statement_alternative
                              { case_statement_alternative }
                              "end" "case" ";"
case_statement_alternative   ::=   "when"   discrete_choice_list   "=>"
sequence_of_statements
discrete_choice_list ::= discrete_choice { "|" discrete_choice }
discrete_choice ::= expression | discrete_range | "others"
```

- L'instruction accept :

Il s'agit d'une instruction d'acceptation d'un rendez-vous. Elle est utilisée au sein de la partie exécutive d'une tâche. Dans notre cas nous nous intéressons à une simple instruction accept. Son écriture BNF se présente comme suit :

```
accept_statement ::= "accept" direct_name
 [ "(" entry_index ")" ] parameter_profile
[ "do"  handled_sequence_of_statements "end" [ entry_identifier ] ] ";"
```

- L'instruction select:

Il s'agit d'une instruction utilisée au sein de la partie exécutive d'une tâche. Elle favorise le non déterminisme lors de l'acceptation des rendez-vous éventuellement gardés.

Dans notre cas nous intéressons à un simple select_or sans garde et sans alternative d'attente. Son écriture BNF se présente comme suit :

```
selective_accept ::= "select"[ guard ] select_alternative
                     { "or"        [ guard ]    select_alternative }
                     [ "else"  sequence_of_statements ]
                     "end" "select" ";"
guard ::= "when" condition "=>"
select_alternative                                          ::=
accept_alternative|delay_alternative|terminate_alternative
accept_alternative ::=  accept_statement [ sequence_of_statements ]
terminate_alternative ::= "terminate" ";"
```

- L'instruction loop :

Il s'agit de l'instruction itérative de base offerte par Ada. Dans notre cas nous nous intéressons à une simple instruction loop. Son écriture BNF se présente comme suit :

```
loop_statement ::= [ statement_identifier ":" ]
[ ( "while" condition ) |
("for" defining_identifier "in" ["reverse"] discrete_subtype_definition)
]
"loop"  sequence_of_statements  "end" "loop" [ statement_identifier] ";"
```

Le méta-modèle qui représente les instructions composées est présenté par la figure 35.

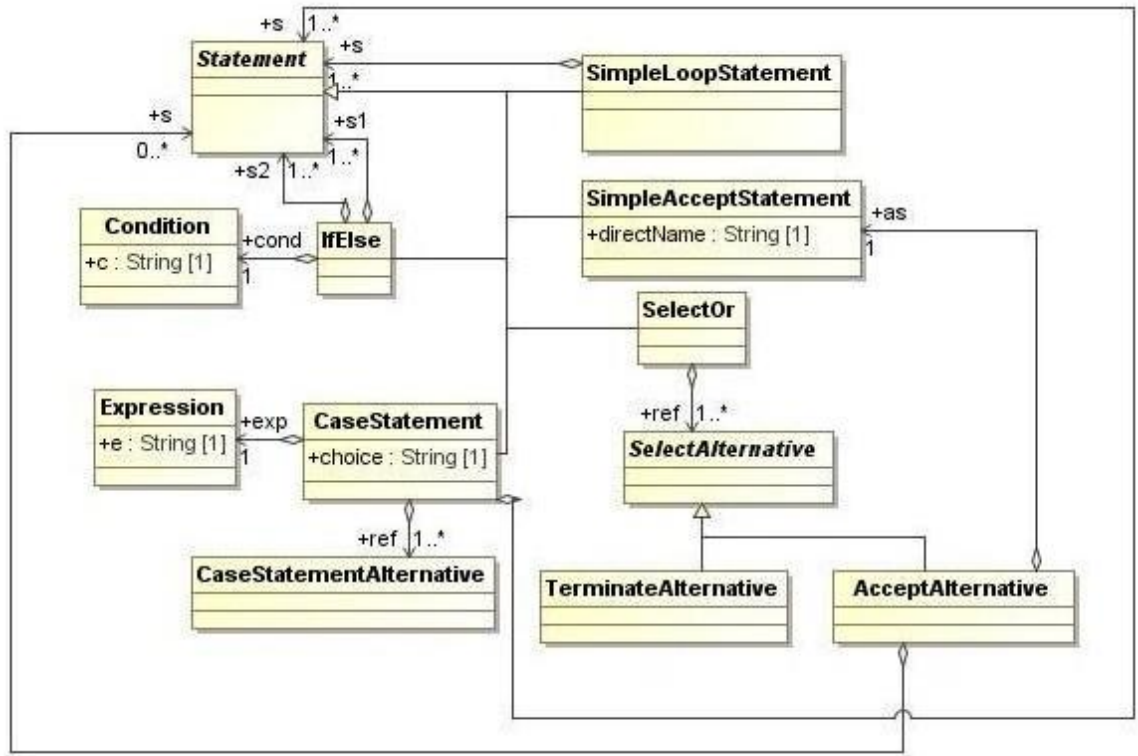

**Figure 35**: *Le méta-modèle des instructions composées*

La structure des instructions composites est définie d'une façon récursive. Par exemple, la méta-classe IfElse descend de Statement et regroupe plusieurs instructions dans les deux parties then et else. Ceci est traduit par les deux méta-agrégations orientées s1 et s2 entre IfElse et Statement.

## 4.4 Vue d'ensemble sur le méta-modèle partiel d'Ada

Le méta-modèle partiel d'Ada utilisé comme méta-modèle cible pour notre approche de transformation de Wright vers Ada est donné par la figure 36.

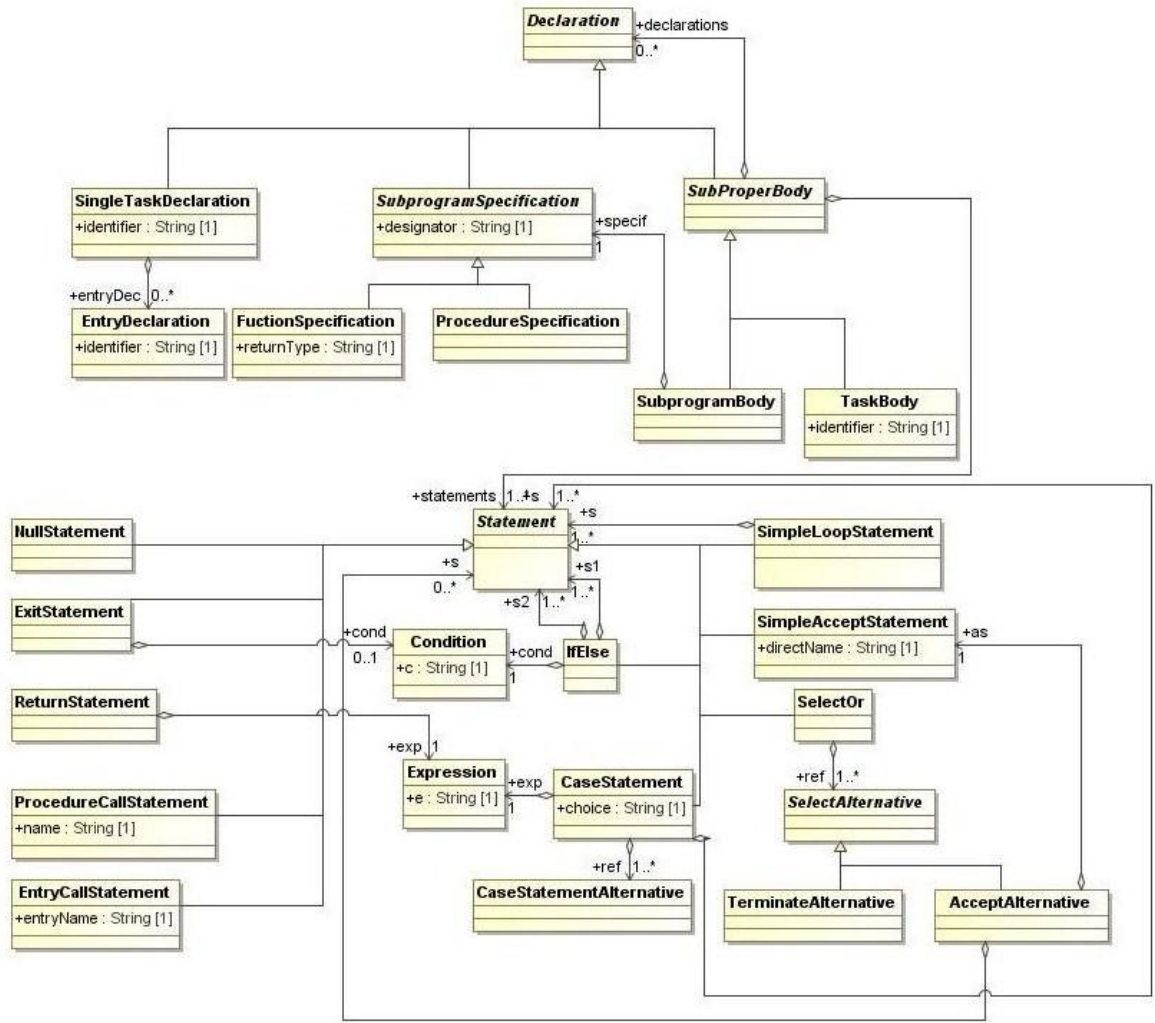

**Figure 36**: *Méta-modèle partiel d'Ada*

## 4.5 Les contraintes OCL

Nous avons établi plusieurs propriétés décrivant des contraintes d'utilisation des constructions d'Ada. De telles propriétés sont décrites d'une façon informelle et formelle en se servant d'OCL.

### 4.5.1 La sémantique statique de la partie stucturelle d'Ada

- Propriété 1 :

Au sein de la partie déclarative d'un sous-programme les noms des tâches (partie spécification implémentation) et des sous-programmes (partie spécification et implémentation) doivent être deux à deux différents.

```
context SubprogramBody
def:       col1:Sequence(String)     =     self.declarations     ->
select(e:Declaration|e.oclIsKindOf(SubprogramSpecification))     ->
collect(e:SubprogramSpecification|e.designator)
def:       col2:Sequence(String)     =     self.declarations     ->
select(e:Declaration|e.oclIsTypeOf(SingleTaskDeclaration))     ->
collect(e:SingleTaskDeclaration|e.identifier)
def:       col3:Sequence(String)     =     self.declarations     ->
select(e:Declaration|e.oclIsTypeOf(TaskBody))     ->
collect(e:TaskBody|e.identifier)
def:       col4:Sequence(String)     =     self.declarations     ->
select(e:Declaration|e.oclIsTypeOf(SubprogramBody))     ->
collect(e:SubprogramBody|e.specif.designator)
inv: col1 -> excludesAll(col2)
inv: col1 -> excludesAll(col3)
inv: col2 -> excludesAll(col4)
inv: col3 -> excludesAll(col4)
inv: col2->includesAll(col3) and col2->size()=col3->size()
```

- Propriété 2 :

Au sein de la partie déclarative d'un sous-programme, les identificateurs des sous-programmes doivent être différents.

```
context SubprogramBody
inv:                     self.declarations                     ->
select(e:Declaration|e.oclIsKindOf(SubprogramSpecification))     ->
forAll(e1:SubprogramSpecification,   e2:SubprogramSpecification|   e1<>e2
implies e1.designator<>e2.designator)
inv:                     self.declarations                     ->
select(e:Declaration|e.oclIsTypeOf(SubprogramBody))     ->
forAll(e1:SubprogramBody,     e2:SubprogramBody|     e1<>e2     implies
e1.specif.designator<>e2.specif.designator)
```

- Propriété 3 :

  Au sein de la partie déclarative d'un sous-programme les identificateurs des tâches doivent être différents.

```
context SubprogramBody
inv:                          self.declarations                          ->
select(e:Declaration|e.oclIsTypeOf(SingleTaskDeclaration))               ->
forAll(e1:SingleTaskDeclaration, e2:SingleTaskDeclaration| e1<>e2 implies
e1.identifier<>e2.identifier)
inv: self.declarations -> select(e:Declaration|e.oclIsTypeOf(TaskBody)) -
>        forAll(e1:TaskBody,      e2:TaskBody|      e1<>e2      implies
e1.identifier<>e2.identifier)
```

- Propriété 4 :

  Au sein de la partie déclarative d'une tâche les identificateurs des tâches (partie spécification implémentation) et des sous-programmes (partie spécification implémentation) doivent être deux à deux différents.

```
context TaskBody
def:       col1:Sequence(String)       =       self.declarations       ->
select(e:Declaration|e.oclIsKindOf(SubprogramSpecification))             ->
collect(e:SubprogramSpecification|e.designator)
def:       col2:Sequence(String)       =       self.declarations       ->
select(e:Declaration|e.oclIsTypeOf(SingleTaskDeclaration))               ->
collect(e:SingleTaskDeclaration|e.identifier)
def:       col3:Sequence(String)       =       self.declarations       ->
select(e:Declaration|e.oclIsTypeOf(TaskBody))                            ->
collect(e:TaskBody|e.identifier)
def:       col4:Sequence(String)       =       self.declarations       ->
select(e:Declaration|e.oclIsTypeOf(SubprogramBody))                      ->
collect(e:SubprogramBody|e.specif.designator)
inv: col1 -> excludesAll(col2)
inv: col1 -> excludesAll(col3)
inv: col2 -> excludesAll(col4)
```

```
inv: col3 -> excludesAll(col4)
inv: col2 -> includesAll(col3) and col2->size()=col3->size()
```

- Propriété 5 :

    Au sein de la partie déclarative d'une tâche les identificateurs des sous-programmes doivent être différents.

```
context TaskBody
inv:                          self.declarations                          ->
select(e:Declaration|e.oclIsKindOf(SubprogramSpecification))             ->
forAll(e1:SubprogramSpecification,  e2:SubprogramSpecification|  e1<>e2
implies e1.designator<>e2.designator)
inv:                          self.declarations                          ->
select(e:Declaration|e.oclIsTypeOf(SubprogramBody))                      ->
forAll(e1:SubprogramBody,      e2:SubprogramBody|      e1<>e2      implies
e1.specif.designator<>e2.specif.designator)
```

- Propriété 6 :

    Au sein de la partie déclarative d'une tâche les identificateurs des tâches être différents.

```
context TaskBody
inv:                          self.declarations                          ->
select(e:Declaration|e.oclIsTypeOf(SingleTaskDeclaration))               ->
forAll(e1:SingleTaskDeclaration, e2:SingleTaskDeclaration| e1<>e2 implies
e1.identifier<>e2.identifier)
inv: self.declarations -> select(e:Declaration|e.oclIsTypeOf(TaskBody)) -
>          forAll(e1:TaskBody,         e2:TaskBody|         e1<>e2      implies
e1.identifier<>e2.identifier)
```

## 4.5.2 La sémantique statique de la partie comportementale d'Ada

- Propriété 7 :

    Une fonction contient au moins une instruction return.

```
context SubprogramBody
inv: specif.oclIsTypeOf(FunctionSpecification) implies statements ->
collect(s:Statement|s.oclIsTypeOf(ReturnStatement)) -> size()>=1
```

- Propriété 8 :

  Un sous-programme ne contient pas d'instruction accept.

```
context SubprogramBody
inv:        statements        ->        forAll(s:Statement        |        not
s.oclIsTypeOf(SimpleAcceptStatement))
```

- Propriété 9 :

  Un sous-programme ne contient pas d'instruction select.

```
context SubprogramBody
inv: statements -> forAll(s:Statement | not s.oclIsTypeOf(SelectOr))
```

- Propriété 10 :

  Une tâche ne contient pas d'instruction return.

```
context TaskBody
inv:        statements        ->        forAll(s:Statement        |        not
s.oclIsTypeOf(ReturnStatement))
```

- Propriété 11 :

  Une tâche ne peut accepter des rendez-vous que sur ses propres entrées (entry).

```
context TaskBody
def:        c1:Sequence(String)        =        self.statements        ->
select(e:Statement|e.oclIsTypeOf(SimpleAcceptStatement))        ->
collect(e:SimpleAcceptStatement|e.direct_name)
def:        c2:Sequence(String)        =        self.declarations        ->
collect(e:SingleTaskDeclaration|e.entryDec)        ->
collect(e:EntryDeclaration|e.identifier)
inv: c2 -> includesAll(c1)
```

## 4.6 Conclusion

Dans ce chapitre, nous avons proposé un méta-modèle partiel d'Ada issu de la description BNF de ce langage. Notre méta-modèle comporte la structure des constructions utilisées dans la transformation de Wright vers Ada. Ces constructions sont : sous-programmes non paramétrés, tâches ayant des entrées non paramétrées, structures de contrôle, demande d'un rendez-vous, acceptation d'un rendez-vous, et instruction de non déterminisme. Dans le chapitre suivant, nous proposons un programme Wright2Ada en ATL permettant de transformer Wright vers Ada en utilisant les deux méta-modèles Wright et Ada (voir chapitres 3 et 4) et les règles de transformation systématique de Wright vers Ada proposées dans le chapitre 1.

# Chapitre 5: Transformation de Wright vers Ada : le programme Wright2Ada en ATL

Dans ce chapitre, nous présentons d'une façon détaillée le programme Wright2Ada écrit en ATL permettant de transformer une architecture logicielle décrite en Wright vers un programme concurrent Ada. Afin de concevoir et développer notre programme Wright2Ada, nous avons utilisé avec profit les constructions : règle standard, règle paresseuse, helper attribut, helper opération et helper opération défini dans le contexte d'éléments de modèles favorisant les appels polymorphiques offertes par le langage de transformation de modèles ATL. En outre, nous avons utilisé la récursivité afin de programmer les helpers fournis par notre programme Wright2Ada. Ce dernier est purement déclaratif. En effet, nous avons réussi à éviter l'utilisation des constructions impératives offertes pour ATL telles que : bloc impératif et règles appelées. L'annexe A présente en entier le programme Wright2Ada et l'annexe B donne un exemple d'utilisation de ce programme.

Ce chapitre comporte trois sections. La première section donne le schéma d'utilisation du programme Wright2Ada. Les deux autres sections décrivent d'une façon assez détaillée le programme Wright2Ada en traitant respectivement les aspects structuraux et comportementaux de traduction de Wright vers Ada.

## 5.1 Vue d'ensemble sur le programme Wright2Ada

La figure 37 donne le contexte de notre programme Wright2Ada permettant de transformer une architecture logicielle décrite en Wright vers un programme concurrent Ada.

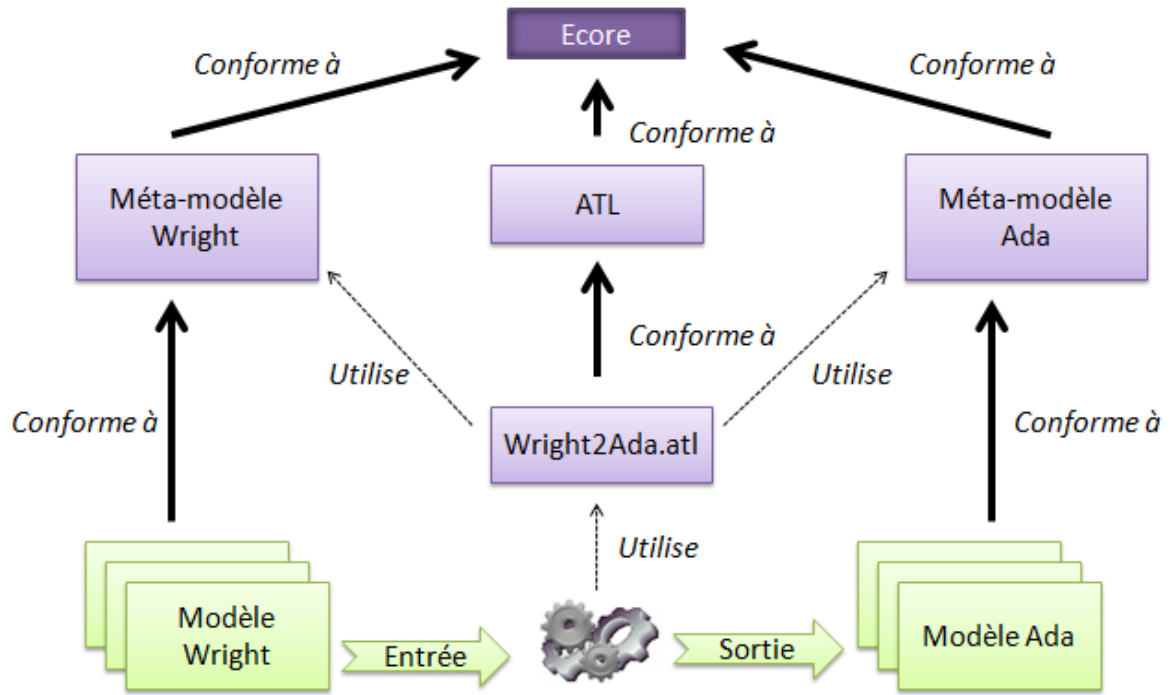

**Figure 37**: *Contexte général du programme Wright2Ada*

Les modèles source et cible (architecture logicielle en Wright et programme concurrent en Ada) ainsi que le programme Wright2Ada sont conforme à leurs méta-modèles Wright, Ada et ATL. Ces méta-modèles sont conformes au méta-modèle Ecore.

Le méta-modèle source de Wright, respectivement cible d'Ada, est représenté par un diagramme Ecore donné par la figure 38, respectivement par la figure 39.

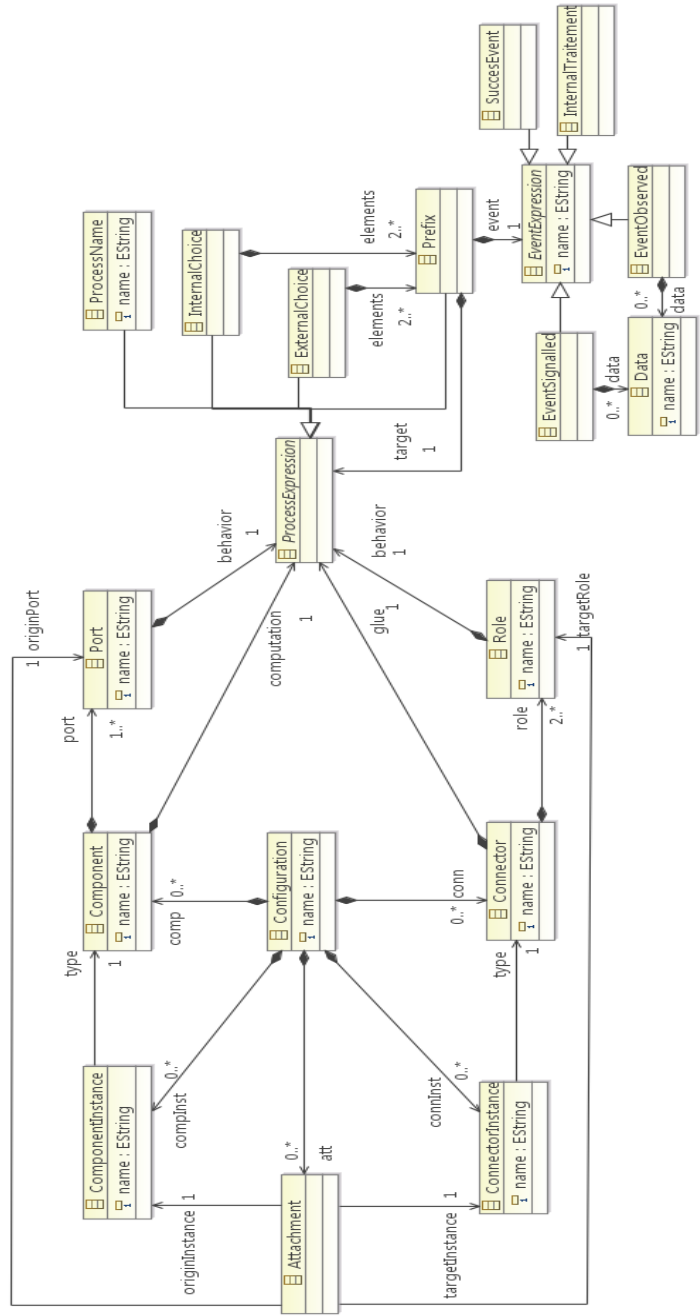

**Figure 38**:*Le méta-modèle Wright en diagramme Ecore*

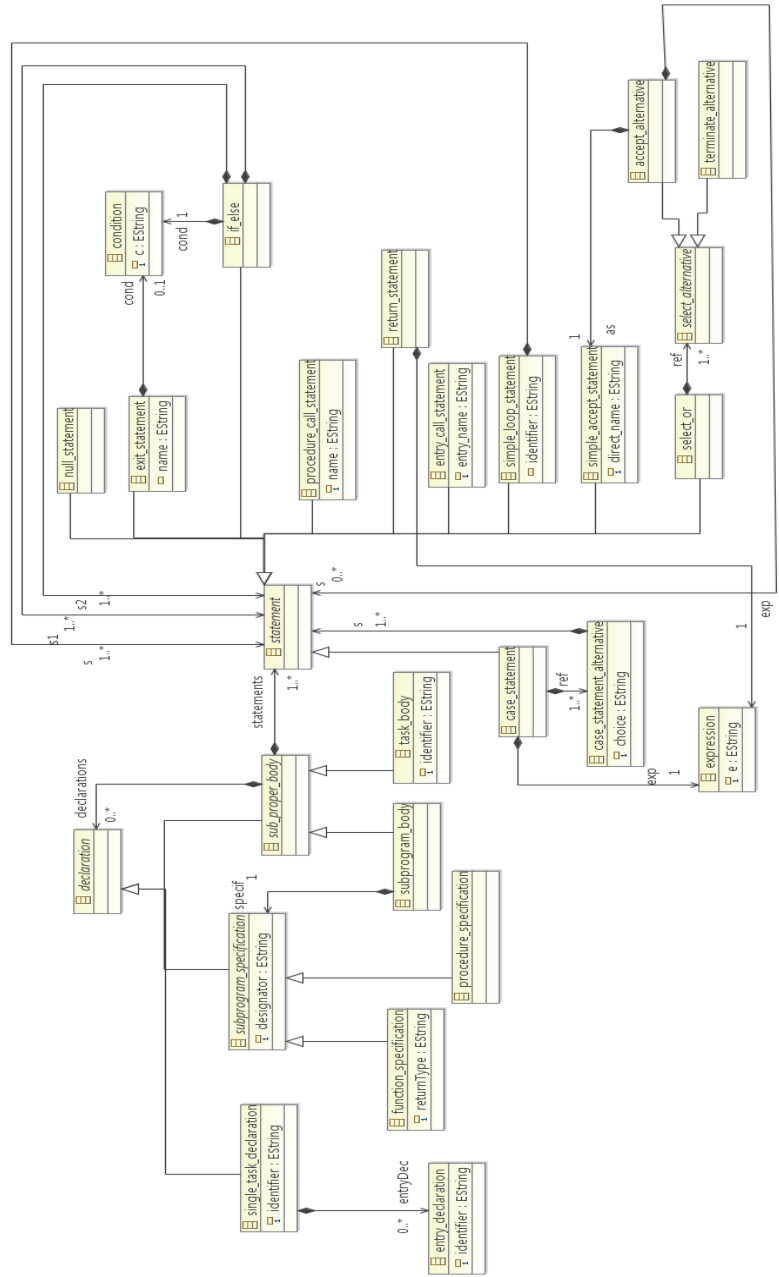

**Figure 39**: Le méta-modèle partiel d'Ada en diagramme ecore

L'entête du programme Wright2Ada stocké dans le fichier Wright2Ada.atl se présente comme suit :

```
module WrightToAda;
create exampleAda : Ada from exampleWright : Wright;
```

Dans notre programme le modèle cible est représenté par la variable exampleAda à partir du modèle source représenté par exampleWright. Les modèles source et cible sont respectivement conformes aux méta-modèles Wright et Ada. Notre programme Wright2Ada opère sur le modèle source exampleWright **en lecture seule** et produit le modèle cible exampleAda **en écriture seule**.

Dans la suite, nous allons présenter progressivement les helpers et les règles standard et paresseuses formant notre programme Wright2Ada écrit en ATL. Notre transformation de Wright vers Ada est basée sur les règles issues de (Bhiri, 2008).

## 5.2 Traduction de la partie structurelle de l'ADL Wright

Dans cette section, nous présentons la traduction des aspects structuraux de Wright. Chaque règle de transformation est présentée informellement et illustrée sur un exemple avant même de passer à sa formalisation en ATL. Les règles de transformation de la partie structurelle de Wright vers Ada sont illustrées sur l'architecture Client-serveur donnée dans la figure 40.

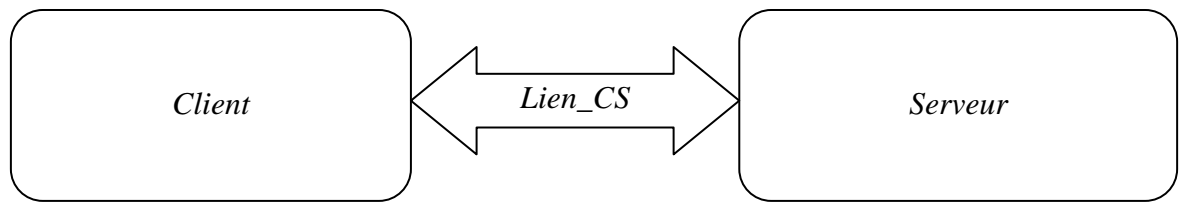

**Figure 40**: *Exemple Client-Serveur*

Dans ce type d'architecture le composant *Client* envoie une requête au composant *Serveur* et attend sa réponse. Le composant *Serveur* quant à lui attend la requête pour répondre. Le connecteur *Lien_CS* joue le rôle d'intermédiaire entre le composant *Client* et le composant *Serveur*.

### 5.2.1 Traduction d'une configuration Wright

Une configuration Wright est traduite en Ada par une procédure. Cette tâche ne fait rien (corps vide); elle constitue une structure d'accueil.

▪ Illustration sur l'exemple Client-Serveur :

| Modélisation en Wright | Modélisation en Ada |
|---|---|
| Configuration Client_Serveur<br><br>...<br><br>End Configuration | procedure Client_Serveur is<br><br>...<br><br>begin<br><br>     null;<br><br>end Client_Serveur; |

▪ Traduction en ATL :

```
rule Configuration2subprogram{
    from c: Wright!Configuration
    to      sb: Ada!subprogram_body (
        specif <- sp ,
        statements <- st ,
        declarations <- ...
        )      ,
        sp: Ada!procedure_specification(    designator <- c.name),
        st: Ada!null_statement
}
```

Dans cette règle nous créeons la procédure qui constitue la structure d'accueil de notre configuration. Dans sa spécification, elle porte le nom de la configuration en question, soit *c.name*, et elle contiendra l'instruction nulle. Sa partie déclarative sera fournie ultérieurement.

## 5.2.2 Traduction de la partie structurelle d'une instance de composant et de connecteur

Chaque instance de type composant est traduite par une tâche Ada portant le nom *Component_nomInstanceComposant*.

Chaque instance de type connecteur est traduite par une tâche Ada portant le nom *Connector_nomInstanceConnecteur*.

Les noms sont conservés pour des raisons de traçabilité.

▪ Illustration sur l'exemple Client-Serveur :

| Modélisation en Wright | Modélisation en Ada |
|---|---|
| Configuration Client_Serveur | procedure Client_Serveur is |
| Component Client | task Component_client1 is |
| ... | ... |
| Component Serveur | end Component_client1; |
| ... | task Component_seveur1 is |
| Connector Lien_CS | ... |
| ... | end Component_serveur1; |
| Instances | task Connector_ appel_cs is |
| client1: Client | ... |
| serveur1: Serveur | end Connector_ appel_cs; |
| appel_cs: Lien_CS | task body Component_client1 is |
| Attachments | begin |
| ... | ... |
| End Configuration | end Component_client1; |
| | |
| | task body Component_seveur1 is |
| | begin |
| | ... |
| | end Component_serveur1; |
| | |
| | task body Connector_ appel_cs is |
| | begin |
| | ... |
| | end Connector_ appel_cs; |
| | begin |
| | null; |
| | end Client_Serveur ; |

▪ Traduction en ATL :

```
rule Configuration2subprogram{
    from c: Wright!Configuration
    to      sb: Ada!subprogram_body (
        specif <- sp ,
        statements <- st ,
        declarations           <-           c.compInst          ->
collect(e|thisModule.ComponentInstance2single_task_declaration(e))
        ->union(c.connInst                                      ->
collect(e|thisModule.ConnectorInstance2single_task_declaration(e)))
        ->union(c.compInst                                     ->
collect(e|thisModule.ComponentInstance2task_body(e)))
```

```
              ->union(c.connInst                        ->
collect(e|thisModule.ConnectorInstance2task_body(e)))
              ...
              )          ,
              sp: Ada!procedure_specification(   designator <- c.name),
              st: Ada!null_statement
}
```

La partie déclarative et le corps des tâches font parties de la partie déclarative de la procédure qui joue le rôle de structure d'accueil. Cette règle déclenche les règles paresseuses correspondantes à la partie déclarative et le corps des tâches des instances de composants et de connecteurs.

```
lazy rule ComponentInstance2single_task_declaration{
      from ci:Wright!ComponentInstance
      to std:Ada!single_task_declaration(
            identifier <- 'Component_'+ci.name,
            entryDec <-...
            )
}
lazy rule ComponentInstance2task_body{
      from ci:Wright!ComponentInstance
      to  tb:Ada!task_body(
            identifier <-'Component_'+ ci.name,
            statements <- ...
            )
}
```

Dans la partie déclarative et dans le corps des tâches qui représent les instances de composants nous préservons le nom de l'instance *ci.name* suivi par le préfixe *Component_*. Les instructions des tâches seront fournies ultérieurement.

```
lazy rule ConnectorInstance2single_task_declaration{
      from ci:Wright!ConnectorInstance
      to std:Ada!single_task_declaration(
            identifier <- 'Connector_'+ci.name,
            entryDec <-...
            )
}
lazy rule ConnectorInstance2task_body{
      from ci:Wright!ConnectorInstance
      to  tb:Ada!task_body(
            identifier <-'Connector_'+ ci.name,
            statements <- ...
            )
}
```

Dans la partie déclarative et dans le corps des tâches qui représent les instances de connecteurs nous préservons le nom de l'instance *ci.name* suivi par le préfixe *Connector_*. Les instructions des tâches représentant les connecteurs seront fournis ultérieurement.

## 5.3 Traduction de la partie comportementale de l'ADL Wright

Cette section présente la traduction des aspects comportementaux de Wright décrits en CSP.

### 5.3.1 Elaboration de la partie déclarative des tâches représentant les instances de composants et de connecteurs

Les événements observés de la partie calcul (Computation) d'un composant, ainsi que de la glu (Glue) d'un connecteur représentent les entrées des tâches qui les matérialisent. Afin d'identifier ces entrées (entry), nous nous inspirons des deux algorithmes décrits dans (Bhiri, 2008) mais pour plus de facilité dans l'automatisation, nous raisonnons sur la partie calcul au lieu des ports respectivement glu au lieu des rôles.

✓ Algorithme d'élaboration de la partie déclarative des tâches représentant les instances de composants :

Pour chaque instance de type composant
Faire
        Pour chaque événement appartenant à Computation
        Faire
                Si événement est un événement observé de la forme «nomPort.événement»
                Alors
                        Créer une entrée portant le nom  nomPort_événement
                        Soit *entry nomPort_événement ;*
                Fin Si
        Fin Faire
Fin Faire

✓ Algorithme d'élaboration de la partie déclarative des tâches représentant les instances de connecteurs :

Pour chaque instance de type connecteur
Faire
        Pour chaque événement appartenant à Glue
        Faire
                Si événement est un événement observé de la forme «nomRôle.événement»
                Alors
                        Créer une entrée portant le nom  nomRôle_événement
                        Soit *entry nomRôle_événement ;*
                Fin Si
        Fin Faire

Fin Faire

- ▪ Illustration sur l'exemple Client-Serveur :

| Modélisation en Wright | Modélisation en Ada |
|---|---|
| Configuration Client_Serveur | procedure Client_Serveur is |
| Component Client | task Component_client1 is |
| ... | entry port_Client_reponse; |
| Computation= traitement_interne -> | end Component_client1; |
| _port_Client.requete -> port_Client.reponse | task Component_seveur1 is |
| -> Computation \|~\| § | entry port_Serveur_requete; |
| Component Serveur | end Component_serveur1; |
| ... | task Connector_ appel_cs is |
| Computation= traitement_interne -> | entry Appelant_requete; |
| port_Serveur.requete -> | entry Appele_reponse; |
| _port_Serveur.reponse -> Computation \|~\| § | end Connector_ appel_cs; |
| Connector Lien_CS | task body Component_client1 is |
| ... | begin |
| Glue= Appelant.requete -> | ... |
| _Appele.requete -> Glue | end Component_client1; |
| [] Appele.reponse -> | task body Component_seveur1 is |
| _Appelant.reponse -> Glue | begin |
| [] § | ... |
| Instances | end Component_serveur1; |
| client1: Client | task body Connector_ appel_cs is |
| serveur1: Serveur | begin |
| appel_cs: Lien_CS | ... |
| Attachments | end Connector_ appel_cs; |
| ... | begin |
| End Configuration | null; |
| | end Client_Serveur ; |

▪ Traduction en ATL :

Afin d'élaborer la partie déclarative des tâches représentant les instances de composants et de connecteurs, un parcours du processus CSP Wright représentant la partie calcul d'un composant et la glu d'un connecteur est indispensable. Le helper *getEventObserved* fourni ci-dessous permet de faire le parcours nécessaire du processus CSP Wright, à la recherche des événements observés. Il retourne à la règle appelante un ensemble *Set* contenant les événements observés rencontrés lors de son parcours.

```
helper    context    Wright!ProcessExpression    def:    getEventObserved():
Set(Wright!EventObserved) =
          if self.oclIsTypeOf(Wright!Prefix)then
                if self.event.oclIsTypeOf(Wright!EventObserved) then
                      Set{self.event}-
>union(self.target.getEventObserved())
                else
                      self.target.getEventObserved()
                endif
          else
              if      self.oclIsTypeOf(Wright!InternalChoice)      or
self.oclIsTypeOf(Wright!ExternalChoice) then
                      self.elements->iterate(  child1  ;  elements1  :
Set(Wright!EventObserved) = Set{} | elements1-
>union(child1.getEventObserved()))
                else
                      Set{}
                endif
          endif;
```

La règle paresseuse *ComponentInstance2single_task_declaration* fournie ci-dessous, correspond à la traduction de la partie déclarative des tâches représentant les instances de composants. Elle comporte un appel au helper *getEventObserved*, qui retourne l'ensemble des événements observés dans la partie calcul du type composant de l'instance de composant, et déclenche la règle paresseuse qui transforme un événement observé en une entrée *EventObserved2entry_declaration*.

```
lazy rule ComponentInstance2single_task_declaration{
      from ci:Wright!ComponentInstance
      to std:Ada!single_task_declaration(
            identifier <- 'Component_'+ci.name,
            entryDec           <-ci.type.computation.getEventObserved()-
>collect(e|thisModule.EventObserved2entry_declaration(e))
            )
}
```

La règle paresseuse *ConnectorInstance2single_task_declaration* fournie ci-dessous, correspond à la traduction de la partie déclarative des tâches représentant les instances de connecteurs. Elle comporte un appel au helper *getEventObserved*, qui retourne l'ensemble des événements observés dans la glu du type connecteur de l'instance de connecteur, et déclenche la règle paresseuse qui transforme un événement observé en une entrée *EventObserved2entry_declaration*.

```
lazy rule ConnectorInstance2single_task_declaration{
    from ci:Wright!ConnectorInstance
    to std:Ada!single_task_declaration(
        identifier <- 'Connector_'+ci.name,
        entryDec            <-ci.type.glue.getEventObserved()-
>collect(e|thisModule.EventObserved2entry_declaration(e))
        )
}
```

La règle paresseuse qui transforme un événement observé en une entrée à la tâche se présente comme suit :

```
lazy rule EventObserved2entry_declaration{
    from eo:Wright!EventObserved
    to ed:Ada!entry_declaration(
        identifier<- eo.name.replaceAll('.','_')
    )
}
```

### 5.3.2 Traduction des événements internes

Les événements internes contenus dans une configuration, c'est-à-dire dans la description des comportements de ses composants ou de ses connecteurs, sont traduits par des procédures dont le corps est à raffiner. Dans cette traduction, le corps de ces procédures contiendra l'instruction nulle.

▪ Illustration sur l'exemple Client-Serveur :

| Modélisation en Wright | Modélisation en Ada |
|---|---|
| Configuration Client_Serveur | procedure Client_Serveur is |
| Component Client | procedure traitement_interne is |
| ... | begin |
| Computation= traitement_interne -> | null; |
| _port_Client.requete -> port_Client.reponse -> Computation |~| § | end traitement_interne; |
| Component Serveur | task Component_client1 is |
| ... | entry port_Client_reponse; |
| Computation= traitement_interne -> port_Serveur.requete -> _port_Serveur.reponse -> Computation |~| § | end Component_client1; |
| | task Component_seveur1 is |
| | entry port_Serveur_requete; |
| Connector Lien_CS | end Component_serveur1; |
| ... | task Connector_ appel_cs is |
| Glue= Appelant.requete -> _Appele.requete -> Glue | entry Appelant_requete; |
| | entry Appele_reponse; |
| [] Appele.reponse -> _Appelant.reponse -> Glue | end Connector_ appel_cs; |
| | task body Component_client1 is |
| [] § | begin |
| Instances | ... |
| client1: Client | end Component_client1; |
| serveur1: Serveur | task body Component_seveur1 is |
| appel_cs: Lien_CS | begin |
| Attachments | ... |
| ... | end Component_serveur1; |
| End Configuration | task body Connector_ appel_cs is |
| | begin |
| | ... |
| | end Connector_ appel_cs; |
| | begin |
| | null; |
| | end Client_Serveur ; |

Traduction en ATL :

Pour ajouter les procédures représentant l'ensemble des événements internes contenus dans une configuration, un parcours des parties calcul (Computation) des composants et des parties glu (Glue) des connecteurs contenus dans cette configuration est indispensable.

```
helper    context    Wright!ProcessExpression    def:    getInternalTrait():
Set(Wright!InternalTraitement) =
            if self.oclIsTypeOf(Wright!Prefix)then
                    if    self.event.oclIsTypeOf(Wright!InternalTraitement)
then
                        Set{self.event}-
>union(self.target.getInternalTrait())
                    else
                        self.target.getInternalTrait()
                    endif
            else
                    if      self.oclIsTypeOf(Wright!InternalChoice)    or
self.oclIsTypeOf(Wright!ExternalChoice)  then
                        self.elements->iterate(  child1  ;  elements1  :
Set(Wright!InternalTraitement) = Set{} | elements1-
>union(child1.getInternalTrait()))
                    else
                        Set{}
                    endif
            endif;
```

Le helper *getInternalTrait* fait le parcours d'un processus CSP Wright à la recherche des événements internes.

```
helper    context    Wright!Configuration    def:    getInternalTraitement:
Set(Wright!InternalTraitement) =
        self.conn->iterate(         child1         ;         elements1         :
Set(Wright!InternalTraitement) = Set{} | elements1-
>union(child1.glue.getInternalTrait()))
        ->union(self.comp->iterate(       child2       ;       elements2       :
Set(Wright!InternalTraitement) = Set{} | elements2-
>union(child2.computation.getInternalTrait())));
```

Le helper *getInternalTraitement* permet de collecter les traitements internes contenus dans la partie *Computation* des composants et dans la partie *Glue* des connecteurs de la configuration. Pour y parvenir, ce helper fait appel au helper *getInternalTrait*.

Une mise à jour est apportée à la règle de transformation de la configuration en une procédure. Cette règle contiendra, de plus, un appel au helper *getInternalTraitement* qui collecte l'ensemble des événements internes dans la configuration pour déclencher ensuite la règle paresseuse qui transforme un traitement interne en une procédure. La mise à jour, ainsi que la règle paresseuse déclenchées sont présentées comme suit :

```
rule Configuration2subprogram{
        from c: Wright!Configuration
        to      sb: Ada!subprogram_body (
            specif <- sp ,
            statements <- st ,
```

```
            declarations              <-c.getInternalTraitement           ->
collect(e|thisModule.InternalTraitement2subprogram(e))
            ->union(c.compInst                                           ->
collect(e|thisModule.ComponentInstance2single_task_declaration(e)))
            ->union(c.connInst                                           ->
collect(e|thisModule.ConnectorInstance2single_task_declaration(e)))
            ->union(c.compInst                                           ->
collect(e|thisModule.ComponentInstance2task_body(e)))
            ->union(c.connInst                                           ->
collect(e|thisModule.ConnectorInstance2task_body(e)))),
            sp: Ada!procedure_specification(   designator <- c.name),
            st: Ada!null_statement
}
```

La règle paresseuse *InternalTraitement2subprogram* fournie ci-dessous traduit un événement interne en une procédure dont le corps est vide, a priori, et dont le nom est celui de l'événement interne en question.

```
lazy rule InternalTraitement2subprogram{
    from i:Wright!InternalTraitment
    to sb: Ada!subprogram_body(
        specif <- ps,
        statements <-ns),
        ns:Ada!null_statement,
        ps: Ada!procedure_specification(   designator <- i.name)
}
```

### 5.3.3 Traduction de l'opérateur de récursivité

Tous les processus relatifs à la description des composants et des connecteurs ont un aspect récursif. Dans notre cas, nous nous intéressons plus particulièrement au processus de description de la partie calcul d'un composant et de la glu d'un connecteur. L'opérateur de récursivité est traduit par l'instruction *loop* d'Ada.

- ▪ Traduction en ATL :

En tenant compte du fait que les processus représentant la *computation* d'un composant et la *glue* d'un connecteur sont délimités par l'opérateur de récursivité de CSP Wright, il en sera de même pour la traduction en Ada qui commencent par l'instruction *loop*. Ceci peut être traduit par les deux règles paresseuses suivantes :

```
lazy rule ComponentInstance2task_body{
    from ci:Wright!ComponentInstance
    to  tb:Ada!task_body(
        identifier <-'Component_'+ ci.name,
        statements <- ls
        ),
        ls : Ada!simple_loop_statement(
```

```
                        s<- ci.type.computation.transformation(ci.name)
                        )
}
lazy rule ConnectorInstance2task_body{
      from ci:Wright!ConnectorInstance
      to    tb:Ada!task_body(
            identifier <-'Connector_'+ ci.name,
            statements <- ls
            ),
            ls : Ada!simple_loop_statement(
                  s<- ci.type.glue.transformation(ci.name))
}
```

L'élaboration du corps de la boucle *loop* se fait par l'intermédiaire du helper *transformation(instance : String)* Celui-ci est redéfini plusieurs fois selon le contexte dans lequel il est appelé, il prend comme paramètre le nom de l'instance de composant ou de connecteur qui l'appelle. Le nom de l'instance passée en paramètre effectif est passé de ce niveau vers le niveau inférieur.

### 5.3.4 Traduction de l'opérateur de choix externe

L'opérateur de choix externe ou de choix déterministe est traduit en Ada par l'instruction *select*.

| Modélisation en CSP Wright | Modélisation en Ada |
|---|---|
| a ->P1 [] b->P2 [] V->STOP [] c->Q | *Select*<br><br>          traduction de a puis de P1<br>*or*<br><br>          traduction de b puis de P2<br>*or*<br><br>          traduction de c<br>*or*<br><br>          *terminate ;*<br>*end select ;* |
| Les a, b et c sont des événements observés.<br>Le « V » est l'événement succès.<br>Le Pi peut être : préfixe ou un opérateur de choix externe ou un opérateur de choix interne.<br>Le Q et le STOP sont des processus. | Nous commençons par la traduction des préfixes qui commencent par les événements observés suivis de la traduction du préfixe |

| | qui commence par l'événement succès « V » s'il existe. |
|---|---|

- Traduction en ATL :

Le helper *getPrefixInOrder* permet de réordonner les préfixes contenus dans l'operateur de choix externe de façon à avoir les préfixes qui commencent par un événement observé suivi du préfixe qui commence par l'événement succès s'il existe. Ce helper retourne un ensemble ordonné contenant l'ensemble des préfixes directement accessibles par l'opérateur de choix externe.

```
helper           context        Wright!ExternalChoice                def:
getPrefixInOrder():OrderedSet(Wright!Prefix) =
     self.elements->select(c                                           |
c.event.oclIsTypeOf(Wright!EventObserved))
     ->union(self.elements->select(c                                   |
c.event.oclIsTypeOf(Wright!SuccesEvent)));
```

Le helper redéfini, qui permet de déclencher la règle paresseuse responsable de la transformation d'un opérateur de choix externe en une instruction select se présente comme suit :

```
helper   context  Wright!ExternalChoice  def:  transformation(instance  :
String):Ada!select_or=
     thisModule.ExternalChoice2select_or(self,instance);
```

Le déclenchement est délégué au helper transformation pour pouvoir profiter des facilités offertes par la propriété de redéfinition.

La règle paresseuse responsable de la traduction d'un opérateur de choix externe en une instruction *select* se présente comme suit :

```
lazy rule ExternalChoice2select_or{
     from p:Wright!ExternalChoice,
          instance : String
     to s:Ada!select_or(
          ref <- p.getPrefixInOrder()->collect(e|
               if e.event.oclIsTypeOf(Wright!EventObserved)then
                    if e.target.oclIsTypeOf(Wright!ProcessName)  then

     thisModule.Prefix2accept_alternative1(e,instance)
                    else

     thisModule.Prefix2accept_alternative2(e,instance)
                    endif
               else
```

```
                    thisModule.SuccesEvent2terminate_alternative(e)
            endif ))
}
```

Cette règle paresseuse fait appel au helper *getPrefixInOrder* pour ordonner les préfixes directement accessibles puis déclenche la règle paresseuse adéquate. Si le préfixe commence par un événement observé suivi du nom d'un processus un déclenchement de la règle paresseuse *Prefix2accept_alternative1* aura lieu, si le préfixe commence par un événement observé suivi d'un opérateur de choix externe ou interne ou un autre préfixe, un déclenchement de la règle paresseuse *Prefix2accept_alternative2* aura lieu. Si le préfixe commence par l'événement succès un déclenchement de la règle paresseuse *SuccesEvent2terminate_alternative* aura lieu.

```
lazy rule Prefix2accept_alternative1{
    from p:Wright!Prefix,
        instance : String
    to a:Ada!accept_alternative(
    as <- thisModule.EventObserved2simple_accept_statement(p.event)
        )
}
```

Cette règle paresseuse correspondant à une alternative de l'instruction *select*. Elle déclenche la règle paresseuse permetant de traduire un événement observé.

```
lazy rule Prefix2accept_alternative2{
    from p:Wright!Prefix,
        instance : String
    to a:Ada!accept_alternative(
    as <- thisModule.EventObserved2simple_accept_statement(p.event),
        s<- p.target.transformation(instance)
        )
}
```

Cette règle paresseuse admet le même comportement que la règle précédente, avec en plus, l'appel au helper *transformation* afin de traduire la cible du préfixe.

```
lazy rule SuccesEvent2terminate_alternative{
    from p:Wright!SuccesEvent
    to a:Ada!terminate_alternative
}
```

Cette règle paresseuse correspondant à l'alternative *terminate* de l'instruction *select*.

## 5.3.5 Traduction de l'opérateur de choix interne

Dans la traduction de l'opérateur de choix interne ou de choix non déterministe, on distingue deux cas : la traduction de l'opérateur de choix interne binaire et la traduction de l'opérateur de choix interne généralisé (n-aire avec n>2).

La traduction de l'operateur de choix interne se fait par l'intermédiaire du helper redéfini *transformation* suivant :

```
helper context Wright!InternalChoice def: transformation(instance :
String):Ada!statement=
        if self.elements->size()=2 then
                thisModule.InternalChoice2if_else(self,instance)
        else
                thisModule.InternalChoice2case_statement(self,instance)
        endif;
```

### 5.3.5.1 Traduction de l'opérateur de choix interne binaire

L'opérateur de choix interne binaire ou de choix non déterministe binaire est traduit en Ada par l'instruction *if*.

| Modélisation en CSP Wright | Modélisation en Ada |
|---|---|
| P1 \|~\| P2<br><br>P1 et P2 sont des préfixes. | *if condition_interne then*<br>            traduction de P1<br>*else*<br>            traduction de P2<br>*end if ;*<br>« *condition_interne* » est une fonction qui retourne un booléen. |

- Traduction en ATL :

La règle paresseuse suivante permet de traduire un opérateur de choix interne binaire en instruction *if*. Cette règle fait appel au helper redéfini *transformation* qui va déclencher la règle adéquate à la traduction du premier et deuxième préfixe directement accessibles.

```
lazy rule InternalChoice2if_else{
    from p:Wright!InternalChoice,
        instance : String
    to ls:Ada!if_else(
            s1 <- p.elements->at(1).transformation(instance),
            s2 <- p.elements->at(2).transformation(instance),
            cond<-c
            ),
            c:Ada!condition(c<-'condition_interne')
}
```

Une mise à jour est nécessaire pour la règle traduisant la structure d'accueil qui doit, de plus, inclure la fonction *condition_interne*. Le corps de cette fonction est à raffiner.

```
helper context Wright!Configuration def: ICBin:Wright!InternalChoice=
```

```
        Wright!InternalChoice.allInstances() ->select(e |e.elements-
>size()=2)->at(1);

helper context Wright!Configuration def: existICBin:Boolean=
        if(Wright!InternalChoice.allInstances() ->select(e |e.elements-
>size()=2)->isEmpty() )then false else true endif;

rule Configuration2subprogram{
        from c: Wright!Configuration
        to      sb: Ada!subprogram_body (
            specif <- sp ,
            statements <- st ,
            declarations <- c.getInternalTraitement ->
collect(e|thisModule.InternalTraitement2subprogram(e))
            ->union(if (c.existICBin)then
OrderedSet{thisModule.InternalChoiceB2function(c.ICBin)} else
OrderedSet{}endif)
            ->union(c.compInst ->
collect(e|thisModule.ComponentInstance2single_task_declaration(e)))
            ->union(c.connInst ->
collect(e|thisModule.ConnectorInstance2single_task_declaration(e)))
            ->union(c.compInst ->
collect(e|thisModule.ComponentInstance2task_body(e)))
            ->union(c.connInst ->
collect(e|thisModule.ConnectorInstance2task_body(e)))),
            sp: Ada!procedure_specification(   designator <- c.name),
            st: Ada!null_statement
}

lazy rule InternalChoiceB2function{
        from i:Wright!InternalChoice(not i.OclUndefined())
        to   s:Ada!subprogram_body(specif <- fs, statements <- r),
            fs: Ada!function_specification(    designator <-
'condition_interne', returnType<-'Boolean'),
            r:Ada!return_statement(exp<-e1),
            e1:Ada!expression(e<-'true')
}
```

## 5.3.5.2 Traduction de l'opérateur de choix interne généralisé

L'opérateur de choix interne généralisé ou de choix non déterministe généralisé est traduit en Ada par l'instruction *case*.

| Modélisation en CSP Wright | Modélisation en Ada |
|---|---|
| P1 \|~\| P2 \|~\| P3 \|~\| … <br><br><br> Les Pi sont des préfixes. | *case condition_interne1 is* <br><br> *when* 1 => traduction de P1 <br><br> *when* 2 => traduction de P2 <br><br> *when* 3 => traduction de P3 |

| | … |
|---|---|
| | *end case ;* |
| | *« condition_interne1 »* est une fonction qui retourne un entier. |

- ▪ Traduction en ATL :

La règle paresseuse suivante permet de traduire l'opérateur de choix interne généralisé en une instruction *case*. Cette règle déclenche la règle paresseuse qui traduit les préfixes directement accessibles en passant, entre autre, le numéro d'ordre du préfixe courant dans l'ensemble ordonné en paramètre.

```
lazy rule InternalChoice2case_statement{
    from p:Wright!InternalChoice,
        instance : String
    to ls:Ada!case_statement(
        ref                        <-                   p.elements-
>collect(e|thisModule.Prefix2case_statement_alternative(e,p.elements.inde
xOf(e),instance)),
        exp<-c
        ),
        c:Ada!Expression(e<-'condition_interne1')
}
```

La règle paresseuse fournie ci-dessous permet de traduire les préfixes directement accessibles en leurs donnant l'indice passé en paramètre. Cette règle fait appel au helper redéfini *transformation* qui va déclencher la règle adéquate à la traduction de ces préfixes.

```
lazy rule Prefix2case_statement_alternative{
    from p:Wright!Prefix,
        index:Integer,
        instance: String
    to cs: Ada!case_statement_alternative(
        choice<-index.toString(),
        s<- p.transformation(instance)
        )
}
```

Une mise à jour est nécessaire pour la règle traduisant la structure d'accueil qui doit, de plus, inclure la fonction *condition_interne1*. Le corps de cette fonction est à raffiner.

```
helper context Wright!Configuration def: ICGen:Wright!InternalChoice=
    Wright!InternalChoice.allInstances() ->select(e |e.elements-
>size()>2)->at(1);

helper context Wright!Configuration def: existICGen:Boolean=
    if(Wright!InternalChoice.allInstances() ->select(e |e.elements-
>size()>2)->isEmpty() )then false else true endif;
```

```
rule Configuration2subprogram{
      from c: Wright!Configuration
      to      sb: Ada!subprogram_body (
          specif <- sp ,
          statements <- st ,
          declarations <- c.getInternalTraitement ->
collect(e|thisModule.InternalTraitement2subprogram(e))
          ->union(if (c.existICGen)then
OrderedSet{thisModule.InternalChoiceG2function(c.ICGen)} else
OrderedSet{}endif)
          ->union(if (c.existICBin)then
OrderedSet{thisModule.InternalChoiceB2function(c.ICBin)} else
OrderedSet{}endif)
          ->union(c.compInst ->
collect(e|thisModule.ComponentInstance2single_task_declaration(e)))
          ->union(c.connInst ->
collect(e|thisModule.ConnectorInstance2single_task_declaration(e)))
          ->union(c.compInst ->
collect(e|thisModule.ComponentInstance2task_body(e)))
          ->union(c.connInst ->
collect(e|thisModule.ConnectorInstance2task_body(e)))),
          sp: Ada!procedure_specification(   designator <- c.name),
          st: Ada!null_statement
}

lazy rule InternalChoiceG2function{
      from i:Wright!InternalChoice(not i.OclUndefined())
      to    s:Ada!subprogram_body(specif <- fs, statements <- r),
          fs: Ada!function_specification(   designator <-
'condition_interne1', returnType<-'Integer'),
          r:Ada!return_statement(exp<-e1),
          e1:Ada!expression(e<-'1')
}
```

## 5.3.6 Traduction de l'opérateur préfixe

L'opérateur préfixe a la forme suivante *EventExpression -> ProcessExpression*.

La traduction de l'opérateur préfixe consiste à traduire *EventExpression* puis à traduire *ProcessExpression*. Le helper redéfini *transformation* permet de le faire.

```
helper    context    Wright!Prefix    def:   transformation(instance   :
String):Sequence(Ada!statement)=
          if self.target.oclIsTypeOf(Wright!Prefix)   then
              Sequence{self.event.event_transform(instance)}-
>union(self.target.transformation(instance))
          else
              if  self.target.oclIsTypeOf(Wright!InternalChoice)   or
self.target.oclIsTypeOf(Wright!ExternalChoice) then
                  Sequence{self.event.event_transform(instance)}-
>union(Sequence {self.target.transformation(instance)})
              else
                  Sequence{self.event.event_transform(instance)}
              endif
```

        **endif**;

Ce helper fait appel à un autre helper redéfini *event_transform* qui permet de transformer un événement selon son contexte. De plus, il fait appel au helper redéfini transformation pour transformer la cible du préfixe.

## 5.3.7 Traduction des événements

Dans la suite, nous présentons la traduction des événements observés, initialisés, traitements internes et l'événement succès.

### 5.3.7.1 Traduction des événements observés et initialisés

Nous rappelons que les attachements sont de la forme « nomInstanceComposant . nomPort As  nomInstanceConnecteur . nomRôle »

- ▪ Traduction des événements observés et initialisés attachés à une instance de composant :

Si événement est un événement observé de la forme «nomPort.événement».

Alors

Accepter le rendez vous portant le nom nomPort_événement ; soit *accept nomPort_ événement ;*

Fin Si

Si événement est un événement initialisé de la forme « _nomPort.événement ».

Alors

Voir le nomRôle qui est attaché à nomPort de l'instance courante, demander un rendez vous comme suit : *Connector_nomInstanceConnecteur.nomRôle_événement ;*

Fin Si

- ▪ Traduction des événements observés et initialisés attachés à une instance de connecteur :

Si événement est un événement observé de la forme «nomRôle.événement».

Alors

Accepter le rendez vous portant le nom nomRôle_événement; soit *accept nomRôle_*

*événement ;*

Fin Si

Si événement est un événement initialisé de la forme « _nomRôle.événement ».

Alors

Voir le nomPort qui est attaché à nomRôle de l'instance courante, demander un rendez

vous comme suit: *Component_nomInstanceComposant.nomPort_événement ;*

Fin Si

- ▪ Traduction en ATL :

La traduction des événements observés et initialisés se fait par l'intermédiaire des deux

helpers redéfinis *event_transform* suivants :

```
helper context Wright!EventObserved def: event_transform(instance :
String):Ada!simple_accept_statement=
      thisModule.EventObserved2simple_accept_statement(self);

helper context Wright!EventSignalled def: event_transform(instance :
String):Ada!entry_call_statement=
      thisModule.EventSignalled2entry_call_statement(self,instance);
```

Ces helpers déclenchent respectivement les règles paresseuses permettant de traduire un

événement observé et un événement initialisé. Ces deux règles paresseuses sont présentées

comme suit :

```
lazy rule EventObserved2simple_accept_statement{
      from e:Wright!EventObserved
      to s:Ada!simple_accept_statement(
            direct_name<- e.name.replaceAll('.','_')
            )
}
```

La règle paresseuse *EventObserved2simple_accept_statement* transforme un événement observé en une instruction *accept* portant le nom de l'événement en remplaçant le point par un tiret bas.

```
lazy rule EventSignalled2entry_call_statement{
      from e:Wright!EventSignalled,
            instance : String
      to ec:Ada!entry_call_statement(
            entry_name<-if(Wright!Attachment.allInstances()-
>select(a|a.originPort.name=e.name.substring(1,e.name.indexOf('.')))-
>select(a|a.originInstance.name=instance)->notEmpty())then
                              'Connector_'
```

```
                                    +Wright!Attachment.allInstances()-
>select(a|a.originPort.name=e.name.substring(1,e.name.indexOf('.')))-
>select(a|a.originInstance.name=instance).at(1).targetInstance.name
                                +'.'+Wright!Attachment.allInstances()-
>select(a|a.originPort.name=e.name.substring(1,e.name.indexOf('.')))-
>select(a|a.originInstance.name=instance).at(1).targetRole.name

        +'_'+e.name.substring(e.name.indexOf('.')+2,e.name.size()))
                            else
                                'Component_'
                                +Wright!Attachment.allInstances()-
>select(a|a.targetRole.name=e.name.substring(1,e.name.indexOf('.')))-
>select(a|a.targetInstance.name=instance).at(1).originInstance.name
                                +'.'+Wright!Attachment.allInstances()-
>select(a|a.targetRole.name=e.name.substring(1,e.name.indexOf('.')))-
>select(a|a.targetInstance.name=instance).at(1).originPort.name

        +'_'+e.name.substring(e.name.indexOf('.')+2,e.name.size()))
                            endif
            )
}
```

La règle paresseuse *EventSignalled2entry_call_statement* transforme un événement initialisé en une instruction *entry* dont le nom dépend de l'instance de composant ou de connecteur à laquelle cet événement appartient. C'est la raison pour laquelle le passage du paramètre instance a eu lieu tout au long de la traduction du processus CSP Wright. De plus, ce nom dépend de l'attachement dans lequel le port ou le rôle est impliqué.

### 5.3.7.2 Traduction des événements succès

L'événement succès « V » qui est toujours suivi du processus STOP représente le processus SKIP ou encore §. Ceci correspond à la terminaison avec succès. Ce cas est traduit dans Ada par l'instruction exit.

Le helper redéfini *event_transform* suivant permet le déclenchement de la règle paresseuse traduisant l'événement succès.

```
helper  context  Wright!SuccesEvent  def:  event_transform(instance  :
String):Ada!exit_statement =
    thisModule.SuccesEvent2exit_statement(self);
```

La règle paresseuse permettant la traduction de l'événement succès se presente comme suit :

```
lazy rule SuccesEvent2exit_statement{
    from p:Wright!SuccesEvent
    to e:Ada!exit_statement
}
```

### 5.3.7.3 Traduction d'un événement interne

Comme déjà cité, un événement interne est traduit en Ada par une procédure dont le corps est à raffiner. L'appel de cette procédure se fait par la règle paresseuse suivante :

```
lazy rule InternalTraitement2procedure_call_statement{
     from e:Wright!InternalTraitement
     to p:Ada!procedure_call_statement(
          name<-e.name
     )
}
```

Le helper redéfini *event_transform* suivant permet le déclenchement de la règle paresseuse traduisant l'appel de la procédure du traitement interne.

```
helper context Wright!InternalTraitement def: event_transform(instance :
String):Ada!procedure_call_statement=
     thisModule.InternalTraitement2procedure_call_statement(self);
```

## 5.4 Conclusion

Dans ce chapitre nous avons présenté d'une façon assez détaillée le programme Wright2Ada conçu et réalisé dans le cadre de ce mémoire afin de transformer une architecture logicielle décrite en Wright vers un programme concurrent Ada. Notre programme Wright2Ada est purement déclaratif et comporte :

- ✓ 1 règle standard (ou matched rule)
- ✓ 19 règles paresseuses (ou lazy rules)
- ✓ 3 helpers attributs
- ✓ 12 helpers opérations
- ✓ 3  helpers polymorphiques

Dans le chapitre suivant, nous apportons des interfaces conviviales afin d'utiliser notre programme Wright2Ada dans un contexte réel. Ces interfaces permettent d'utiliser le programme Wright2Ada en introduisant du code Wright et en produisant du code Ada.

# Chapitre 6: Interfaces conviviales d'utilisation de Wright2Ada

La transformation accomplie par le programme Wright2Ada, présentée dans le chapitre précédent, suppose une compréhension des méta-modèles source et cible par l'utilisateur, un certain savoir faire pour produire le modèle source et comprendre le modèle cible généré. De plus, il est souvent difficile de produire l'entrée de la transformation lorsque la spécification Wright est complexe. En effet, l'utilisateur est censé se servir du navigateur de modèles et de l'éditeur de propriétés afin d'introduire le texte Wright sous format XMI. Ceci est non convivial et sujet à des erreurs potentielles. En outre, il est censé transformer manuellement le modèle Ada au format XMI en code Ada afin de se servir des outils associés à Ada tels que: compilateur et model-checker.

Dans ce chapitre, nous nous proposons de traiter les deux étapes d'injection et d'extraction. L'injection prend un modèle exprimé dans la syntaxe concrète textuelle de Wright et génère un modèle conforme au méta-modèle Wright dans l'espace technique de l'ingénierie des modèles. L'extraction travaille sur la représentation interne des modèles exprimés en Ada et crée la représentation textuelle (code Ada).

Pour ce faire nous nous proposons d'utiliser en premier lieu les possibilités fournies par l'outil Xtext de oAW (open Architecture Ware), afin de procéder à la transformation du texte Wright vers son modèle. Ensuite, nous poposons une transformation du modèle d'Ada vers son code avec l'outil Xpand de oAW. Enfin, nous présentons un exemple d'utilisation.

## 6.1 Texte Wright vers modèle Wright

Cette section présente la validation et la transformation d'un texte Wright vers modèle Wright conforme au méta-modèle Wright proposé dans la section 5.1. Le schéma général de cette transformation est donné dans la figure 41.

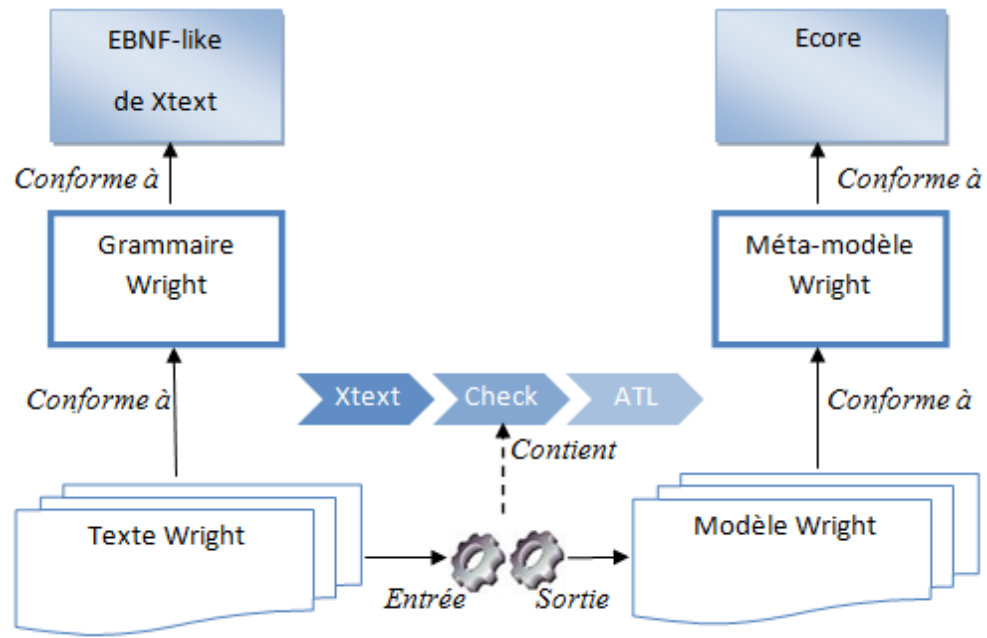

**Figure 41** *: Vue d'ensemble sur la transformation texte vers modèle Wright*

La transformation proposée comporte trois étapes. La première étape a pour objectif de créer le méta-modèle Ecore appelé Grammaire Wright à partir d'une description en Xtext de la grammaire de Wright et de produire l'analyseur lexico-syntaxique de Wright via Xtext. La deuxième étape a pour objectif de valider (via Check) et transformer le modèle exprimé dans la syntaxe concrète textuelle de Wright en un modèle XMI conforme au méta-modèle Grammaire Wright. La troisième étape a pour objectif de transformer le modèle XMI conforme au méta-modèle Grammaire Wright vers un modèle XMI conforme au méta-modèle Wright via ATL.

## 6.1.1 Injection via Xtext

Dans cette sous section, nous commençons par présenter la création du projet Xtext. Puis, nous passerons à présenter la grammaire du langage Wright avec Xtext. Enfin, nous donnerons un aperçu sur le méta-modèle de Wright généré avec Xtext.

### 6.1.1.1 Création du projet xtext

Pour commencer nous allons créer un nouveau projet Xtext avec l'option de création de générateur de projet, car nous allons utiliser ce générateur plus tard. Ceci est présenté par la figure 42.

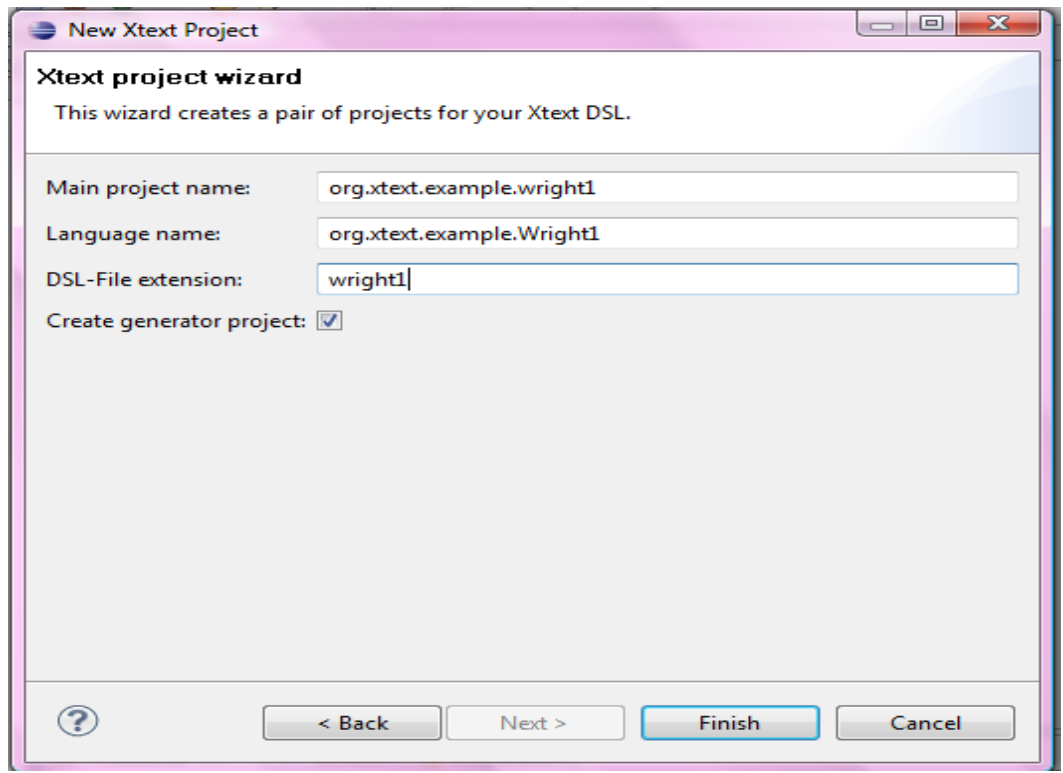

**Figure 42**: *Création du projet xtext*

Nous devons avoir après cette étape de création, trois projets dans notre espace de travail. Le premier projet est le projet principal de Xtext où nous allons définir la grammaire de l'ADL Wright. Le second projet est l'éditeur de projet, il contiendra l'éditeur Xtext généré automatiquement à base de notre DSL Wright. Enfin, le troisième projet fournit l'interface d'utilisation (User Interface).

### 6.1.1.2 La grammaire de l'ADL Wright

Dans le premier projet, et dans le fichier d'extension .xtext, nous allons créer la grammaire de notre DSL Wright. Ce fichier contient au premier niveau les deux lignes suivantes :

```
grammar org.xtext.example.Wright1 with org.eclipse.xtext.common.Terminals

generate wright1 "http://www.xtext.org/example/Wright1"
```

La première ligne déclare l'identificateur du modèle et la base des déclarations. La deuxième ligne est la directive de création du méta-modèle Ecore généré avec son emplacement.

✓ Création de l'entité Configuration :

La première entité de notre grammaire est la configuration. Une configuration Wright a le format suivant :

| |
|---|
| Configuration *nom_de _la configuration* |
| *L'ensemble des définitions de composants et de connecteurs* |
| Instances |
| *L'ensemble de déclaration des instances* |
| Attachment |
| *L'ensemble des attachements* |
| End Configuration |

La méta-classe Configuration dans la méta-modèle Wright est présentée par la figure 43.

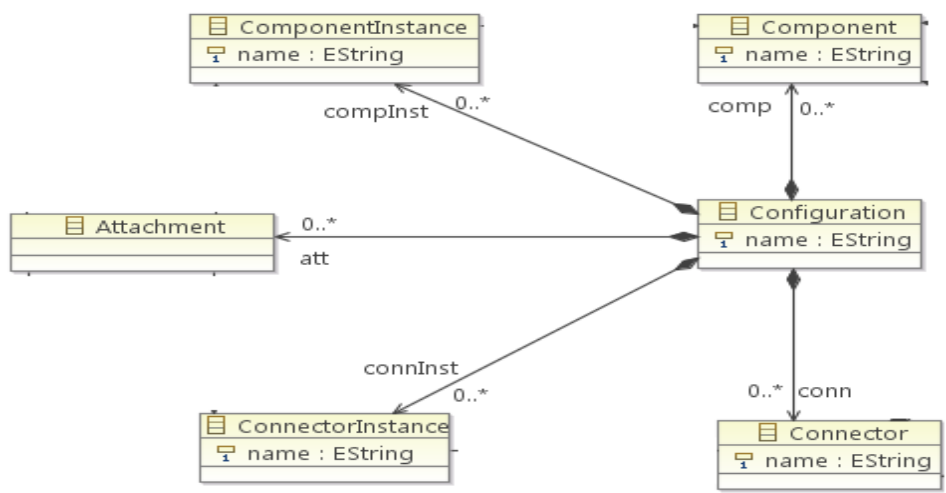

**Figure 43**: *La méta-classe Configuration*

L'entité Configuration peut être traduite par la règle de production ci-dessous.

```
Configuration : "Configuration" name=ID
                        ( TypeList+=Type )*
                "Instances"
                        ( InstanceList+=Instance)*
                "Attachments"
                        ( att+=Attachment )*
                "End Configuration";

Instance: ComponentInstance | ConnectorInstance ;

Type: Component| Connector;
```

Nous venons de définir une configuration avec un nom, l'ensemble de types qui peuvent être des composants ou des connecteurs, l'ensemble d'instances qui peuvent être des instances de composants ou de connecteurs, et enfin, l'ensemble des attachements.

Le symbole « += » signifie que la variable contient un ensemble du type correspondant. Le symbole « * » signifie la cardinalité zéro ou plusieurs.

Il y a une différence entre le méta-modèle présenté et la grammaire de l'entité Configuration. Cette différence est dûe au fait que nous ne pouvons pas imposer un ordre de déclaration pour les composants et les connecteurs. Ceci est également vrai pour les instances de composants et de connecteurs.

✓ Création des entités Component et Port:

Un composant est défini comme suit :

| |
|---|
| Component *nom_du _composant* |
| *L'ensemble des définitions de ces ports* |
| Computation = |
| *L'expression du processus compuation* |

Les méta-classes Component et Port sont présentées par la figure 44.

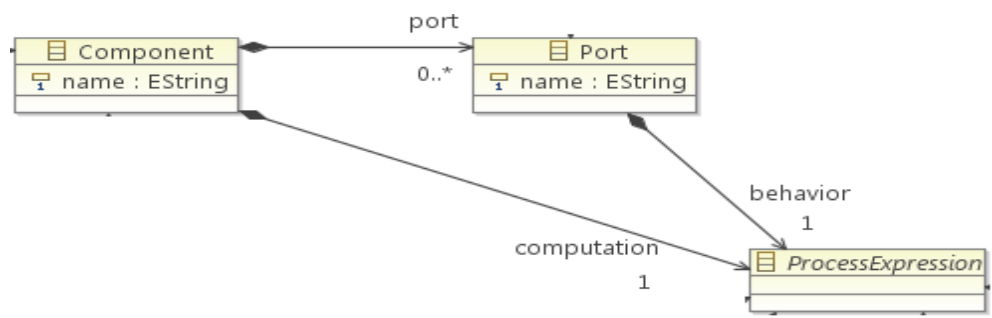

**Figure 44***: Les méta-classes Component et Port*

L'entité Component peut être traduite par la règle de production ci-dessous.

```
Component : "Component" name=ID
                       ( port+=Port )+
                       "Computation" '='  computation=ProcessExpression
;
```

Nous venons de définir un composant avec un nom, l'ensemble de ses ports et son processus compuation.

Un port est défini comme suit :

> Port *nom_du _port*
>
>  =
>
> *Le comportement du port ;l'expression du processus nom_port*

L'entité port peut être traduite par la règle de production ci-dessous.

```
Port : "Port" name=ID '=' behavior=ProcessExpression;
```

Un port a un nom et un comportement décrit par l'expression d'un processus

✓ Création des entités Connector et Role:

Un connecteur est défini comme suit :

> Connector *nom_du _connecteur*
>
> *L'ensemble des définitions de ces rôles*
>
> Glue =
>
> *L'expression du processus glue*

Les méta-classes Connector et Role sont présentées par la figure 45.

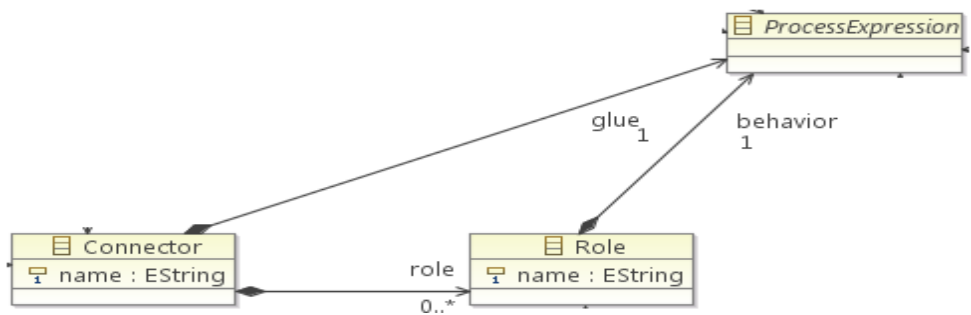

**Figure 45***: Les méta-classes Connector et Role*

L'entité Connector peut être traduite par la règle de production ci-dessous.

```
Connector : "Connector" name=ID
                        ( role+=Role )+
                        "Glue" '=' glue=ProcessExpression ;
```

Nous venons de définir un connecteur avec un nom, l'ensemble de ses rôles et son processus glue.

Un rôle est défini comme suit :

> Role *nom_du _rôle*
>
> =
>
> *Le comportement du rôle ;l'expression du processus nom_rôle*

L'entité rôle peut être traduite par la règle de production ci-dessous.

```
Role : "Role" name=ID '=' behavior=ProcessExpression;
```

Un rôle a un nom et un comportement décrit par l'expression d'un processus

✓ Création des entités ComponentInstance et ConnectorInstance :

Une instance de composant est définie comme suit :

> *Nom_ de_l'instance_de_composant* : *nom_du_composant_type*

La méta-classe ComponentInstance est présentée par la figure 46.

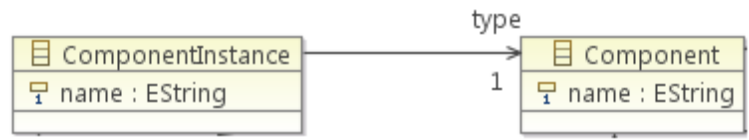

**Figure 46***: La méta-classe ComponentInstance*

Cette instance peut être traduite par la règle de production ci-dessous.

```
ComponentInstance : name=ID ':' type=[Component];
```

Une instance de composant a un nom et une référence vers le composant type.

Une instance de connecteur est définie comme suit :

> *Nom_ de_l'instance_de_connecteur* : *nom_du_connecteur_type*

Un raisonnement similaire donne la règle de production de l'instance de connecteur :

```
ConnectorInstance : name=ID ':' type=[Connector];
```

Une instance de connecteur a un nom et une référence vers le connecteur type.

Les deux règles présentées ci-dessus posent un problème avec l'analyseur lexico-syntaxique de Xtext, car elles sont similaires. L'analyseur va avoir une confusion sur l'alternative qui va choisir ; celle de l'instance de composant ou de l'instance de connecteur.

Une solution pour remédier à ce problème, est de changer la règle de production de ces deux entités comme suit :

```
ComponentInstance : name=ID ':' "Component" type=[Component];

ConnectorInstance : name=ID ':' "Connector" type=[Connector];
```

La définition de ces deux instances devient alors :

| | |
|---|---|
| *Nom_ de l'instance_de_composant* : Component | *nom_du_composant_type* |
| *Nom_ de l'instance_de_connecteur* : Connector | *nom_du_connecteur_type* |

✓ Création de l'entité Attachment :

Un attachement est défini comme suit :

| |
|---|
| *le_nom_d'une_instance_de_composant* |
| `'.'` |
| *le_nom_du_port_d'origine* |
| `"As"` |
| *le_nom_d'une_instance_de_connecteur* |
| `'.'` |
| *le_nom_du_role_cible* |

La méta-classe Attachment est présentée par la figure 47.

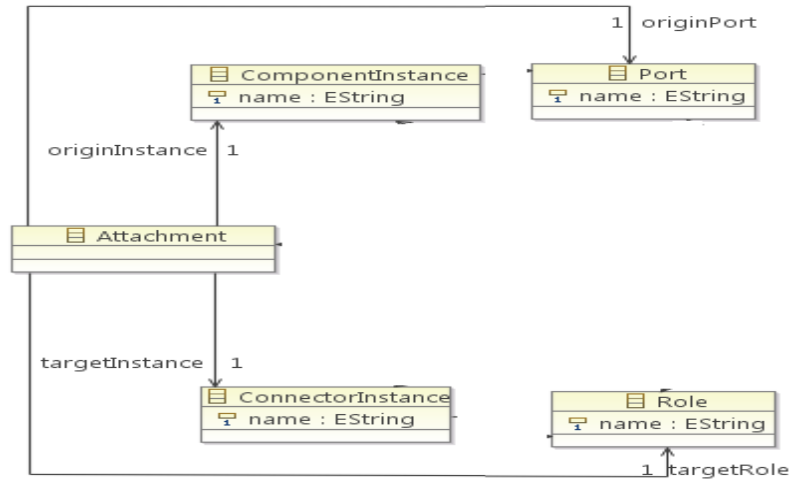

**Figure 47**: *La méta-classe Attachment*

L'entité Attachment peut être traduite par la règle de production ci-dessous.

```
Attachment : originInstance=[ComponentInstance] '.' originPort=[Port]
"As" targetInstance=[ConnectorInstance] '.' targetRole=[Role] ;
```

Un attachement est composé de quatres références qui ont pour cible les méta-classes Component, Port, Connector et Role.

La règle présentée ci-dessus pose problème avec le « . » car il se trouve dans la règle terminale ID qui a son tour se présente comme suit ::

```
terminal ID: ('a'..'z'|'A'..'Z') ('a'..'z'|'A'..'Z'|'_'|'.'|'0'..'9')*;
```

Une solution pour remédier à ce problème est de le remplacer par « - » .

La règle de production de l'entité Attachment devient :

```
Attachment : originInstance=[ComponentInstance] '-' originPort=[Port]
"As" targetInstance=[ConnectorInstance] '-' targetRole=[Role] ;
```

La définition d'un attachement devient :

> *le_nom_d'une_instance_de_composant*
>
> '-'
>
> *le_nom_du_port_d'origine*
>
> "As"
>
> *le_nom_d'une_instance_de_connecteur*

| |
|---|
| '-' |
| *le_nom_du_role_cible* |

✓  Création des entités des événements Wright:

Les événements sont présentés par le méta-modèle de la figure 48.

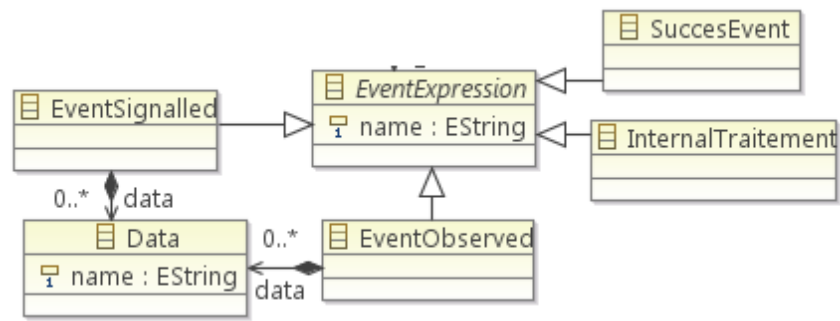

**Figure 48**: *Le méta-modèle des événements*

Pour faire la distinction entre les événements observés, les événements initialisés et les traitements internes les événements initialisés doivent être obligatoirement préfixés par « _ » et le traitement interne par « - ». L'événement succès est toujours nommé √ soit « V ».

Ci-dessous nous présentons les entités des événements :

```
EventExpression : EventSignalled | EventObserved | InternalTraitement |
SuccesEvent;

EventSignalled:  '_' name=ID (data+=Data)*;

EventObserved: name=ID (data+=Data)*;

InternalTraitement: '-' name=ID;

SuccesEvent: name='V';
```

Les événements observés et initialisés peuvent transporter des données préfixées par « ! » ou par « ? » ce qui représentent respectivement des données en sortie et en entrée.

```
Data : ('?' | '!') name=ID;
```
✓  Création des opérateurs du processus CSP Wright:

Le processus CSP Wright est décrit par le méta-modèle de la figure 49.

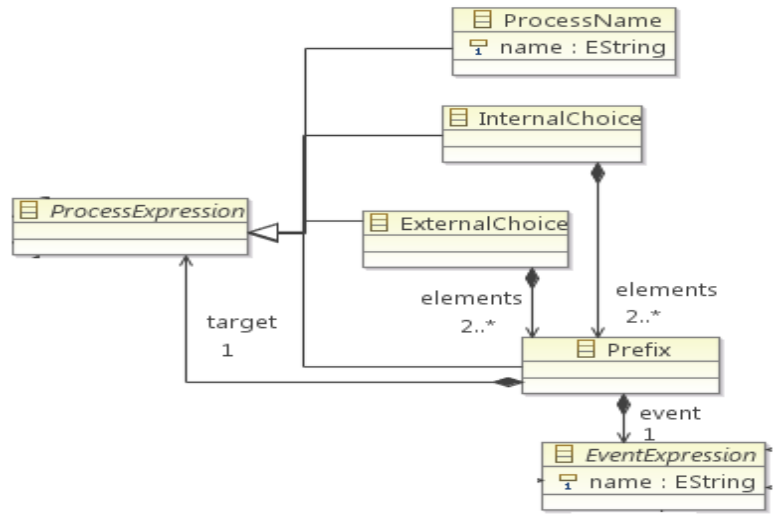

**Figure 49:** *Le méta-modèle du processus CSP Wright*

La méta-classe *ProcessName* est traduite par la règle de production ci-dessous.

```
ProcessName:  name=ID ;
```

Un opérateur de préfixe peut être décrit comme suit:

| *EventExpression -> ProcessExpression* | *EventExpression -> (ProcessExpression)* |
| --- | --- |

Un opérateur de choix externe est décrit comme suit :

| *Préfixe1* [] *Préfixe2* [] … |
| --- |

Un opérateur de choix interne est décrit comme suit :

| *Préfixe1* \|~\| *Préfixe2* \|~\| … |
| --- |

Ces derniers peuvent êtres traduits par les règles de production ci-dessous :

```
ProcessExpression: InternalChoice | ExternalChoice | ProcessName | Prefix
|
Parentheses;

Parentheses: '(' p=ProcessExpression ')';

Prefix: event=EventExpression '->' target=ProcessExpression;

InternalChoice: p=Prefix ('|~|' e+=Prefix)+;
```

```
ExternalChoice: p=Prefix ('[]' e+=Prefix)+;
```

Mais cette solution pose malheureusement un problème, car elle n'est pas LL(*). Xtext fonctionne avec l'analyseur syntaxique ANTLR qui est basé sur les algorithmes LL(*). Le problème peut être résolu par une factorisation gauche. Nos règles de productions deviennent :

```
Prefix: event=EventExpression '->' target=TargetPrefix;
TargetPrefix: Parentheses | Prefix | ProcessName;
Parentheses: '(' p=ProcessExpression ')';
ProcessExpression  :  right=Prefix  (('[]'  ECLeft+=Prefix)+|('|~|'
ICLeft+=Prefix)+)?;
```

De plus, le symbole § ou encore SKIP désigne V -> STOP, donc la règle de production de préfix devient : `Prefix: event=EventExpression '->' target=TargetPrefix | name='§'| name='SKIP';`

Dans les règles présentées ci-dessus l'opérateur de préfixe, l'opérateur de choix interne et externe sont traduites dans une même règle de grammaire nommée ici *ProcessExpression*.

La grammaire de l'ADL Wright décrite en Xtext est fournie dans l'annexe C.

### 6.1.1.3 Le méta-modèle de Wright générer avec xtext

L'exécution du moteur workflow qui existe par defaut dans le premier projet permet, entre autre, de générer le diagramme Ecore présenté dans la figure 50. Le diagramme Ecore généré correspond à la grammaire de l'ADL Wright en Xtext (voir annexe C).

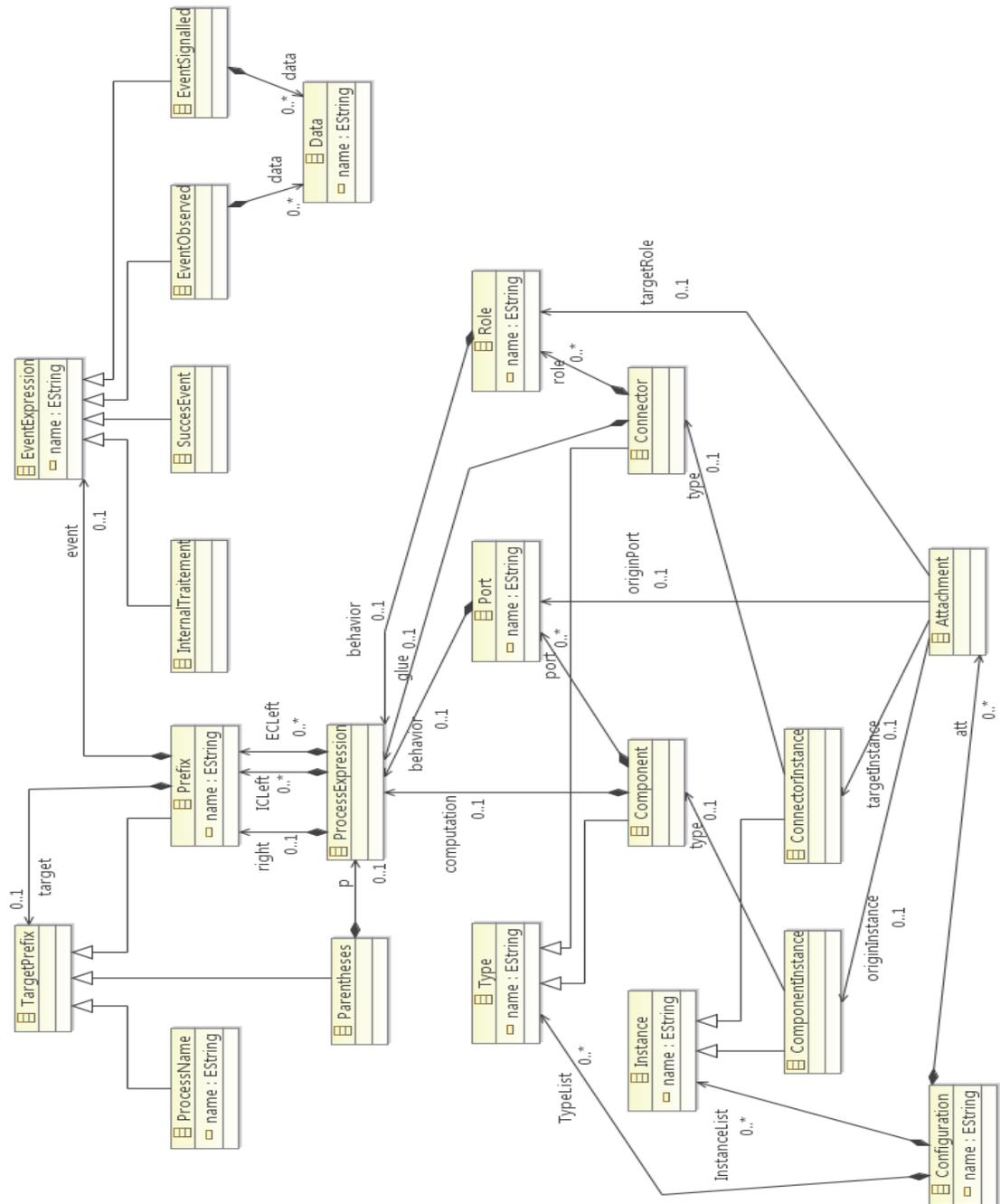

**Figure 50**: *Le diagramme Ecore du méta-modèle Grammaire Wright généré -Wright1-*

## 6.1.2 Validation et génération du modèle Wright en XMI

### 6.1.2.1 Sémantique statique de Wright

La sémantique statique de Wright est décrite à l'aide des contraintes OCL attachées au méta-modèle Wright (voir chapitre 3). Ces contraintes sont réécrites en Check et attachées au méta-modèle Grammaire Wright -appelé Wright1- généré par l'outil Xtext.

Les contraintes Check données ci-dessous seront évaluées sur les textes Wright. Ensuite, ces textes Wright seront transformés en XMI conformes au méta-modèle Grammaire Wright -appelé Wright1- moyennant l'utilisation des plugins :

- org.xtext.example.wright1 que nous avons développé dans la section 6.1.1. et le plugin org.eclipse.xtext. Ces plugins permettent d'interpréter le texte Wright comme étant un modèle conforme au méta-modèle Grammaire Wright –appelé Wright1-.

- org.eclipse.xtend pour le langage Check.

- org.eclipse.emf.mwe.utils de l'EMF

```
import wright1;

context Attachment ERROR
"Un attachement n'est pas valide: le port "+originPort.name+" ne peut pas
être attaché à l'instance "+originInstance.name+".":
originInstance.type.port.contains(originPort);

context Attachment ERROR
"Un attachement n'est pas valide: le rôle "+targetRole.name+" ne peut pas
être attaché à l'instance "+targetInstance.name+".":
targetInstance.type.role.contains(targetRole);

context Configuration ERROR
"Chaque instance de composant déclarée au sein de la configuration
"+name+" doit utiliser un type composant déclaré au sein de la même
configuration.":
((List[wright1::ComponentInstance])InstanceList.typeSelect(wright1::Compo
nentInstance)).forAll( i | TypeList.contains( i.type));

context Configuration ERROR
"Chaque instance de connecteur déclarée au sein d'une configuration
"+name+" doit utiliser un type connecteur déclaré au sein de la même
configuration.":
((List[wright1::ConnectorInstance])InstanceList.typeSelect(wright1::Conne
ctorInstance)).forAll( i | TypeList.contains( i.type));

context Attachment ERROR
"Un attachement utilise une instance de connecteur
"+targetInstance.name+" non déclarée dans la configuration que lui.":
```

```
targetInstance.eContainer==this.eContainer;
```

**context** Attachment **ERROR**
"Un attachement utilise une instance de composant "+originInstance.name+"
non déclarée dans la configuration que lui.":
```
originInstance.eContainer==this.eContainer;
```

**context** Component **ERROR**
"Le composant "+name+" utilise plusieurs fois le même nom pour ses
ports.":
```
port.forAll( p1 | port.notExists(p2| p1!=p2 && p1.name == p2.name));
```

**context** Connector **ERROR**
"Le connecteur "+name+" utilise plusieurs fois le même nom pour ses
rôles.":
```
role.forAll( r1 | role.notExists(r2| r1!=r2 && r1.name == r2.name));
```

**context** Configuration **ERROR**
"Dans la configuration "+name+" les types n'ont pas des noms deux à deux
différents.":
```
TypeList.forAll( t1 | TypeList.notExists(t2| t1!=t2 && t1.name ==
t2.name));
```

**context** Configuration **ERROR**
"Dans la configuration "+name+" les instances n'ont pas des noms deux à
deux différents.":
```
InstanceList.forAll( i1 | InstanceList.notExists(i2| i1!=i2 && i1.name ==
i2.name));
```

**context** Configuration **ERROR**
"Dans la configuration "+name+" les instances et les types n'ont pas des
noms deux à deux différents.":
```
InstanceList.forAll( i | TypeList.notExists(t| t.name == i.name));
```

**context** Configuration **ERROR**
"Une configuration privée de types n'admet ni instances ni attachement.":
```
TypeList.size>0 || InstanceList.isEmpty && att.isEmpty;
```

**context** Connector **WARNING**
"Le connecteur "+name+" a un nombre de rôles inférieur à deux":
```
role.size>=2;
```

**context** ProcessExpression **if** !(ECLeft.isEmpty) **ERROR**
"Un choix externe doit être basé uniquement sur des événements observés,
l'événement succés ou le processus SKIP.":
```
ECLeft.forAll(p|p.event.metaType==wright1::EventObserved ||
p.event.metaType==wright1::SuccesEvent || p.name=='$'|| p.name=='SKIP')
&& right.event.metaType==wright1::EventObserved ||
right.event.metaType==wright1::SuccesEvent|| right.name=='$'||
right.name=='SKIP';
```

## 6.1.2.2 Le moteur de validation et de génération

Le moteur workflow du deuxième projet doit être modifié comme suit :

```xml
<workflow>
        <property name="modelFile"
value="classpath:/model/MyModel1.wright1"/>
        <property name="targetDir" value="src-gen/example1"/>

        <bean class="org.eclipse.emf.mwe.utils.StandaloneSetup"
platformUri=".."/>

        <component class="org.eclipse.emf.mwe.utils.DirectoryCleaner"
directory="${targetDir}"/>

        <component class="org.eclipse.xtext.MweReader" uri="${modelFile}">
                <!--Cette classe est générée par le générateur de xtext -->
                <register class="org.xtext.example.Wright1StandaloneSetup"/>
        </component>

        <!--valider le modèle -->
        <component class="org.eclipse.xtend.check.CheckComponent">
        <metaModel
class="org.eclipse.xtend.typesystem.emf.EmfRegistryMetaModel"/>
                <checkFile value="model::CheckFile" />
                <emfAllChildrenSlot value="model" />
        </component>

        <!--générer le modèle -->
        <component class="org.eclipse.emf.mwe.utils.Writer">
                <modelSlot value="model"/>
                <uri value="${targetDir}/exampleWright1.xmi"/>
        </component>

</workflow>
```

L'exécution de ce workflow permet la génération du modèle xmi conforme au méta-modèle Grammaire Wright –appelé Wright1- relatif au texte Wright écrit dans le fichier d'extension wright1. Le fichier d'extension wright1 se trouve dans le dossier src deuxième projet. Le modèle XMI généré se trouve dans le dossier src-gen du deuxième projet. Cette étape est présentée par la figure 51.

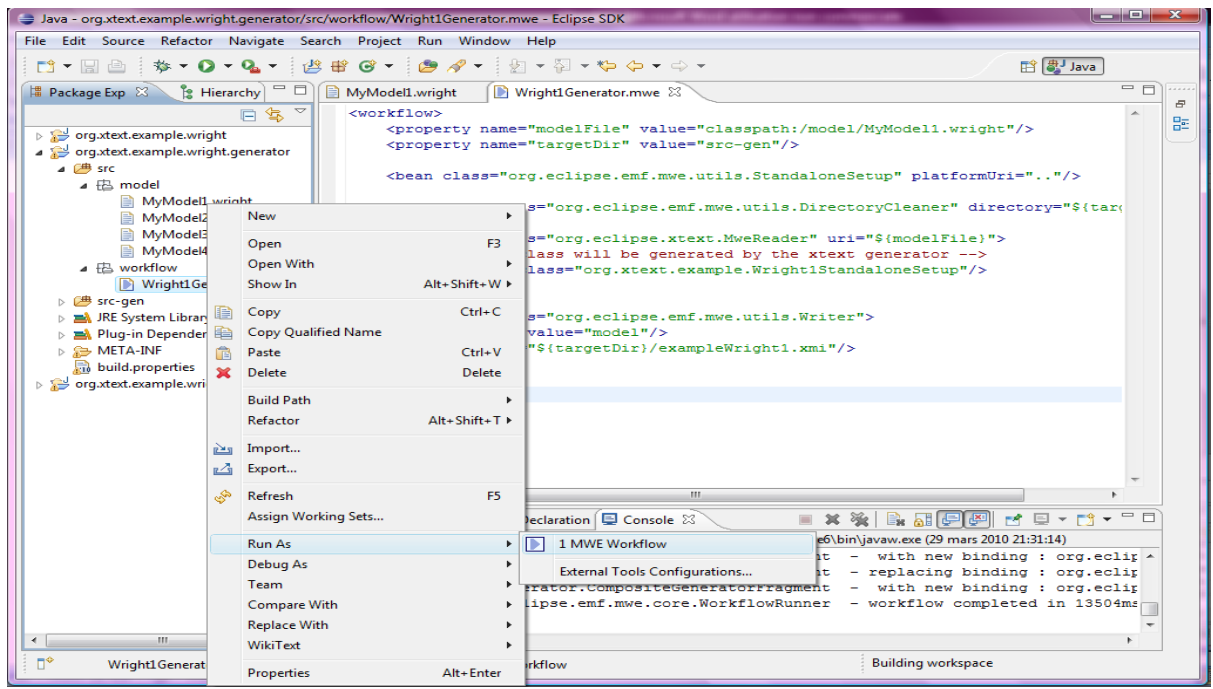

**Figure 51:** *Capture d'écran de l'exécution workflow du deuxième projet*

### 6.1.2.3 Exemple Client-Serveur

Dans ce qui suit, nous allons donner une illustration sur l'exemple Client-Serveur.

L'architecture client-serveur conforme à notre grammaire se présente comme suit :

```
Configuration Client_Serveur
Connector Lien_CS
        Role Appelant= _requete -> reponse -> Appelant |~| V -> STOP
        Role Appele= requete -> _reponse -> Appele [] V -> STOP
        Glue = Appelant.requete -> _Appele.requete -> glue
               [] Appele.reponse -> _Appelant.reponse -> glue
               [] V -> STOP

Component Client
        Port port_Client= _requete -> reponse -> port_Client |~| V -> STOP
        Computation=  -traitement_interne1 -> _port_Client.requete ->
port_Client.reponse -> computation |~| V -> STOP
Component Serveur
        Port port_Serveur= requete -> _reponse -> port_Serveur |~| V ->
STOP
        Computation=  -traitement_interne2 -> port_Serveur.requete ->
_port_Serveur.reponse -> computation |~| V -> STOP

Instances
        client1: Component Client
        serveur1: Component Serveur
```

```
        appel_cs: Connector Lien_CS
Attachments
        client1-port_Client As appel_cs-Appelant
        serveur1-port_Serveur As appel_cs-Appele
End Configuration
```

Après avoir vérifié les propriétés syntaxiques et sémantiques en passant par l'analyseur lexico-syntaxique généré par Xtext et l'évaluation des contraintes Check, le modèle correspondant à la configuration Client_Serveur est généré (voir annexe D). Un tel modèle XMI est conforme au méta-modèle Grammaire Wright -appelé Wright1-.

### 6.1.3 De Grammaire Wright vers Wright

Dans cette partie, nous allons présenter le programme GrammaireWright2Ada écrit en ATL permettant la transformation des modèles sources conformes au méta-modèle Grammaire Wright –appelé Wright1- vers des modèles cibles conformes au méta-modèle Wright.

L'en-tête de ce fichier ATL se présente comme suit :

```
module Wright1ToWright;
create exampleWright : Wright from exampleWright1 : Wright1;
```

✓ Transformation de la méta-classe *Configuration* de Wright1 vers la méta-classe *Configuration* de Wright:

Dans le méta-modèle de Wright1 une configuration se présente comme le montre la figure 52.

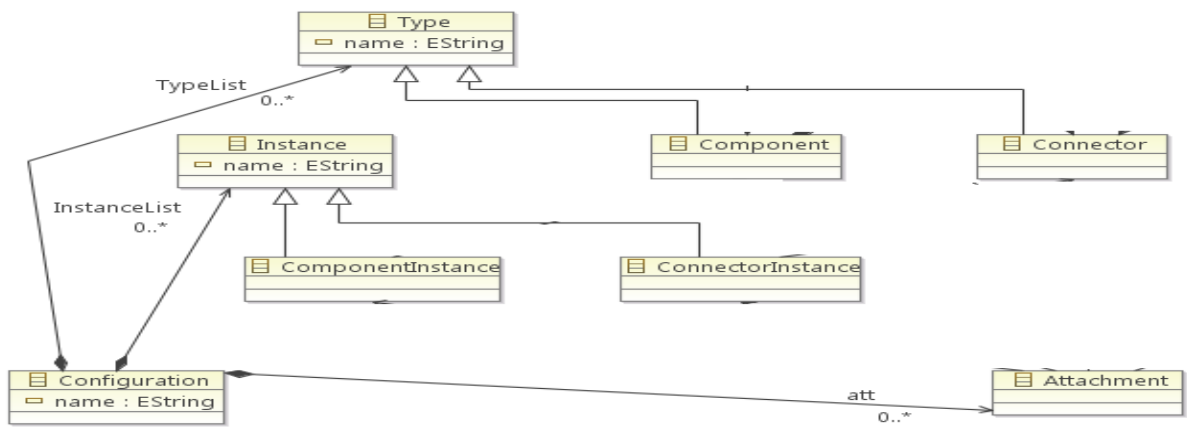

**Figure 52**: *La méta-classe Configuration du méta-modèle Wright1*

Dans le méta-modèle de Wright une configuration se présente comme le montre la figure 53.

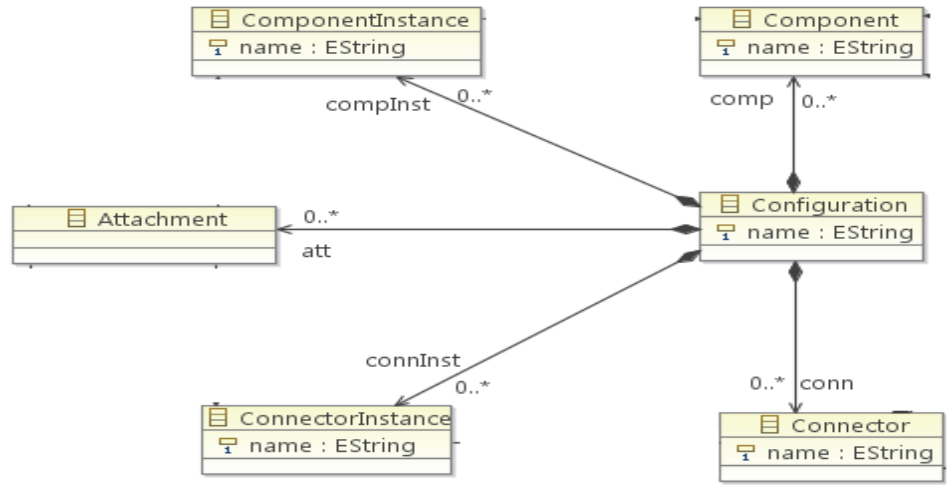

**Figure 53:** *La méta-classe Configuration du méta-modèle Wright*

La règle de transformation de la configuration se présente comme suit :

```
rule Configuration2Configuration{
    from c1:Wright1!Configuration
    to c:Wright!Configuration(
        name<-c1.name,
        comp<-c1.TypeList-
>select(e|e.oclIsTypeOf(Wright1!Component)),
        conn<-c1.TypeList-
>select(e|e.oclIsTypeOf(Wright1!Connector)),
        compInst<-c1.InstanceList-
>select(e|e.oclIsTypeOf(Wright1!ComponentInstance)),
        connInst<-c1.InstanceList-
>select(e|e.oclIsTypeOf(Wright1!ConnectorInstance)),
        att<-c1.att)
}
```

La référence *comp* prend l'ensemble des éléments de la méta-classe Component référencé par *TypeList*. Réciproquement, la référence *conn* prend l'ensemble des éléments de la méta-classe *Connector* référencé par *TypeList*. Et la référence *compInst* prend l'ensemble des éléments de la méta-classe *ComponentInstance* référencé par *InstanceList*. Réciproquement, la référence *connInst* prend l'ensemble des éléments de la méta-classe *ConnectorInstance* référencé par *InstanceList*. Le nom *name* et les attachements *att* restent inchangés.

✓ Transformation des méta-classes *ComponentInstance* et *ConnectorInstance* :

Pour les méta-classes *ComponentInstance* et *ConnectorInstance* : on signale aucun changement. Leurs règles de transformation se présentent comme suit :

```
rule ComponentInstance2ComponentInstance{
    from i1:Wright1!ComponentInstance
    to i:Wright!ComponentInstance(
        name<-i1.name,
        type<-i1.type
        )
 }
rule ConnectorInstance2ConnectorInstance{
    from i1:Wright1!ConnectorInstance
    to i:Wright!ConnectorInstance(
        name<-i1.name,
        type<-i1.type
        )
 }
```

✓ Transformation de la méta-classe *Attachment* :

Pour la méta-classe *Attachment,* il ne va y avoir aucun changement. Sa règle de transformation se présente comme suit :

```
rule Attachment2Attachment{
    from a1:Wright1!Attachment
    to a:Wright!Attachment(
        originInstance<-a1.originInstance,
        targetInstance<-a1.targetInstance,
        originPort<-a1.originPort,
        targetRole<-a1.targetRole
    )
 }
```

✓ Transformation des méta-classes *Component* et *Connector* :

Pour les méta-classes *Component* et *Connector*, le seul changement est dans l'expression du processus CSP Wright référencé par *computation* respectivement *glue*. Leurs règles de transformation se présentent comme suit :

```
rule Component2Component{
    from c1:Wright1!Component
    to c:Wright!Component(
        name<-c1.name,
        port<-c1.port,
        computation<-c1.computation.transformation()
        )
 }
rule Connector2Connector{
    from c1:Wright1!Connector
    to c:Wright!Connector(
        name<-c1.name,
        role<-c1.role,
```

```
            glue<-c1.glue.transformation()
            )
    }
```

Le helper *transformation* est un helper redéfini. Ce helper se charge de la transformation du processus CSP Wright selon le contexte dans lequel il est appelé.

   ✓ Transformation des méta-classes *Port* et *Role* :

Pour les méta-classes *Port* et *Role*, le seul changement est dans l'expression du processus CSP Wright référencé par *behavior*. Leurs règles de transformation se présentent comme suit :

```
rule Port2Port{
      from p1:Wright1!Port
      to p:Wright!Port(
            name<-p1.name,
            behavior<-p1.behavior.transformation()
      )
}
rule Role2Role{
      from r1:Wright1!Role
      to r:Wright!Role(
            name<-r1.name,
            behavior<-r1.behavior.transformation()
      )
}
```

Le helper *transformation* est un helper redéfini. Ce helper se charge de la transformation du processus CSP Wright selon le contexte dans lequel il est appelé.

   ✓ Transformation du processus CSP Wright:

Le processus CSP dans le méta-modèle Wright1 se présente comme le montre la figure 54.

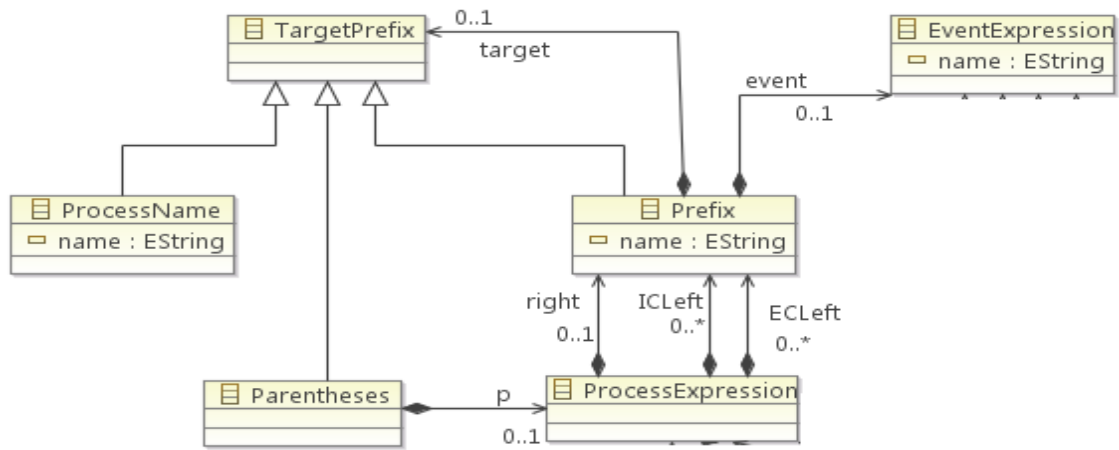

**Figure 54**: *Le processus CSP dans le méta-modèle Wright1*

Le processus CSP dans le méta-modèle Wright se présente comme le montre la figure 55.

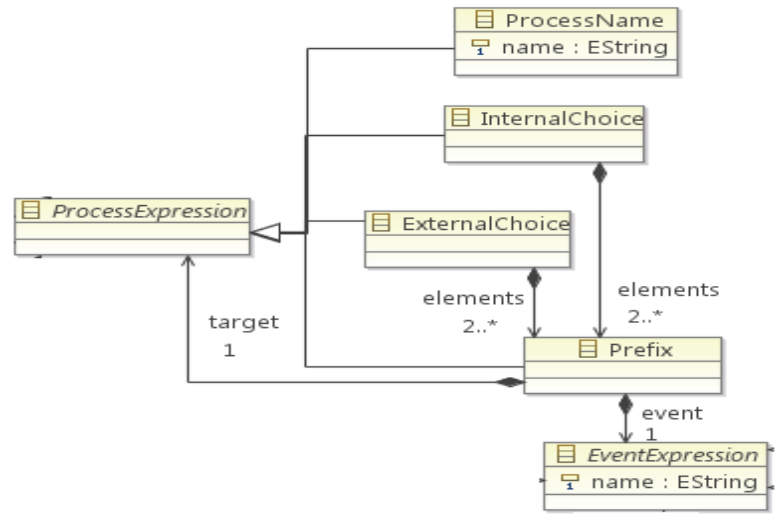

**Figure 55:** *Le méta-modèle du processus CSP Wright*

✓ Transformation de la méta-classe *ProcessExpression* du méta-modèle Wright1 :

Lorsque les références *ICLeft* et *ECLeft* de la méta-classe *ProcessExpression* de Wright1 sont vides, cette méta-classe est transformée en méta-classe P*refix* de Wright. Lorsque la référence *ICLeft* de la méta-classe *ProcessExpression* de Wright1 est non vide et la référence *ECLeft* est vide, la méta-classe *ProcessExpression* de Wright1 est transformée en méta-classe *InternalChoice* de Wright. Lorsque la référence *ECLeft* de la méta-classe *ProcessExpression* de Wright1 est non vide et la référence *ICLeft* est vide, la méta-classe *ProcessExpression* de Wright1 est transformée en méta-classe *ExternalChoice* de Wright. Ceci peut être traduit par le helper *transformation* suivant :

```
helper                    context                 Wright1!ProcessExpression
def:transformation():Wright!ProcessExpression=
     if(self.ECLeft->notEmpty())then
          thisModule.ProcessExpression2ExternalChoice(self)
     else
          if(self.ICLeft->notEmpty())then
               thisModule.ProcessExpression2InternalChoice(self)
          else
               if(self.ICLeft->isEmpty()        and        self.ECLeft-
>isEmpty())then
                    thisModule.ProcessExpression2Prefix(self)
               else
                    false
```

```
                    endif
                endif
        endif;
```

Dans la règle paresseuse *ProcessExpression2Prefix* fournie ci-dessous le nouveau *Prefix* crée prend les attributs du préfixe qui se trouve dans la référence *right* de la méta-classe source *ProcessExpression*.

```
lazy rule ProcessExpression2Prefix{
        from e:Wright1!ProcessExpression
        to p:Wright!Prefix(
                event<-e.right.event,
                target<-e.right.target.transformation()
        )
}
```

Dans la règle paresseuse *ProcessExpression2ExternalChoice* fournie ci-dessous la méta-classe *ExternalChoice* prend dans sa référence *elements* la réunion d'un nouveau préfixe nouvellement créé avec les préfixes qui sont dans la référence *ECLeft* de la méta-classe source *ProcessExpression*. Le nouveau préfixe créé prend les références du préfixe référencé par *right* de la méta-classe source *ProcessExpression*.

```
lazy rule ProcessExpression2ExternalChoice{
        from e:Wright1!ProcessExpression
        to pe:Wright!ExternalChoice(
                elements<-Set{p}->union(e.ECLeft-
>collect(e|e.transformation()))),
                p:Wright!Prefix(event<-e.right.event,            target<-
e.right.target.transformation())
}
```

Dans la règle paresseuse *ProcessExpression2ExternalChoice* la méta-classe *InternalChoice* prend dans sa référence *elements* la réunion d'un nouveau préfixe nouvellement créé avec les préfixes qui sont dans la référence *ICLeft* de la méta-classe source *ProcessExpression*. Le nouveau préfixe créé prend les références du préfixe référencé par *right* de la méta-classe source *ProcessExpression*.

```
lazy rule ProcessExpression2InternalChoice{
        from e:Wright1!ProcessExpression
        to pe:Wright!InternalChoice(
                elements<-Set{p}->union(e.ICLeft-
>collect(e|e.transformation()))),
                p:Wright!Prefix(event<-e.right.event,            target<-
e.right.target.transformation())
}
```

✓ Transformation de la méta-classe *Parentheses* :

La transformation de la méta-classe *Parentheses* dépend du type de sa référence p. Ceci est traduit par le helper transformation redéfini *transformation* suivant :

```
helper                         context                Wright1!Parentheses
def:transformation():Wright!ProcessExpression=
     self.p.transformation();
```

✓ Transformation de la méta-classe *Prefix*:

La méta-classe *Prefix* présente un changement dans le cas du processus SKIP (§), elle sera transformée en V -> STOP. Ceci est traduit par le code ATL suivant :

```
helper context Wright1!Prefix def:transformation():Wright!Prefix=
     if(self.name='§' or self.name='SKIP') then
          thisModule.SKIP2Prefix(self)
     else
          thisModule.Prefix2Prefix(self)
     endif;
lazy rule Prefix2Prefix{
     from p1:Wright1!Prefix
     to p:Wright!Prefix(
          event<-p1.event,
          target<-p1.target.transformation())
}
lazy rule SKIP2Prefix{
     from p1:Wright1!Prefix
     to p:Wright!Prefix(
          event<-e,
          target<-t),
          e:Wright!SuccesEvent(name<-'V'),
          t:Wright!ProcessName(name<-'STOP')
}
```

✓ Transformation de la méta-classe *ProcessName* :

Pour la méta-classe *ProcessName*, il ne va y avoir aucun changement. Ceci est traduit par le code ATL suivant :

```
helper                         context                Wright1!ProcessName
def:transformation():Wright!ProcessName=
     thisModule.ProcessName2ProcessName(self);

lazy rule ProcessName2ProcessName{
     from p1:Wright1!ProcessName
     to p:Wright!ProcessName(name<-p1.name)
}
```

✓ Transformation des événements :

Pour les événements, il ne va y avoir aucun changement. Ceci est traduit par les règles suivantes :

```
rule EventSignalled2EventSignalled{
    from e1:Wright1!EventSignalled
    to e:Wright!EventSignalled(name<-e1.name)
}
rule EventObserved2EventObserved{
    from e1:Wright1!EventObserved
    to e:Wright!EventObserved(name<-e1.name)
}
rule InternalTraitement2InternalTraitement{
    from e1:Wright1!InternalTraitement
    to e:Wright!InternalTraitement(name<-e1.name)
}
rule SuccesEvent2SuccesEvent{
    from e1:Wright1!SuccesEvent
    to e:Wright!SuccesEvent(name<-e1.name)
}
```

Le programme GrammaireWright2Wright en entier est fourni dans l'annexe E.

## 6.2. Modèle Wright vers texte Ada : extraction via Xpand

Cette section présente la validation du modèle Ada conforme au méta-modèle partiel d'Ada et la transformation de ce modèle vers un texte Ada. Le principe d'extraction est par la figure 56. Pour y parvenir, nous avons utilisé avec profit les outils Check pour la validation et Xpand pour la transformation.

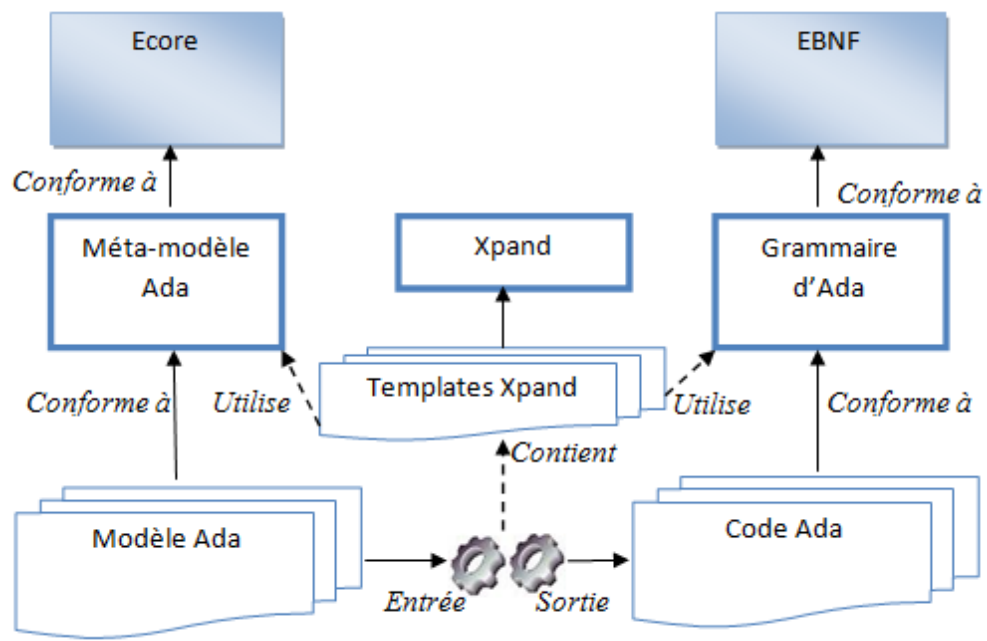

**Figure 56**: *Schéma de transformation de modèle Ada vers texte ada*

## 6.2.1 La sémantique statique d'Ada

La sémantique statique d'Ada est décrite à l'aide des contraintes OCL attachées au méta-modèle partiel d'Ada (voir chapitre 4). Ces contraintes sont réécrites en Check et attachées au méta-modèle partiel d'Ada.

Les contraintes Check données ci-dessous seront évaluées sur les modèles Ada conformes au méta-modèle Ada. Ensuite, ces modèles seront transformés en code Ada moyennant le moteur workflow (voir section 6.2.3) qui utilise les templates Xpand (voir section 6.2.2).

🔸 Les contraintes relatives à la partie déclarative d'un sous-programme en Ada :

```
context subprogram_body  if ! declarations.isEmpty ERROR
"Au sein de la partie déclarative de "+specif.designator+" les
identificateurs des spécifications des tâches et des sous-programmes ne
sont pas deux à deux différents":
declarations.typeSelect(subprogram_specification).collect(e|e.designator)
.intersect(declarations.typeSelect(single_task_declaration).collect(e|e.i
dentifier)).isEmpty;

context subprogram_body  if ! declarations.isEmpty ERROR
"Au sein de la partie déclarative de "+specif.designator+" les
identificateurs des implémentations des tâches et les spécifications des
sous-programmes ne sont pas deux à deux différents":
declarations.typeSelect(subprogram_specification).collect(e|e.designator)
.intersect(declarations.typeSelect(task_body).collect(e|e.identifier)).is
Empty;

context subprogram_body  if ! declarations.isEmpty ERROR
"Au sein de la partie déclarative de "+specif.designator+" les
identificateurs des spécifications des tâches et des implémentations des
sous-programmes ne sont pas deux à deux différents":
declarations.typeSelect(subprogram_body).collect(e|e.specif.designator).i
ntersect(declarations.typeSelect(single_task_declaration).collect(e|e.ide
ntifier)).isEmpty;

context subprogram_body  if ! declarations.isEmpty ERROR
"Au sein de la partie déclarative de "+specif.designator+" les
identificateurs des implémentations des tâches et des sous-programmes ne
sont pas deux à deux différents":
declarations.typeSelect(subprogram_body).collect(e|e.specif.designator).i
ntersect(declarations.typeSelect(task_body).collect(e|e.identifier)).isEm
pty;

context subprogram_body  if ! declarations.isEmpty ERROR
```

"Au sein de la partie déclarative de "+specif.designator+" les
identificateurs des implémentations des tâches et les spécifications des
tâches ne sont pas égaux":
```
declarations.typeSelect(single_task_declaration).collect(e|e.identifier)=
=declarations.typeSelect(task_body).collect(e|e.identifier);
```

**context** subprogram_body **if** ! declarations.isEmpty **ERROR**
"Au sein de la partie déclarative de "+specif.designator+" les
spécifications des sous-programmes doivent avoir des identificateurs
différents" :
```
declarations.typeSelect(subprogram_specification).forAll(e1|
declarations.typeSelect(subprogram_specification).notExists(e2|e1!=e2 &&
e1.designator==e2.designator));
```

**context** subprogram_body **if** ! declarations.isEmpty **ERROR**
"Au sein de la partie déclarative de "+specif.designator+" les
implémentations des sous-programmes doivent avoir des identificateurs
différents" :
```
declarations.typeSelect(subprogram_body).forAll(e1|
declarations.typeSelect(subprogram_body).notExists(e2|e1!=e2 &&
e1.specif.designator==e2.specif.designator));
```

**context** subprogram_body **if** ! declarations.isEmpty **ERROR**
"Au sein de la partie déclarative de "+specif.designator+" les
spécifications des tâches doivent avoir des identificateurs différents" :
```
declarations.typeSelect(single_task_declaration).forAll(e1|
declarations.typeSelect(single_task_declaration).notExists(e2|e1!=e2 &&
e1.identifier==e2.identifier));
```

**context** subprogram_body **if** ! declarations.isEmpty **ERROR**
"Au sein de la partie déclarative de "+specif.designator+" les
implémentations des tâches doivent avoir des noms différents" :
```
declarations.typeSelect(task_body).forAll(e1|
declarations.typeSelect(task_body).notExists(e2|e1!=e2 &&
e1.identifier==e2.identifier));
```

    ✦  Les contraintes relatives à la partie déclarative d'une tâche en Ada :

**context** task_body  **if** ! declarations.isEmpty **ERROR**
"Au sein de la partie déclarative de "+identifier+" les identificateurs
des spécifications des tâches et des sous-programmes ne sont pas deux à
deux différents":
```
declarations.typeSelect(subprogram_specification).collect(e|e.designator)
.intersect(declarations.typeSelect(single_task_declaration).collect(e|e.i
dentifier)).isEmpty;
```

**context** task_body  **if** ! declarations.isEmpty **ERROR**
"Au sein de la partie déclarative de "+identifier+" les identificateurs
des parties implémentations des tâches et les spécifications des sous-
programmes ne sont pas deux à deux différents":
```
declarations.typeSelect(subprogram_specification).collect(e|e.designator)
.intersect(declarations.typeSelect(task_body).collect(e|e.identifier)).is
Empty;
```

```
context task_body  if ! declarations.isEmpty ERROR
"Au sein de la partie déclarative de "+identifier+" les identificateurs
des parties spécifications des tâches et des implémentations des sous-
programmes ne sont pas deux à deux différents":
declarations.typeSelect(subprogram_body).collect(e|e.specif.designator).i
ntersect(declarations.typeSelect(single_task_declaration).collect(e|e.ide
ntifier)).isEmpty;
```

```
context task_body  if ! declarations.isEmpty ERROR
"Au sein de la partie déclarative de "+identifier+" les identificateurs
des implémentations des tâches et des sous-programmes ne sont pas deux à
deux différents":
declarations.typeSelect(subprogram_body).collect(e|e.specif.designator).i
ntersect(declarations.typeSelect(task_body).collect(e|e.identifier)).isEm
pty;
```

```
context task_body  if ! declarations.isEmpty ERROR
"Au sein de la partie déclarative de "+identifier+" les identificateurs
des implémentations des tâches et les spécifications des tâches ne sont
pas égaux":
declarations.typeSelect(single_task_declaration).collect(e|e.identifier)=
=declarations.typeSelect(task_body).collect(e|e.identifier);
```

```
context task_body if ! declarations.isEmpty ERROR
"Au sein de la partie déclarative de "+identifier+" les sous-programmes
doivent avoir des identificateurs différents" :
declarations.typeSelect(subprogram_specification).forAll(e1|
declarations.typeSelect(subprogram_specification).notExists(e2|e1!=e2 &&
e1.designator==e2.designator));
```

```
context task_body if ! declarations.isEmpty ERROR
"Au sein de la partie déclarative de "+identifier+" les déclarations des
tâches doivent avoir des identificateurs différents" :
declarations.typeSelect(single_task_declaration).forAll(e1|
declarations.typeSelect(single_task_declaration).notExists(e2|e1!=e2 &&
e1.identifier==e2.identifier));
```

```
context task_body if ! declarations.isEmpty ERROR
"Au sein de la partie déclarative de "+identifier+" les implémentations
des tâches doivent avoir des identificateurs différents" :
declarations.typeSelect(task_body).forAll(e1|
declarations.typeSelect(task_body).notExists(e2|e1!=e2 &&
e1.identifier!=e2.identifier));
```

```
context task_body if ! declarations.isEmpty ERROR
"Au sein de la partie déclarative de "+identifier+" les implémentations
des sous-programmes doivent avoir des identificateurs différents" :
declarations.typeSelect(subprogram_body).forAll(e1|
declarations.typeSelect(subprogram_body).notExists(e2|e1!=e2 &&
e1.specif.designator==e2.specif.designator));
```

     Les contraintes relatives à la partie comportementale d'un sous-programme en Ada :

```
context subprogram_body if specif.metaType==procedure_specification
WARNING
"La procédure "+specif.designator+" contient une (des) instruction(s)
return.":
statements.notExists(e|e.metaType==return_statement);

context subprogram_body if specif.metaType==function_specification ERROR
"La fonction "+specif.designator+" contient plus qu'une instruction
return.":
statements.collect(e|e.metaType==return_statement).size>=1;

context subprogram_body ERROR
"Le sous-programme "+specif.designator+" contient une (des)
instruction(s) accept.":
statements.notExists(e|e.metaType==simple_accept_statement);

context subprogram_body ERROR
"Le sous-programme "+specif.designator+" contient une (des)
instruction(s) select.":
statements.notExists(e|e.metaType==select_or);
```

🔸 Les contraintes relatives à la partie comportementale d'une tâche en Ada :

```
context task_body ERROR
"La tâche "+identifier+" ne peut pas contenir une (des) instruction(s)
return.":
statements.notExists(e|e.metaType==return_statement);

context task_body ERROR
"la tâche "+identifier+" ne peut pas accepter un rendez-vous non
disponible parmi sa liste d'entrée":
statements.typeSelect(single_task_declaration).collect(e|e.entryDec).coll
ect(e|e.identifier).containsAll(statements.typeSelect(simple_accept_state
ment).collect(e|e.direct_name));
```

## 6.2.2 Génération de code d'un sous programme Ada

Le sous-programme joue le rôle d'une fonction principale. Il est composé d'une spécification, d'un corps composé d'une partie déclarative et d'une partie exécutive. Ceci peut être traduit par le code Xpand suivant :

```
«DEFINE main FOR subprogram_body»
«FILE "adaCode.adb"»
«EXPAND specification FOR this.specif-»
«EXPAND declaration FOREACH this.declarations-»
begin
«EXPAND statement FOREACH this.statements-»
```

```
end «this.specif.designator»;
«ENDFILE»
«ENDDEFINE»
```

Les templates specification, declaration et statement seront redéfinis selon le contexte de leurs appels. Ceci permet de simplifier le code.

```
«DEFINE specification FOR subprogram_specification»
«ENDDEFINE»
«DEFINE declaration FOR declaration»
«ENDDEFINE»
«DEFINE statement FOR statement»
«ENDDEFINE»
```

### 6.2.2.1 La spécification d'un sous-programme Ada

Il existe deux formes de spécification pour les sous-programmes: une procédure et une fonction. Ceci peut être traduit par le code qui suit :

```
«DEFINE specification FOR procedure_specification»
procedure «this.designator» is
«ENDDEFINE»

«DEFINE specification FOR function_specification»
function «this.designator» return «this.returnType» is
«ENDDEFINE»
```

### 6.2.2.2 La partie déclarative d'Ada

La partie déclarative d'un sous-programme Ada peut contenir la déclaration d'autres sous-programmes, d'autres sous-programmes et des tâches.

#### 6.2.2.2.1 La déclaration de sous-programmes

La déclaration de sous-programmes se fait par leurs prototypes. Ceci est traduit par le code suivant :

```
«DEFINE declaration FOR procedure_specification»
procedure «this.designator» ;
«ENDDEFINE»

«DEFINE declaration FOR function_specification»
function «this.designator» return «this.returnType» ;
«ENDDEFINE»
```

#### 6.2.2.2.2 Les sous-programmes

La déclaration d'autres sous-programmes se traduit par le code suivant :

```
«DEFINE declaration FOR subprogram_body»
«EXPAND specification FOR this.specif-»
«EXPAND declaration FOREACH this.declarations-»
```

```
begin
«EXPAND statement FOREACH this.statements-»
end «this.specif.designator»;
«ENDDEFINE»
```

*6.2.2.2.3 Les tâches Ada*

Une tâche Ada est constituée  d'une partie déclarative et d'un corps.

✓ La partie déclarative d'une tâche :

La partie déclarative d'une tâche peut contenir les entrées de cette dernière.

```
«DEFINE declaration FOR single_task_declaration»
task «this.identifier» «IF this.entryDec.isEmpty» ; «ELSE» is
«EXPAND Entry FOREACH this.entryDec»
end «this.identifier»;
«ENDIF»
«ENDDEFINE»
«DEFINE Entry FOR entry_declaration»
entry «this.identifier» ;
«ENDDEFINE»
```

✓ Le corps d'une tâche :

Cette partie est constituée en deux parties : une partie décrivant les éventuelles déclarations de la tâche et une autre décrivant sa partie exécutive. Ceci peut être traduit par le code suivant :

```
«DEFINE declaration FOR task_body»
task body «this.identifier» is
«EXPAND declaration FOREACH this.declarations»
begin
«EXPAND statement FOREACH this.statements»
end «this.identifier»;
«ENDDEFINE»
```

Les deux parties déclarative et exécutive sont les même que celles d'un sous-programme ada.

## 6.2.2.3 La partie exécutive d'Ada

Cette partie concerne les instructions Ada.

✓ L'instruction if :

```
«DEFINE statement FOR if_else»
if «this.cond.c» then
«EXPAND statement FOREACH this.s1»
else
«EXPAND statement FOREACH this.s2»
```

```
end if;
«ENDDEFINE»
```

✓ L'instruction case :

```
«DEFINE statement FOR case_statement»
case «this.exp.e» is
«IF this.ref.notExists(e|e.choice=="others")»
«EXPAND Case FOREACH this.ref.reject(e|e.choice=="others")»
others => null;
«ELSE»
«EXPAND Case FOREACH this.ref.reject(e|e.choice=="others")»
«EXPAND Case FOREACH this.ref.select(e|e.choice=="others")»
«ENDIF»
end case;
«ENDDEFINE»

«DEFINE Case FOR case_statement_alternative»
when «this.choice» => «EXPAND statement FOREACH this.s»
«ENDDEFINE»
```

✓ L'instruction select_or :

```
«DEFINE statement FOR select_or»
select
«EXPAND Alternative FOREACH
this.ref.reject(e|e.metaType==ada::terminate_alternative) SEPARATOR 'or'-
»
«IF !this.ref.select(e|e.metaType==ada::terminate_alternative).isEmpty»
or
«ENDIF»
«EXPAND Alternative FOREACH
this.ref.select(e|e.metaType==ada::terminate_alternative) SEPARATOR 'or'-
»
end select;
«ENDDEFINE»

«DEFINE Alternative FOR select_alternative»
«ENDDEFINE»
```

✓ L'instruction terminate :

```
«DEFINE Alternative FOR terminate_alternative»
terminate;
«ENDDEFINE»
```

✓ L'instruction accept :

```
«DEFINE Alternative FOR accept_alternative»
accept «this.as.direct_name»;
```

```
«EXPAND statement FOREACH this.s»
«ENDDEFINE»
```

✓ L'instruction loop :

```
«DEFINE statement FOR simple_loop_statement»
loop
«EXPAND statement FOREACH this.s»
end loop;
«ENDDEFINE»
```

✓ L'instruction nulle :

```
«DEFINE statement FOR null_statement»
null;
«ENDDEFINE»
```

✓ L'instruction exit :

```
«DEFINE statement FOR exit_statement»
exit;
«ENDDEFINE»
```

✓ L'instruction return :

```
«DEFINE statement FOR return_statement»
return «this.exp.e»;
«ENDDEFINE»
```

✓ L'appel d'une procédure :

```
«DEFINE statement FOR procedure_call_statement»
«this.name»;
«ENDDEFINE»
```

✓ L'appel des entrées :

```
«DEFINE statement FOR entry_call_statement»
«this.entry_name»;
«ENDDEFINE»
```

Le template de génération de code Ada, en entier, est fourni dans l'annexe F.

### 6.2.3 Le moteur de validation et de génération de code Ada

Le workflow donné ci-dessous permet de générer le code Ada relatif au modèle XMI conforme au méta-modèle partiel d'Ada en utilisant les templates Xpand fournis précédemment.

```
<workflow>
        <property name="model"
value="my.generator.ada/src/example1/exampleAda.xmi" />
        <property name="src-gen" value="src-gen/example1" />

        <!-- set up EMF for standalone execution -->
        <bean class="org.eclipse.emf.mwe.utils.StandaloneSetup" >
                <platformUri value=".."/>
                <registerEcoreFile
value="platform:/resource/my.generator.ada/src/metamodel/Ada.ecore" />
        </bean>
        <!-- load model and store it in slot 'model' -->
        <component class="org.eclipse.emf.mwe.utils.Reader">
                <uri value="platform:/resource/${model}"  />
                <modelSlot value="model" />
        </component>

        <!-- check model -->
        <component class="org.eclipse.xtend.check.CheckComponent">
        <metaModel
class="org.eclipse.xtend.typesystem.emf.EmfRegistryMetaModel"/>
                <checkFile value="metamodel::CheckFile" />
                <emfAllChildrenSlot value="model" />
        </component>

        <!--  generate code -->
        <component class="org.eclipse.xpand2.Generator">
                <metaModel
class="org.eclipse.xtend.typesystem.emf.EmfRegistryMetaModel"/>
                <expand
                        value="template::Template::main FOR model" />
                <outlet path="${src-gen}" />
        </component>
</workflow>
```

### 6.2.4 Exemple d'utilisation

En exécutant le workflow sur le modèle d'Ada présenté dans l'annexe B, nous obtenons le code Ada suivant :

| | |
|---|---|
| **procedure** Client_Serveur **is** | |
| **function** condition_interne **return** Boolean **is** | **end if;** |
| **begin** | **end loop;** |

```
return true;
end condition_interne;
procedure traitement_interne1 is
begin
null;
end traitement_interne1;
procedure traitement_interne2 is
begin
null;
end traitement_interne2;
task Component_client1 is
entry port_Client_reponse ;
end Component_client1;
task Component_serveur1 is
entry port_Serveur_requete ;
end Component_serveur1;
task Connector_appel_cs is
entry Appele_reponse ;
entry Appelant_requete ;
end Connector_appel_cs;
task body Component_client1 is
begin
loop
if condition_interne then
exit;
else
traitement_interne1;
Connector_appel_cs.Appelant_requete;
accept port_Client_reponse;

end Component_client1;
task body Component_serveur1 is
begin
loop
if condition_interne then
exit;
else
traitement_interne2;
accept port_Serveur_requete;
Connector_appel_cs.Appele_reponse;
end if;
end loop;
end Component_serveur1;
task body Connector_appel_cs is
begin
loop
select
accept Appelant_requete;
Component_serveur1.port_Serveur_requete;
or
accept Appele_reponse;
Component_client1.port_Client_reponse;
or
terminate;
end select;
end loop;
end Connector_appel_cs;
begin
null;
end Client_Serveur;
```

Nous avons compilé et exécuté ce programme concurrent Ada en utilisant l'environnement (ObjectAda). Un tel programme traduisant l'architecture abstraite en Ada peut être raffiné step-by-step en prenant des décisions conceptuelles et techniques. La correction du raffinement est obtenue par l'utilisation des outils de vérification formelle associés à Ada tel que FLAVERS (Cobleigh, 2002).

## 6.3 Conclusion

Dans ce chapitre, nous avons proposé deux interfaces conviviales permettant d'utiliser le programme Wright2Ada transformant une architecture logicielle écrite en Wright vers un programme concurrent Ada. Pour y parvenir, nous avons utilisé avec profit les trois outils Xtext, Check, ATL et Xpand. Xtext nous a permis de transformer un texte Wright vers un modèle Wright, conforme à Grammaire Wright, respectant des contraintes exprimées en Check. Ensuite, moyennant notre programme GrammaireWright2Wright écrit en ATL, nous pouvons transformer des modèles Wright conforme au méta-modèle Grammaire Wright -obtenu automatiquement par Xtext- vers des modèles conforme à notre méta-modèle Wright. Xpand nous a permis de transformer un modèle XMI Ada vers un code Ada moyennant l'écriture des templates Xpand.

Dans le chapitre suivant, nous nous penchons sur la validation de notre programme Wright2Ada en utilisant une approche basé sur les tests syntaxiques (Syntax-Based Testing) (Xanthakis, 1999).

# Chapitre 7 : Validation

L'objectif de ce chapitre est de valider notre programme Wright2Ada écrit en ATL permettant de transformer une architecture logicielle décrite en Wright vers un programme concurrent Ada comportant plusieurs tâches (task). Pour y parvenir, nous préconisons une activité de vérification de ce programme basée sur le test fonctionnel ou encore boîte noire.

Ce chapitre comporte deux sections. La première section propose une approche orientée tests syntaxiques afin de vérifier notre programme Wright2Ada. La deuxième section propose des données de test (DT) permettant une certaine couverture de l'espace d'entrée de notre programme Wright2Ada.

## 7.1 Tests syntaxiques

Nous considérons le programme Wright2Ada comme boîte noire. Ainsi, nous nous plaçons dans le cadre de test fonctionnel. Notre programme Wright2Ada nécessite des données d'entrée (des spécifications ou des descriptions en Wright) respectant une syntaxe rigide et bien définie : la syntaxe de Wright décrite en Xtext (voir annexe C). Afin de couvrir l'espace de données du programme Wright2Ada, nous retenons les deux critères de couverture suivants :

Critère 1 : Couverture des symboles terminaux. Ils sont au nombre 79 unités lexicales couvrant des mots-clefs et des symboles utilisés par l'ADL Wright tels que : « Configuration », « Component », « Port », « Connector », « Role », « Instances », « Attachments », « As », « : », « ! », « ? », « § », « -> », « [] », « |~| », etc.

Critère 2 : Couverture des règles de productions permettant de définir les constructions syntaxiques offertes par Wright. Elles sont au nombre de 20 règles telles que : ComponentInstance, ConnectorInstance, EventSignalled, InternalTraitement, SuccesEvent, Data, ProcessName, Prefix, ProcessExpression, Port, Role, Component, Connector, Configuration.

Nous avons suivi une approche de prédiction des sorties attendues afin de tester notre programme Wright2Ada. La fonction d'oracle permettant de comparer la sortie observée par rapport à la sortie attendue pour une DT fournie est actuellement manuelle. Une automatisation de celle-ci peut être envisagée en s'inspirant de la commande diff offerte par un système d'exploitation de type Unix.

## 7.2 Les données de test (DT)

Afin de couvrir les deux critères proposés dans la section précédente, nous avons établi les DT suivantes :

### 7.2.1 Exemple dîner des philosophes

Nous testons ci-après l'exemple bien connu du dîner des philosophes. Notre exemple est tiré de (Déplanche, 2005). Nous nous limitons à une configuration de deux philosophes.

Cet exemple couvre toutes les règles de production excepté la règle InternalTraitement. De même, il couvre tous les teminaux exceptés l'opérateur déterministe [].

```
Configuration Diner
Component Philo
Port Gauche = _prendre -> _deposer
-> Gauche |~| §
Port Droite = _prendre -> _deposer
-> Droite |~| §
Computation = -penser ->
_Gauche.prendre -> _Droite.prendre
-> -manger -> _Gauche.deposer ->
_Droite.deposer -> computation |~|
§

Component Fourchette
Port Manche = prend -> depose ->
Manche |~| §
Computation = Manche.prend ->
Manche.depose -> computation |~| §
Connector Main
Role Mangeur = _prendre -> _deposer
-> Mangeur |~| V->STOP
Role Outil = prend -> depose ->
Outil |~| V->STOP

Glue = Mangeur.prendre ->
_Outil.prend -> glue |~|
Mangeur.deposer -> _Outil.depose ->
glue |~| SKIP
Instances
p1: Component Philo
p2: Component Philo
f1: Component Fourchette
f2: Component Fourchette
m11: Connector Main
m12: Connector Main
m21: Connector Main
m22: Connector Main

Attachments
p1-Gauche As m11-Mangeur
p1-Droite As m12-Mangeur
p2-Gauche As m21-Mangeur
p2-Droite As m22-Mangeur
f1-Manche As m11-Outil
f1-Manche As m22-Outil
f2-Manche As m12-Outil
f2-Manche As m21-Outil

End Configuration
```

La traduction Ada correspondante est à cet exemple se présente comme suit :

```ada
        procedure Diner is
        function  condition_interne  return
Boolean is
        begin return true;
        end condition_interne;
        function  condition_interne1  return
Integer is
        begin return 1;
        end condition_interne1;
        procedure penser is
        begin null; end penser;
        procedure manger is
        begin null; end manger;
        task Component_p1;
        task Component_p2;
        task Component_f1 is
        entry Manche_prend ;
        entry Manche_depose ;
        end Component_f1;
        task Component_f2 is
        entry Manche_prend ;
        entry Manche_depose ;
        end Component_f2;
        task Connector_m11 is
        entry Mangeur_prendre ;
        entry Mangeur_deposer ;
        end Connector_m11;
        task Connector_m12 is
        entry Mangeur_prendre ;
        entry Mangeur_deposer ;
        end Connector_m12;
        task Connector_m21 is
        entry Mangeur_prendre ;
```

```ada
        task body Component_f1 is
        begin loop
        if condition_interne then
        accept Manche_prend;
        accept Manche_depose;
        else exit; end if; end loop;
        end Component_f1;
        task body Component_f2 is
        begin loop
        if condition_interne then
        accept Manche_prend;
        accept Manche_depose;
        else exit; end if; end loop;
        end Component_f2;
        task body Connector_m11 is
        begin loop
        case condition_interne1 is
        when 1 => exit;
        when    2    =>    accept
Mangeur_prendre;
        Component_f1.Manche_prend;
        when    3    =>    accept
Mangeur_deposer;
        Component_f1.Manche_depose;
        when others => null;
        end case; end loop;
        end Connector_m11;
        task body Connector_m12 is
        begin loop
        case condition_interne1 is
        when 1 => exit;
        when    2    =>    accept
Mangeur_prendre;
        Component_f2.Manche_prend;
```

```
entry Mangeur_deposer ;
end Connector_m21;
task Connector_m22 is
entry Mangeur_prendre ;
entry Mangeur_deposer ;
end Connector_m22;
task body Component_p1 is
begin loop
if condition_interne then
penser;
Connector_m11.Mangeur_prendre;
Connector_m12.Mangeur_prendre;
manger;
Connector_m11.Mangeur_deposer;
Connector_m12.Mangeur_deposer;
else exit; end if; end loop;
end Component_p1;
task body Component_p2 is
begin loop
if condition_interne then
penser;
Connector_m21.Mangeur_prendre;
Connector_m22.Mangeur_prendre;
manger;
Connector_m21.Mangeur_deposer;
Connector_m22.Mangeur_deposer;
else exit; end if; end loop;
end Component_p2;

when 3 => accept
Mangeur_deposer;
Component_f2.Manche_depose;
when others => null ;
end case; end loop;
end Connector_m12;
task body Connector_m21 is
begin loop
case condition_interne1 is
when 1 => exit;
when 2 => accept
Mangeur_prendre;
Component_f2.Manche_prend;
when 3 => accept
Mangeur_deposer;
Component_f2.Manche_depose;
when others =>null;
end case; end loop;
end Connector_m21;
task body Connector_m22 is
begin loop
case condition_interne1 is
when 1 => exit;
when 2 => accept
Mangeur_prendre;
Component_f1.Manche_prend;
when 3 => accept
Mangeur_deposer;
Component_f1.Manche_depose;
when others =>null;
end case; end loop;
end Connector_m22;
begin
null;
end Diner;
```

### 7.2.2 Exemple gestion de places d'un parking

Nous testons ci-après un exemple d'une configuration pour la gestion de places d'un parking tiré de (Bhiri, 2008).

Cet exemple couvre toutes les règles de production excepté les règles InternalTraitement et SuccesEvent. De même, il couvre tous les teminaux exceptés SKIP et §.

```
Configuration GestionParking
Component Acces
Port r1 = voitureArrive -> (_reservation -> (reponsePositive -> r1[]
reponseNegative -> r1) |~| _liberation -> r1)
Computation = r1.voitureArrive -> (_r1.reservation -> (r1.reponsePositive
-> computation []r1.reponseNegative -> computation) |~| _r1.liberation ->
computation)
Component Afficheur
Port ecran = maj -> ecran
Computation = ecran.maj -> computation
Connector Parking
Role acces1 = voitureArrive -> (_reservation -> (reponsePositive ->
Acces1 [] reponseNegative -> acces1) |~| _liberation -> acces1)
Role acces2 = voitureArrive -> (_reservation -> (reponsePositive ->
Acces2 [] reponseNegative -> acces2) |~| _liberation -> acces2)
Role afficheur = maj -> afficheur
Glue = _acces1.voitureArrive -> (acces1.reservation -
>(_acces1.reponsePositive
-> _afficheur.maj -> glue |~| _acces1.reponseNegative -> glue) []
acces1.liberation ->
_afficheur.maj -> glue)
|~|
_acces2.voitureArrive -> (acces2.reservation ->
(_acces2.reponsePositive -> _afficheur.maj -> glue |~|
_acces2.reponseNegative -> glue)
[] acces2.liberation -> _afficheur.maj -> glue)
Instances
acces1: Component Acces
acces2 : Component Acces
afficheur1 : Component Afficheur
parking1 : Connector Parking
Attachments
acces1-r1 As parking1-acces1
acces2-r1 As parking1-acces2
afficheur1-ecran As parking1-afficheur
End Configuration
```

La traduction Ada qui correspond à cet exemple se présente comme suit :

**procedure** GestionParking **is**

**function** condition_interne **return** Boolean **is**

**begin return** true;

**end** condition_interne;

**task** Component_acces1 **is**

**entry** r1_reponsePositive ;

**entry** r1_reponseNegative ;

**entry** r1_voitureArrive ;

**end** Component_acces1;

**task** Component_acces2 **is**

**entry** r1_reponsePositive ;

**entry** r1_reponseNegative ;

**entry** r1_voitureArrive ;

**end** Component_acces2;

**task** Component_afficheur1 **is**

**entry** ecran_maj ;

**end** Component_afficheur1;

**task** Connector_parking1 **is**

**entry** acces2_reservation ;

**entry** acces1_reservation ;

**entry** acces1_liberation ;

**entry** acces2_liberation ;

**end** Connector_parking1;

**task body** Component_acces1 **is**

**begin loop**

**accept** r1_voitureArrive;

**if** condition_interne **then**

Connector_parking1.acces1_liberation;

**else**

Connector_parking1.acces1_reservation ; **select**

**accept** r1_reponseNegative;

**task body** Component_acces2 **is**

**begin loop**

**accept** r1_voitureArrive;

**if** condition_interne **then**

Connector_parking1.acces2_liberation;

**else**

Connector_parking1.acces2_reservation ; **select accept** r1_reponseNegative;

**or accept** r1_reponsePositive;

**end select; end if; end loop;**

**end** Component_acces2;

**task body** Component_afficheur1 **is**

**begin loop**

**accept** ecran_maj;

**end loop; end** Component_afficheur1;

**task body** Connector_parking1 **is**

**begin loop**

**if** condition_interne **then**

Component_acces1.r1_voitureArrive;

**select accept** acces1_reservation;

**if** condition_interne **then**

Component_acces1.r1_reponsePositive; Component_afficheur1.ecran_maj;

**else**

Component_acces1.r1_reponseNegative ; **end if;**

**or accept** acces1_liberation;

Component_afficheur1.ecran_maj;

**end select;**

**else**

Component_acces2.r1_voitureArrive;

**select accept** acces2_liberation;

Component_afficheur1.ecran_maj;

**or accept** acces2_reservation;

| | |
|---|---|
| **or accept** r1_reponsePositive;<br><br>**end select; end if; end loop;**<br>**end** Component_acces1; | **if** condition_interne **then**<br>    Component_acces2.r1_reponseNegative ;  **else**<br>    Component_acces2.r1_reponsePositive; Component_afficheur1.ecran_maj;<br>    **end if; end select; end if; end loop;**<br>    **end** Connector_parking1;<br>    **begin** null; **end** GestionParking; |

### 7.2.3 Exemple d'architecture client serveur

L'architecture Client-Serveur est déjà présentée dans la section 6.1.2.3. Le code en Ada correspondant est présenté dans la section 6.2.3.

Cet exemple couvre toutes les règles de production. Alors qu'il couvre tous les teminaux exceptés § et SKIP.

En conclusion, les trois exemples (DTs) fournis ci-dessus couvrent les deux critères de génération des DTs retenus dans 7.1.

## 7.3. Conclusion

Dans ce chapitre, nous avons proposé une approche de validation de notre programme Wright2Ada basée sur les tests syntaxiques. Nous avons identifié des données de test permettant une certaine couverture de l'espace de donnée de test permettant une certaine couverture de l'espace de données du programme Wright2Ada. Nous avons appliqué avec succès les tests établis sur le programme Wright2Ada. Ainsi, nous pouvons dire que le programme Wright2Ada est digne de confiance.

# Conclusion

Nous avons proposé une approche IDM permettant de transformer une architecture logicielle décrite à l'aide de l'ADL formel Wright vers un programme concurrent Ada comportant plusieurs tâches exécutées en parallèle. Pour y parvenir, nous avons élaboré deux méta-modèles en Ecore : le méta-modèle de Wright et le méta-modèle partiel d'Ada.

De plus, nous avons conçu et réalisé un programme Wright2Ada permettant de transformer un modèle source Wright conforme au méta-modèle Wright vers, un modèle cible Ada conforme au méta-modèle partiel d'Ada. Notre programme Wright2Ada est purement déclaratif et utilise avec profit les constructions déclaratives fournis par ATL, telles que : règles standard, règles paresseuses, helpers (attribut, opération). En outre, nous avons proposé des interfaces conviviales permettant de transformer du text Wright vers du code Ada en utilisant les outils de modélisation Xtext, Xpand et Check. Enfin, nous avons testé notre programme Wright2Ada en adoptant une approche orientée tests syntaxiques.

Quant aux perspectives de ce travail, nous pourrions envisager les prolongements suivants :

- ✓ Intégrer les facilités syntaxiques offertes par CSP telles que : where, when, processus avec état et opérateur de quantification sur les ensembles.

- ✓ Traiter les composants et les connecteurs composites offerts par Wright.

- ✓ Proposer l'opération de transformation inverse d'Ada vers Wright. Ceci favorise l'extraction d'une architecture logicielle abstraite en Wright à partir d'un programme concurrent Ada.

- ✓ Améliorer éventuellement l'efficacité de notre programme Wright2Ada en étudiant l'apport des patterns d'OCL pour la transformation de modèles (Cuadrado, 2009).

- ✓ Comparer l'approche classique de traduction dirigée par la syntaxe vis-à-vis de la transformation exogène de modèles.

- ✓ Vérifier davantage le programme Wright2Ada en utilisant des techniques de vérification applicables sur des programmes de transformations de modèles : test structurel, analyse de mutation, problème d'oracle et analyse statique (Küster, 2006) (Mottu, 2005) (Mottu, 2008) (Baudry, 2009).

# Bibliographie

# Annexe A : Wright2Ada.atl

```
-- @path Wright=/Wright/model/Wright.ecore
-- @path Ada=/my.generator.ada/src/metamodel/Ada.ecore

module WrightToAda;
create exampleAda : Ada from exampleWright : Wright;

helper context Wright!ProcessExpression def: getEventObserved():
Set(Wright!EventObserved) =
        if self.oclIsTypeOf(Wright!Prefix)then
            if self.event.oclIsTypeOf(Wright!EventObserved) then
                Set{self.event}-
>union(self.target.getEventObserved())
            else
                self.target.getEventObserved()
            endif
        else
            if self.oclIsTypeOf(Wright!InternalChoice) or
self.oclIsTypeOf(Wright!ExternalChoice) then
                self.elements->iterate( child1 ; elements1 :
Set(Wright!EventObserved) = Set{} | elements1-
>union(child1.getEventObserved()))
            else
                Set{}
            endif
        endif;

helper context Wright!ProcessExpression def: getInternalTrait():
Set(Wright!InternalTraitement) =
        if self.oclIsTypeOf(Wright!Prefix)then
            if self.event.oclIsTypeOf(Wright!InternalTraitement)
then
                Set{self.event}-
>union(self.target.getInternalTrait())
            else
                self.target.getInternalTrait()
            endif
        else
            if self.oclIsTypeOf(Wright!InternalChoice) or
self.oclIsTypeOf(Wright!ExternalChoice) then
                self.elements->iterate( child1 ; elements1 :
Set(Wright!InternalTraitement) = Set{} | elements1-
>union(child1.getInternalTrait()))
            else
                Set{}
            endif
        endif;

helper context Wright!Configuration def: getInternalTraitement:
Set(Wright!InternalTraitement)=
    self.conn->iterate( child1 ; elements1 :
Set(Wright!InternalTraitement) = Set{} | elements1-
>union(child1.glue.getInternalTrait()))
```

```
        ->union(self.comp->iterate( child2 ; elements2 :
Set(Wright!InternalTraitement) = Set{} | elements2-
>union(child2.computation.getInternalTrait()))));

helper context Wright!ExternalChoice def:
getPrefixInOrder():OrderedSet(Wright!Prefix) =
        self.elements->select(c |
c.event.oclIsTypeOf(Wright!EventObserved))
        ->union(self.elements->select(c |
c.event.oclIsTypeOf(Wright!SuccesEvent)));

helper context Wright!ExternalChoice def: transformation(instance :
String):Ada!select_or=
        thisModule.ExternalChoice2select_or(self,instance);

helper context Wright!InternalChoice def: transformation(instance :
String):Ada!statement=
        if self.elements->size()=2 then
                thisModule.InternalChoice2if_else(self,instance)
        else
                thisModule.InternalChoice2case_statement(self,instance)
        endif;

helper context Wright!Prefix def: transformation(instance :
String):Sequence(Ada!statement) =
                if self.target.oclIsTypeOf(Wright!Prefix)   then
                    Sequence{self.event.event_transform(instance)}-
>union(self.target.transformation(instance))
                else
                        if self.target.oclIsTypeOf(Wright!InternalChoice) or
self.target.oclIsTypeOf(Wright!ExternalChoice) then
                            Sequence{self.event.event_transform(instance)}-
>union(Sequence {self.target.transformation(instance)})
                        else
                            Sequence{self.event.event_transform(instance)}
                        endif
                endif;

helper context Wright!EventObserved def: event_transform(instance :
String):Ada!simple_accept_statement=
        thisModule.EventObserved2simple_accept_statement(self);

helper context Wright!EventSignalled def: event_transform(instance :
String):Ada!entry_call_statement=
        thisModule.EventSignalled2entry_call_statement(self,instance);

helper context Wright!InternalTraitement def: event_transform(instance :
String):Ada!procedure_call_statement=
        thisModule.InternalTraitement2procedure_call_statement(self);

helper context Wright!SuccesEvent def: event_transform(instance :
String):Ada!exit_statement=
        thisModule.SuccesEvent2exit_statement(self);

helper context Wright!Configuration def: ICBin:Wright!InternalChoice=
        Wright!InternalChoice.allInstances() ->select(e |e.elements-
>size()=2)->at(1);
```

```
helper context Wright!Configuration def: ICGen:Wright!InternalChoice=
      Wright!InternalChoice.allInstances() ->select(e |e.elements-
>size()>2)->at(1);

helper context Wright!Configuration def: existICGen:Boolean=
      if(Wright!InternalChoice.allInstances() ->select(e |e.elements-
>size()>2)->isEmpty() )then false else true endif;

helper context Wright!Configuration def: existICBin:Boolean=
      if(Wright!InternalChoice.allInstances() ->select(e |e.elements-
>size()=2)->isEmpty() )then false else true endif;

rule Configuration2subprogram{
      from c: Wright!Configuration
      to      sb: Ada!subprogram_body (
          specif <- sp ,
          statements <- st ,
          declarations <- c.getInternalTraitement ->
collect(e|thisModule.InternalTraitement2subprogram(e))
          ->union(if (c.existICGen)then
OrderedSet{thisModule.InternalChoiceG2function(c.ICGen)} else
OrderedSet{}endif)
          ->union(if (c.existICBin)then
OrderedSet{thisModule.InternalChoiceB2function(c.ICBin)} else
OrderedSet{}endif)
          ->union(c.compInst ->
collect(e|thisModule.ComponentInstance2single_task_declaration(e)))
          ->union(c.connInst ->
collect(e|thisModule.ConnectorInstance2single_task_declaration(e)))
          ->union(c.compInst ->
collect(e|thisModule.ComponentInstance2task_body(e)))
          ->union(c.connInst ->
collect(e|thisModule.ConnectorInstance2task_body(e)))),
          sp: Ada!procedure_specification(   designator <- c.name),
          st: Ada!null_statement
}

lazy rule InternalChoiceB2function{
      from i:Wright!InternalChoice(not i.OclUndefined())
      to    s:Ada!subprogram_body(specif <- fs, statements <- r),
          fs: Ada!function_specification(   designator <-
'condition_interne', returnType<-'Boolean'),
          r:Ada!return_statement(exp<-e1),
          e1:Ada!expression(e<-'true')
}

lazy rule InternalChoiceG2function{
      from i:Wright!InternalChoice(not i.OclUndefined())
      to    s:Ada!subprogram_body(specif <- fs, statements <- r),
          fs: Ada!function_specification(   designator <-
'condition_interne1', returnType<-'Integer'),
          r:Ada!return_statement(exp<-e1),
          e1:Ada!expression(e<-'1')
}

lazy rule InternalTraitement2subprogram{
      from i:Wright!InternalTraitment
      to sb: Ada!subprogram_body(
```

```
                    specif <- ps,
                    statements <-ns),
                ns:Ada!null_statement,
              ps: Ada!procedure_specification(      designator <- i.name)
}

lazy rule EventObserved2entry_declaration{
        from eo:Wright!EventObserved
        to ed:Ada!entry_declaration(
              identifier<- eo.name.replaceAll('.','_')
        )
}

lazy rule ComponentInstance2single_task_declaration{
        from ci:Wright!ComponentInstance
        to std:Ada!single_task_declaration(
              identifier <- 'Component_'+ci.name,
              entryDec <-ci.type.computation.getEventObserved()-
>collect(e|thisModule.EventObserved2entry_declaration(e))
              )
}

lazy rule ComponentInstance2task_body{
        from ci:Wright!ComponentInstance
        to    tb:Ada!task_body(
              identifier <-'Component_'+ ci.name,
              statements <- ls
              ),
              ls : Ada!simple_loop_statement(
                  s<- ci.type.computation.transformation(ci.name)
                  )
}

lazy rule ConnectorInstance2single_task_declaration{
        from ci:Wright!ConnectorInstance
        to std:Ada!single_task_declaration(
              identifier <- 'Connector_'+ci.name,
              entryDec <-ci.type.glue.getEventObserved()-
>collect(e|thisModule.EventObserved2entry_declaration(e))
              )
}

lazy rule ConnectorInstance2task_body{
        from ci:Wright!ConnectorInstance
        to    tb:Ada!task_body(
              identifier <-'Connector_'+ ci.name,
              statements <- ls
              ),
              ls : Ada!simple_loop_statement(
                  s<- ci.type.glue.transformation(ci.name))
}

lazy rule ExternalChoice2select_or{
        from p:Wright!ExternalChoice,
             instance : String
        to s:Ada!select_or(
              ref <- p.getPrefixInOrder()->collect(e|
                  if e.event.oclIsTypeOf(Wright!EventObserved)then
```

```
                            if e.target.oclIsTypeOf(Wright!ProcessName)  then

        thisModule.Prefix2accept_alternative1(e,instance)
                            else

        thisModule.Prefix2accept_alternative2(e,instance)
                            endif
                    else
                            thisModule.SuccesEvent2terminate_alternative(e)
                    endif ))
}

lazy rule Prefix2accept_alternative1{
        from p:Wright!Prefix,
            instance : String
        to a:Ada!accept_alternative(
        as <- thisModule.EventObserved2simple_accept_statement(p.event)
            )
}

lazy rule Prefix2accept_alternative2{
        from p:Wright!Prefix,
            instance : String
        to a:Ada!accept_alternative(
        as <- thisModule.EventObserved2simple_accept_statement(p.event),
            s<- p.target.transformation(instance)
            )
}

lazy rule SuccesEvent2terminate_alternative{
        from p:Wright!SuccesEvent
        to a:Ada!terminate_alternative
}

lazy rule InternalChoice2if_else{
        from p:Wright!InternalChoice,
            instance : String
        to ls:Ada!if_else(
            s1 <- p.elements->at(1).transformation(instance),
            s2 <- p.elements->at(2).transformation(instance),
            cond<-c
            ),
            c:Ada!condition(c<-'condition_interne')
}

lazy rule InternalChoice2case_statement{
        from p:Wright!InternalChoice,
            instance : String
        to ls:Ada!case_statement(
            ref <- p.elements-
>collect(e|thisModule.Prefix2case_statement_alternative(e,p.elements.inde
xOf(e),instance)),
            exp<-c
            ),
            c:Ada!expression(e<-'condition_interne1')
}

lazy rule Prefix2case_statement_alternative{
```

```
        from p:Wright!Prefix,
             index:Integer,
             instance: String
        to cs: Ada!case_statement_alternative(
             choice<-index.toString(),
             s<- p.transformation(instance)
        )
}

lazy rule SuccesEvent2exit_statement{
        from p:Wright!SuccesEvent
        to e:Ada!exit_statement
}

lazy rule EventObserved2simple_accept_statement{
        from e:Wright!EventObserved
        to s:Ada!simple_accept_statement(
             direct_name<- e.name.replaceAll('.','_')
             )
}

lazy rule EventSignalled2entry_call_statement{
        from e:Wright!EventSignalled,
             instance : String
        to ec:Ada!entry_call_statement(
             entry_name<-if(Wright!Attachment.allInstances()-
>select(a|a.originPort.name=e.name.substring(1,e.name.indexOf('.')))-
>select(a|a.originInstance.name=instance)->notEmpty())then
                              'Connector_'
                              +Wright!Attachment.allInstances()-
>select(a|a.originPort.name=e.name.substring(1,e.name.indexOf('.')))-
>select(a|a.originInstance.name=instance).at(1).targetInstance.name
                              +'.'+Wright!Attachment.allInstances()-
>select(a|a.originPort.name=e.name.substring(1,e.name.indexOf('.')))-
>select(a|a.originInstance.name=instance).at(1).targetRole.name

        +'_'+e.name.substring(e.name.indexOf('.')+2,e.name.size())
                              else
                              'Component_'
                              +Wright!Attachment.allInstances()-
>select(a|a.targetRole.name=e.name.substring(1,e.name.indexOf('.')))-
>select(a|a.targetInstance.name=instance).at(1).originInstance.name
                              +'.'+Wright!Attachment.allInstances()-
>select(a|a.targetRole.name=e.name.substring(1,e.name.indexOf('.')))-
>select(a|a.targetInstance.name=instance).at(1).originPort.name

        +'_'+e.name.substring(e.name.indexOf('.')+2,e.name.size())
                              endif
             )
}

lazy rule InternalTraitement2procedure_call_statement{
        from e:Wright!InternalTraitement
        to p:Ada!procedure_call_statement(
             name<-e.name
             )
}
```

# Annexe B : Exemple d'utilisation de Wright2Ada

- Le modèle source, de l'architecture client serveur, conforme au méta-modèle Wright se présente comme suit :

```xml
<?xml version="1.0" encoding="ISO-8859-1"?>
<wright:Configuration                              xmi:version="2.0"
xmlns:xmi="http://www.omg.org/XMI"
xmlns:xsi="http://www.w3.org/2001/XMLSchema-instance"
xmlns:wright="http://wright/1.0" name="Client_Serveur">
  <comp name="Client">
    <port name="port_Client">
      <behavior xsi:type="wright:InternalChoice">
        <elements>
          <target xsi:type="wright:ProcessName" name="STOP"/>
          <event xsi:type="wright:SuccesEvent" name="V"/>
        </elements>
        <elements>
          <target xsi:type="wright:Prefix">
            <target xsi:type="wright:ProcessName" name="port_Client"/>
            <event xsi:type="wright:EventObserved" name="reponse"/>
          </target>
          <event xsi:type="wright:EventSignalled" name="requete"/>
        </elements>
      </behavior>
    </port>
    <computation xsi:type="wright:InternalChoice">
      <elements>
        <target xsi:type="wright:ProcessName" name="STOP"/>
        <event xsi:type="wright:SuccesEvent" name="V"/>
      </elements>
      <elements>
        <target xsi:type="wright:Prefix">
          <target xsi:type="wright:Prefix">
            <target xsi:type="wright:ProcessName" name="computation"/>
            <event                        xsi:type="wright:EventObserved"
name="port_Client.reponse"/>
          </target>
          <event                          xsi:type="wright:EventSignalled"
name="port_Client.requete"/>
        </target>
        <event                         xsi:type="wright:InternalTraitement"
name="traitement_interne1"/>
      </elements>
    </computation>
  </comp>
  <comp name="Serveur">
    <port name="port_Serveur">
      <behavior xsi:type="wright:InternalChoice">
        <elements>
          <target xsi:type="wright:Prefix">
            <target xsi:type="wright:ProcessName" name="port_Serveur"/>
            <event xsi:type="wright:EventSignalled" name="reponse"/>
          </target>
          <event xsi:type="wright:EventObserved" name="requete"/>
        </elements>
        <elements>
```

```xml
              <target xsi:type="wright:ProcessName" name="STOP"/>
              <event xsi:type="wright:SuccesEvent" name="V"/>
            </elements>
          </behavior>
        </port>
        <computation xsi:type="wright:InternalChoice">
          <elements>
            <target xsi:type="wright:ProcessName" name="STOP"/>
            <event xsi:type="wright:SuccesEvent" name="V"/>
          </elements>
          <elements>
            <target xsi:type="wright:Prefix">
              <target xsi:type="wright:Prefix">
                <target xsi:type="wright:ProcessName" name="computation"/>
                <event                        xsi:type="wright:EventSignalled"
name="port_Serveur.reponse"/>
              </target>
              <event                              xsi:type="wright:EventObserved"
name="port_Serveur.requete"/>
            </target>
            <event                            xsi:type="wright:InternalTraitement"
name="traitement_interne2"/>
          </elements>
        </computation>
      </comp>
      <compInst type="//@comp.0" name="client1"/>
      <compInst type="//@comp.1" name="serveur1"/>
      <conn name="Lien_CS">
        <role name="Appelant">
          <behavior xsi:type="wright:InternalChoice">
            <elements>
              <target xsi:type="wright:Prefix">
                <target xsi:type="wright:ProcessName" name="Appelant"/>
                <event xsi:type="wright:EventObserved" name="reponse"/>
              </target>
              <event xsi:type="wright:EventSignalled" name="requete"/>
            </elements>
            <elements>
              <target xsi:type="wright:ProcessName" name="STOP"/>
              <event xsi:type="wright:SuccesEvent" name="V"/>
            </elements>
          </behavior>
        </role>
        <role name="Appele">
          <behavior xsi:type="wright:ExternalChoice">
            <elements>
              <target xsi:type="wright:ProcessName" name="STOP"/>
              <event xsi:type="wright:SuccesEvent" name="V"/>
            </elements>
            <elements>
              <target xsi:type="wright:Prefix">
                <target xsi:type="wright:ProcessName" name="Appele"/>
                <event xsi:type="wright:EventSignalled" name="reponse"/>
              </target>
              <event xsi:type="wright:EventObserved" name="requete"/>
            </elements>
          </behavior>
        </role>
```

```xml
    <glue xsi:type="wright:ExternalChoice">
      <elements>
        <target xsi:type="wright:Prefix">
          <target xsi:type="wright:ProcessName" name="glue"/>
          <event xsi:type="wright:EventSignalled" name="Appele.requete"/>
        </target>
        <event xsi:type="wright:EventObserved" name="Appelant.requete"/>
      </elements>
      <elements>
        <target xsi:type="wright:ProcessName" name="STOP"/>
        <event xsi:type="wright:SuccesEvent" name="V"/>
      </elements>
      <elements>
        <target xsi:type="wright:Prefix">
          <target xsi:type="wright:ProcessName" name="glue"/>
          <event                         xsi:type="wright:EventSignalled"
name="Appelant.reponse"/>
        </target>
        <event xsi:type="wright:EventObserved" name="Appele.reponse"/>
      </elements>
    </glue>
  </conn>
  <att    originInstance="//@compInst.0"    originPort="//@comp.0/@port.0"
targetRole="//@conn.0/@role.0" targetInstance="//@connInst.0"/>
  <att    originInstance="//@compInst.1"    originPort="//@comp.1/@port.0"
targetRole="//@conn.0/@role.1" targetInstance="//@connInst.0"/>
  <connInst name="appel_cs" type="//@conn.0"/>
</wright:Configuration>
```

- Le modèle ada généré conforme au méta-modèle Ada se présente comme suit :

```xml
<?xml version="1.0" encoding="ISO-8859-1"?>
<ada:subprogram_body xmi:version="2.0"  xmlns:xmi="http://www.omg.org/XMI"
xmlns:xsi="http://www.w3.org/2001/XMLSchema-instance"
xmlns:ada="http://ada/1.0">
  <statements xsi:type="ada:null_statement"/>
  <declarations                          xsi:type="ada:single_task_declaration"
identifier="Component_client1">
    <entryDec identifier="port_Client_reponse"/>
  </declarations>
  <declarations                          xsi:type="ada:single_task_declaration"
identifier="Component_serveur1">
    <entryDec identifier="port_Serveur_requete"/>
  </declarations>
  <declarations xsi:type="ada:task_body" identifier="Component_client1">
    <statements xsi:type="ada:simple_loop_statement">
      <s xsi:type="ada:if_else">
        <cond expression="condition_interne"/>
        <s1 xsi:type="ada:exit_statement"/>
        <s2                     xsi:type="ada:procedure_call_statement"
name="traitement_interne1"/>
        <s2                          xsi:type="ada:entry_call_statement"
entry_name="Connector_appel_cs.Appelant_requete"/>
        <s2                          xsi:type="ada:simple_accept_statement"
direct_name="port_Client_reponse"/>
```

```xml
            </s>
        </statements>
    </declarations>
    <declarations xsi:type="ada:task_body" identifier="Component_serveur1">
        <statements xsi:type="ada:simple_loop_statement">
            <s xsi:type="ada:if_else">
                <cond expression="condition_interne"/>
                <s1 xsi:type="ada:exit_statement"/>
                <s2                          xsi:type="ada:procedure_call_statement"
name="traitement_interne2"/>
                <s2                          xsi:type="ada:simple_accept_statement"
direct_name="port_Serveur_requete"/>
                <s2                          xsi:type="ada:entry_call_statement"
entry_name="Connector_appel_cs.Appele_reponse"/>
            </s>
        </statements>
    </declarations>
    <declarations                         xsi:type="ada:single_task_declaration"
identifier="Connector_appel_cs">
        <entryDec identifier="Appele_reponse"/>
        <entryDec identifier="Appelant_requete"/>
    </declarations>
    <declarations xsi:type="ada:task_body" identifier="Connector_appel_cs">
        <statements xsi:type="ada:simple_loop_statement">
            <s xsi:type="ada:select_or">
                <ref xsi:type="ada:accept_alternative">
                    <as direct_name="Appelant_requete"/>
                    <s                          xsi:type="ada:entry_call_statement"
entry_name="Component_serveur1.port_Serveur_requete"/>
                </ref>
                <ref xsi:type="ada:accept_alternative">
                    <as direct_name="Appele_reponse"/>
                    <s                          xsi:type="ada:entry_call_statement"
entry_name="Component_client1.port_Client_reponse"/>
                </ref>
                <ref xsi:type="ada:terminate_alternative"/>
            </s>
        </statements>
    </declarations>
    <declarations xsi:type="ada:subprogram_body">
        <statements xsi:type="ada:return_statement" expression="true"/>
        <specif                          xsi:type="ada:function_specification"
designator="condition_interne" returnType="Boolean"/>
    </declarations>
    <declarations xsi:type="ada:subprogram_body">
        <statements xsi:type="ada:return_statement" expression="1"/>
        <specif                          xsi:type="ada:function_specification"
designator="condition_interne1" returnType="Integer"/>
    </declarations>
    <declarations xsi:type="ada:subprogram_body">
        <statements xsi:type="ada:null_statement"/>
        <specif                          xsi:type="ada:procedure_specification"
designator="traitement_interne2"/>
    </declarations>
    <declarations xsi:type="ada:subprogram_body">
        <statements xsi:type="ada:null_statement"/>
        <specif                          xsi:type="ada:procedure_specification"
designator="traitement_interne1"/>
```

```
  </declarations>
  <specif                            xsi:type="ada:procedure_specification"
designator="Client_Serveur"/>
</ada:subprogram_body>
```

# Annexe C : Grammaire de Wright en Xtext

```
grammar org.xtext.example.Wright1 with org.eclipse.xtext.common.Terminals

generate wright1 "http://www.xtext.org/example/Wright1"

Configuration : "Configuration" name=ID
                          ( TypeList+=Type )*
                          "Instances"
                                  ( InstanceList+=Instance)*
                          "Attachments"
                                  ( att+=Attachment )*
                          "End Configuration";

Instance: ComponentInstance | ConnectorInstance ;

Type: Component| Connector;

Component : "Component" name=ID
                          ( port+=Port )+
                          "Computation" '=' computation=ProcessExpression
;

Port : "Port" name=ID '=' behavior=ProcessExpression;

Connector : "Connector" name=ID
                          ( role+=Role )+
                          "Glue" '=' glue=ProcessExpression ;

Role : "Role" name=ID '=' behavior=ProcessExpression;

ComponentInstance : name=ID ':' "Component" type=[Component];

ConnectorInstance : name=ID ':' "Connector" type=[Connector];

Attachment : originInstance=[ComponentInstance] '-' originPort=[Port]
"As" targetInstance=[ConnectorInstance] '-' targetRole=[Role] ;

EventExpression : EventSignalled | EventObserved | InternalTraitement |
SuccesEvent;

EventSignalled:  '_' name=ID (data+=Data)*;

EventObserved: name=ID (data+=Data)*;

InternalTraitement: '-' name=ID;

SuccesEvent: name='V';

Data : ('?' | '!') name=ID;

Prefix: event=EventExpression '->' target=TargetPrefix | name='§'|
name='SKIP';

TargetPrefix: Parentheses | Prefix | ProcessName ;
```

```
ProcessName:  name=ID ;

Parentheses: '(' p=ProcessExpression ')';

ProcessExpression : right=Prefix (('[]' ECLeft+=Prefix)+|('|~|'
ICLeft+=Prefix)+)?;

terminal ID: ('a'..'z'|'A'..'Z') ('a'..'z'|'A'..'Z'|'_'|'.'|'0'..'9')*;
```

# Annexe D : Le modèle XMI conforme à la grammaire de Wright

```xml
<?xml version="1.0" encoding="ASCII"?>
<wright1:Configuration                                    xmi:version="2.0"
xmlns:xmi="http://www.omg.org/XMI"
xmlns:xsi="http://www.w3.org/2001/XMLSchema-instance"
xmlns:wright1="http://www.xtext.org/example/Wright1"
xsi:schemaLocation="http://www.xtext.org/example/Wright1
java://org.xtext.example.wright1.Wright1Package" name="Client_Serveur">
  <TypeList xsi:type="wright1:Connector" name="Lien_CS">
    <role name="Appelant">
      <behavior>
        <right>
          <event xsi:type="wright1:EventSignalled" name="requete"/>
          <target xsi:type="wright1:Prefix">
            <event xsi:type="wright1:EventObserved" name="reponse"/>
            <target xsi:type="wright1:ProcessName" name="Appelant"/>
          </target>
        </right>
        <ICLeft>
          <event xsi:type="wright1:SuccesEvent" name="V"/>
          <target xsi:type="wright1:ProcessName" name="STOP"/>
        </ICLeft>
      </behavior>
    </role>
    <role name="Appele">
      <behavior>
        <right>
          <event xsi:type="wright1:EventObserved" name="requete"/>
          <target xsi:type="wright1:Prefix">
            <event xsi:type="wright1:EventSignalled" name="reponse"/>
            <target xsi:type="wright1:ProcessName" name="Appele"/>
          </target>
        </right>
        <ECLeft>
          <event xsi:type="wright1:SuccesEvent" name="V"/>
          <target xsi:type="wright1:ProcessName" name="STOP"/>
        </ECLeft>
      </behavior>
    </role>
    <glue>
      <right>
        <event xsi:type="wright1:EventObserved" name="Appelant.requete"/>
        <target xsi:type="wright1:Prefix">
          <event                       xsi:type="wright1:EventSignalled"
name="Appele.requete"/>
          <target xsi:type="wright1:ProcessName" name="glue"/>
        </target>
      </right>
      <ECLeft>
        <event xsi:type="wright1:EventObserved" name="Appele.reponse"/>
        <target xsi:type="wright1:Prefix">
          <event                       xsi:type="wright1:EventSignalled"
name="Appelant.reponse"/>
          <target xsi:type="wright1:ProcessName" name="glue"/>
```

```xml
            </target>
          </ECLeft>
          <ECLeft>
            <event xsi:type="wright1:SuccesEvent" name="V"/>
            <target xsi:type="wright1:ProcessName" name="STOP"/>
          </ECLeft>
        </glue>
      </TypeList>
      <TypeList xsi:type="wright1:Component" name="Client">
        <port name="port_Client">
          <behavior>
            <right>
              <event xsi:type="wright1:EventSignalled" name="requete"/>
              <target xsi:type="wright1:Prefix">
                <event xsi:type="wright1:EventObserved" name="reponse"/>
                <target xsi:type="wright1:ProcessName" name="port_Client"/>
              </target>
            </right>
            <ICLeft>
              <event xsi:type="wright1:SuccesEvent" name="V"/>
              <target xsi:type="wright1:ProcessName" name="STOP"/>
            </ICLeft>
          </behavior>
        </port>
        <computation>
          <right>
            <event                        xsi:type="wright1:InternalTraitement"
name="traitement_interne1"/>
            <target xsi:type="wright1:Prefix">
              <event                          xsi:type="wright1:EventSignalled"
name="port_Client.requete"/>
              <target xsi:type="wright1:Prefix">
                <event                        xsi:type="wright1:EventObserved"
name="port_Client.reponse"/>
                <target xsi:type="wright1:ProcessName" name="computation"/>
              </target>
            </target>
          </right>
          <ICLeft>
            <event xsi:type="wright1:SuccesEvent" name="V"/>
            <target xsi:type="wright1:ProcessName" name="STOP"/>
          </ICLeft>
        </computation>
      </TypeList>
      <TypeList xsi:type="wright1:Component" name="Serveur">
        <port name="port_Serveur">
          <behavior>
            <right>
              <event xsi:type="wright1:EventObserved" name="requete"/>
              <target xsi:type="wright1:Prefix">
                <event xsi:type="wright1:EventSignalled" name="reponse"/>
                <target xsi:type="wright1:ProcessName" name="port_Serveur"/>
              </target>
            </right>
            <ICLeft>
              <event xsi:type="wright1:SuccesEvent" name="V"/>
              <target xsi:type="wright1:ProcessName" name="STOP"/>
            </ICLeft>
```

```
            </behavior>
        </port>
        <computation>
            <right>
                <event                         xsi:type="wright1:InternalTraitement"
name="traitement_interne2"/>
                <target xsi:type="wright1:Prefix">
                    <event                          xsi:type="wright1:EventObserved"
name="port_Serveur.requete"/>
                    <target xsi:type="wright1:Prefix">
                        <event                      xsi:type="wright1:EventSignalled"
name="port_Serveur.reponse"/>
                        <target xsi:type="wright1:ProcessName" name="computation"/>
                    </target>
                </target>
            </right>
            <ICLeft>
                <event xsi:type="wright1:SuccesEvent" name="V"/>
                <target xsi:type="wright1:ProcessName" name="STOP"/>
            </ICLeft>
        </computation>
    </TypeList>
    <InstanceList     xsi:type="wright1:ComponentInstance"      name="client1"
type="//@TypeList.1"/>
    <InstanceList     xsi:type="wright1:ComponentInstance"      name="serveur1"
type="//@TypeList.2"/>
    <InstanceList     xsi:type="wright1:ConnectorInstance"      name="appel_cs"
type="//@TypeList.0"/>
    <att                                  originInstance="//@InstanceList.0"
originPort="//@TypeList.1/@port.0"       targetInstance="//@InstanceList.2"
targetRole="//@TypeList.0/@role.0"/>
    <att                                  originInstance="//@InstanceList.1"
originPort="//@TypeList.2/@port.0"       targetInstance="//@InstanceList.2"
targetRole="//@TypeList.0/@role.1"/>

        </wright1:Configuration>
```

# Annexe E : GrammaireWright2Wright en ATL

```
-- @path Wright=/Wright/model/Wright.ecore
-- @path Wright1=/org.xtext.example.wright1/src-
gen/org/xtext/example/Wright1.ecore

module Wright1ToWright;
create exampleWright : Wright from exampleWright1 : Wright1;

helper context Wright1!ProcessExpression
def:transformation():Wright!ProcessExpression=
    if(self.ECLeft->notEmpty())then
        thisModule.ProcessExpression2ExternalChoice(self)
    else
        if(self.ICLeft->notEmpty())then
            thisModule.ProcessExpression2InternalChoice(self)
        else
            if(self.ICLeft->isEmpty() and self.ECLeft-
>isEmpty())then
                thisModule.ProcessExpression2Prefix(self)
            else
                false
            endif
        endif
    endif;
helper context Wright1!Parentheses
def:transformation():Wright!ProcessExpression=
    self.p.transformation();
helper context Wright1!Prefix def:transformation():Wright!Prefix=
    if(self.name='$' or self.name='SKIP') then
        thisModule.SKIP2Prefix(self)
    else
        thisModule.Prefix2Prefix(self)
    endif;
helper context Wright1!ProcessName
def:transformation():Wright!ProcessName=
    thisModule.ProcessName2ProcessName(self);

rule Configuration2Configuration{
    from c1:Wright1!Configuration
    to c:Wright!Configuration(
        name<-c1.name,
        comp<-c1.TypeList-
>select(e|e.oclIsTypeOf(Wright1!Component)),
        conn<-c1.TypeList-
>select(e|e.oclIsTypeOf(Wright1!Connector)),
        compInst<-c1.InstanceList-
>select(e|e.oclIsTypeOf(Wright1!ComponentInstance)),
        connInst<-c1.InstanceList-
>select(e|e.oclIsTypeOf(Wright1!ConnectorInstance)),
        att<-c1.att)
}
rule Component2Component{
    from c1:Wright1!Component
    to c:Wright!Component(
        name<-c1.name,
```

```
                port<-c1.port,
                computation<-c1.computation.transformation()
                )
    }
    rule Connector2Connector{
        from c1:Wright1!Connector
        to c:Wright!Connector(
                name<-c1.name,
                role<-c1.role,
                glue<-c1.glue.transformation()
                )
    }
    rule ComponentInstance2ComponentInstance{
        from i1:Wright1!ComponentInstance
        to i:Wright!ComponentInstance(
                name<-i1.name,
                type<-i1.type
                )
    }
    rule ConnectorInstance2ConnectorInstance{
        from i1:Wright1!ConnectorInstance
        to i:Wright!ConnectorInstance(
                name<-i1.name,
                type<-i1.type
                )
    }
    rule Attachment2Attachment{
        from a1:Wright1!Attachment
        to a:Wright!Attachment(
                originInstance<-a1.originInstance,
                targetInstance<-a1.targetInstance,
                originPort<-a1.originPort,
                targetRole<-a1.targetRole
        )
    }
    rule Port2Port{
        from p1:Wright1!Port
        to p:Wright!Port(
                name<-p1.name,
                behavior<-p1.behavior.transformation()
        )
    }
    rule Role2Role{
        from r1:Wright1!Role
        to r:Wright!Role(
                name<-r1.name,
                behavior<-r1.behavior.transformation()
        )
    }
    lazy rule ProcessExpression2Prefix{
        from e:Wright1!ProcessExpression
        to p:Wright!Prefix(
                event<-e.right.event,
                target<-e.right.target.transformation()
        )
    }
    lazy rule ProcessExpression2ExternalChoice{
        from e:Wright1!ProcessExpression
```

```
        to pe:Wright!ExternalChoice(
            elements<-Set{p}->union(e.ECLeft-
>collect(e|e.transformation()))),
            p:Wright!Prefix(event<-e.right.event, target<-
e.right.target.transformation())
}
lazy rule ProcessExpression2InternalChoice{
    from e:Wright1!ProcessExpression
        to pe:Wright!InternalChoice(
            elements<-Set{p}->union(e.ICLeft-
>collect(e|e.transformation()))),
            p:Wright!Prefix(event<-e.right.event, target<-
e.right.target.transformation())
}
lazy rule Prefix2Prefix{
    from p1:Wright1!Prefix
        to p:Wright!Prefix(
            event<-p1.event,
            target<-p1.target.transformation())
}
lazy rule SKIP2Prefix{
    from p1:Wright1!Prefix
        to p:Wright!Prefix(
            event<-e,
            target<-t),
            e:Wright!SuccesEvent(name<-'V'),
            t:Wright!ProcessName(name<-'STOP')
}
lazy rule ProcessName2ProcessName{
    from p1:Wright1!ProcessName
        to p:Wright!ProcessName(name<-p1.name)
}
rule EventSignalled2EventSignalled{
    from e1:Wright1!EventSignalled
        to e:Wright!EventSignalled(name<-e1.name)
}
rule EventObserved2EventObserved{
    from e1:Wright1!EventObserved
        to e:Wright!EventObserved(name<-e1.name)
}
rule InternalTraitement2InternalTraitement{
    from e1:Wright1!InternalTraitement
        to e:Wright!InternalTraitement(name<-e1.name)
}
rule SuccesEvent2SuccesEvent{
    from e1:Wright1!SuccesEvent
        to e:Wright!SuccesEvent(name<-e1.name)
}
```

# Annexe F : Le template de génération de code Ada en Xpand

```
«IMPORT ada»

«DEFINE main FOR subprogram_body»
«FILE "adaCode.adb"»
«EXPAND specification FOR this.specif-»
«EXPAND declaration FOREACH this.declarations-»
begin
«EXPAND statement FOREACH this.statements-»
end «this.specif.designator»;
«ENDFILE»
«ENDDEFINE»

«DEFINE specification FOR subprogram_specification»
«ENDDEFINE»

«DEFINE specification FOR procedure_specification»
procedure «this.designator» is
«ENDDEFINE»

«DEFINE specification FOR function_specification»
function «this.designator» return «this.returnType» is
«ENDDEFINE»

«DEFINE declaration FOR declaration»
«ENDDEFINE»

«DEFINE declaration FOR procedure_specification»
procedure «this.designator» ;
«ENDDEFINE»

«DEFINE declaration FOR function_specification»
function «this.designator» return «this.returnType» ;
«ENDDEFINE»

«DEFINE declaration FOR subprogram_body»
«EXPAND specification FOR this.specif-»
«EXPAND declaration FOREACH this.declarations-»
begin
«EXPAND statement FOREACH this.statements-»
end «this.specif.designator»;
«ENDDEFINE»

«DEFINE declaration FOR single_task_declaration»
task «this.identifier» «IF this.entryDec.isEmpty» ; «ELSE» is
«EXPAND Entry FOREACH this.entryDec-»
end «this.identifier»;
«ENDIF»
«ENDDEFINE»

«DEFINE Entry FOR entry_declaration»
entry «this.identifier» ;
«ENDDEFINE»
```

```
«DEFINE declaration FOR task_body»
task body «this.identifier» is
«EXPAND declaration FOREACH this.declarations-»
begin
«EXPAND statement FOREACH this.statements-»
end «this.identifier»;
«ENDDEFINE»

«DEFINE statement FOR statement»
«ENDDEFINE»

«DEFINE statement FOR null_statement»
null;
«ENDDEFINE»

«DEFINE statement FOR simple_loop_statement»
loop
«EXPAND statement FOREACH this.s-»
end loop;
«ENDDEFINE»

«DEFINE statement FOR if_else»
if «this.cond.c» then
«EXPAND statement FOREACH this.s1-»
else
«EXPAND statement FOREACH this.s2-»
end if;
«ENDDEFINE»

«DEFINE statement FOR select_or»
select
«EXPAND Alternative FOREACH
this.ref.reject(e|e.metaType==ada::terminate_alternative) SEPARATOR 'or'-
»
«IF !this.ref.select(e|e.metaType==ada::terminate_alternative).isEmpty»
or
«ENDIF»
«EXPAND Alternative FOREACH
this.ref.select(e|e.metaType==ada::terminate_alternative) SEPARATOR 'or'-
»
end select;
«ENDDEFINE»

«DEFINE statement FOR case_statement»
case «this.exp.e» is
«IF this.ref.notExists(e|e.choice=="others")»
«EXPAND Case FOREACH this.ref.reject(e|e.choice=="others")-»
when others => null;
«ELSE»
«EXPAND Case FOREACH this.ref.reject(e|e.choice=="others")-»
«EXPAND Case FOREACH this.ref.select(e|e.choice=="others")-»
«ENDIF»
end case;
«ENDDEFINE»

«DEFINE statement FOR return_statement»
return «this.exp.e»;
«ENDDEFINE»
```

```
«DEFINE statement FOR exit_statement»
exit;
«ENDDEFINE»

«DEFINE statement FOR procedure_call_statement»
«this.name»;
«ENDDEFINE»

«DEFINE statement FOR entry_call_statement»
«this.entry_name»;
«ENDDEFINE»

«DEFINE statement FOR simple_accept_statement»
accept «this.direct_name»;
«ENDDEFINE»

«DEFINE Alternative FOR select_alternative»
«ENDDEFINE»

«DEFINE Alternative FOR terminate_alternative»
terminate;
«ENDDEFINE»

«DEFINE Alternative FOR accept_alternative»
accept «this.as.direct_name»;
«EXPAND statement FOREACH this.s-»
«ENDDEFINE»

«DEFINE Case FOR case_statement_alternative»
when «this.choice» => «EXPAND statement FOREACH this.s-»
«ENDDEFINE»
```

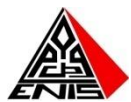

# Une approche IDM de transformation exogène de Wright vers Ada

## Farah FOURATI

**Résumé** : L'ADL formel Wright permet de décrire les aspects structuraux et comportementaux d'une architecture logicielle abstraite. Les aspects comportementaux sont décrits en CSP et vérifiés avec le model-checker FDR. Alors que, l'ADL Wright n'offre aucun moyen qui permet de concrétiser de telles architectures abstraites. L'objectif de ce mémoire est d'ouvrir l'ADL Wright sur Ada en suivant une approche d'automatisation de type IDM. Pour y parvenir, nous avons élaboré deux méta-modèles en Ecore : le méta-modèle Wright et le méta-modèle partiel d'Ada. De plus, nous avons conçu, réalisé et testé notre programme Wright2Ada en ATL permettant de transformer une architecture logicielle décrite en Wright vers un programme concurrent en Ada.

**Mots clés**: IDM, Opération de transformation exogène, ATL, Injection, Xtext, Wright, Extraction, Xpand, Ada et Check.

**Abstract:** The formal ADL Wright allows to describe the structural and behavioral aspects of abstract software architecture. The behavioral aspects are described in CSP and checked using the model-checker FDR. While the ADL Wright does not offer any means to realize such architectures abstract. The objective of this work is to open up the ADL Wright for Ada through an automated approach based on MDE. To achieve this, we have developed two Ecore meta-models: the meta-model Wright and the partial meta-model of Ada. Moreover, we have designed, built and tested our program Wright2Ada written in ATL to transform software architecture described in Wright to a concurrent program in Ada.

**Key-words**: MDE, Exogenous transformation, ATL, Injection, Xtext, Wright, Extraction, Xpand, Ada and Check.